\documentclass[12pt,a4paper,american]{article}

\usepackage[latin9]{inputenc}
\usepackage{geometry}
\usepackage{color}
\usepackage{babel}
\usepackage{algorithm2e}
\usepackage{amsmath}
\usepackage{amsthm}
\usepackage{setspace}
\usepackage[export]{adjustbox}
\usepackage[colorinlistoftodos,prependcaption,textsize=tiny]{todonotes}
\usepackage{float}
\usepackage{graphicx}
\usepackage{epstopdf}
\usepackage{epsfig}
\usepackage[section]{placeins}
\usepackage{amsfonts}
\usepackage{natbib}
\usepackage{microtype, booktabs}
\usepackage{lscape}
\usepackage{pdfpages}
\usepackage{bm}
\usepackage[framemethod=tikz]{mdframed}
\usepackage{authblk}

\usepackage{geometry}

\usepackage[%
    font={small,sf},
    labelfont=bf,
    format=hang,
    format=plain,
    margin=0pt,
    width=0.8\textwidth,
]{caption}
\usepackage[list=true]{subcaption}

\definecolor{winered}{rgb}{0.5,0,0}
\usepackage[bookmarks=true, bookmarksnumbered=true, allbordercolors={1 1 1}]{hyperref}
\hypersetup{
  colorlinks   = true, 
  urlcolor     = blue, 
  linkcolor    = blue, 
  citecolor    = winered,
}
\makeatletter

\newcommand\Autoref[1]{\@first@ref#1,@}
\def\@throw@dot#1.#2@{#1}
\def\@set@refname#1{
    \edef\@tmp{\getrefbykeydefault{#1}{anchor}{}}%
    \xdef\@tmp{\expandafter\@throw@dot\@tmp.@}%
    \ltx@IfUndefined{\@tmp autorefnameplural}%
         {\def\@refname{\@nameuse{\@tmp autorefname}s}}%
         {\def\@refname{\@nameuse{\@tmp autorefnameplural}}}%
}
\def\@first@ref#1,#2{%
  \ifx#2@\autoref{#1}\let\@nextref\@gobble
  \else%
    \@set@refname{#1}
    \@refname~\ref{#1}
    \let\@nextref\@next@ref
  \fi%
  \@nextref#2%
}
\def\@next@ref#1,#2{%
   \ifx#2@ and~\ref{#1}\let\@nextref\@gobble
   \else, \ref{#1}
   \fi%
   \@nextref#2%
}

 \theoremstyle{definition}
 
  \theoremstyle{plain}
  
  \theoremstyle{plain}

\usepackage{babel}

  \addto\captionsamerican{\renewcommand{\definitionname}{Definition}}
  \addto\captionsamerican{\renewcommand{\lemmaname}{Lemma}}
  \addto\captionsamerican{\renewcommand{\propositionname}{Proposition}}
  \addto\captionsenglish{\renewcommand{\definitionname}{Definition}}
  \addto\captionsenglish{\renewcommand{\lemmaname}{Lemma}}
  \addto\captionsenglish{\renewcommand{\propositionname}{Proposition}}
  \providecommand{\definitionname}{Definition}
  \providecommand{\lemmaname}{Lemma}
  \providecommand{\propositionname}{Proposition}
 \addto\captionsamerican{\renewcommand{\lemmaname}{Lemma}}
 \addto\captionsamerican{\renewcommand{\propositionname}{Proposition}}
  \addto\captionsenglish{\renewcommand{\lemmaname}{Lemma}}
  \addto\captionsenglish{\renewcommand{\propositionname}{Proposition}}
  \providecommand{\lemmaname}{Lemma}
  \providecommand{\propositionname}{Proposition}

  \addto\captionsamerican{\renewcommand{\lemmaname}{Lemma}}
  \addto\captionsamerican{\renewcommand{\propositionname}{Proposition}}
  \addto\captionsenglish{\renewcommand{\lemmaname}{Lemma}}
  \addto\captionsenglish{\renewcommand{\propositionname}{Proposition}}
  \providecommand{\lemmaname}{Lemma}
  \providecommand{\propositionname}{Proposition}

  \providecommand{\lemmaname}{Lemma}
  \providecommand{\propositionname}{Proposition}

\makeatother

  \providecommand{\lemmaname}{Lemma}
  \providecommand{\propositionname}{Proposition}

\begin{document}

\title{\textbf{Agreed and Disagreed Uncertainty}\footnote{We would like to thank Philippe Andrade, George-Marios Angeletos, Fabio Canova, Joshua Chan, Eric Eisenstat, Laura G\'{a}ti, Yuriy Gorodnichenko, Refet G\"{u}rkaynak, Luigi Iovino, George Kapetanios, Riccardo Masolo, Franck Portier, Barbara Rossi, Ozge Senay, Ulf S\"{o}derstr\"{o}m, participants to the Bank of Finland ``2018 Workshop on Empirical Macroeconomics,'' ``2022 Workshop on Heterogeneity'' at the University of Exeter, and seminar participants at the Banque de France, EDHEC Business School, European Central Bank, King's College London, Shanghai University of Finance and Economics (SUFE), Sveriges Riksbank, Sveriges Riksbank, and Universities of Birmingham, Glasgow, Oxford and St. Andrews for valuable comments.}\vspace{0.5cm}}
\author[1,2]{Luca Gambetti}
\author[3]{Dimitris Korobilis}
\author[3]{John D. Tsoukalas}
\author[4]{Francesco Zanetti}
\affil[1]{{\footnotesize Universitat Aut\`{o}noma de Barcelona, 08193 Bellaterra, Barcelona, Spain}}
\affil[2]{{\footnotesize Universit\'{a} di Torino, 10134, Torino, Italy}}
\affil[3]{{\footnotesize University of Glasgow, Adam Smith Business School, Glasgow, G12 8QQ, U.K.}}
\affil[4]{{\footnotesize University of Oxford, Department of Economics, Oxford, OX1 3UQ, U.K.}}
\date{February 2023}
\maketitle

\begin{abstract}


\noindent When agents' information is imperfect and dispersed, existing measures of macroeconomic uncertainty based on the forecast error variance have two distinct drivers: the variance of the economic shock and the variance of the information dispersion. The former driver increases uncertainty and reduces agents' disagreement (agreed uncertainty). The latter increases both uncertainty and disagreement (disagreed uncertainty). We use these implications to identify empirically the effects of agreed and disagreed uncertainty shocks, based on a novel measure of consumer disagreement derived from survey expectations. Disagreed uncertainty has no discernible economic effects and is benign for economic activity, but agreed uncertainty exerts significant depressing effects on a broad spectrum of macroeconomic indicators.

\vspace{1.5cm}

\newline
\bigskip


\end{abstract}

\thispagestyle{empty}

\newpage


\setcounter{page}{1}
\onehalfspacing

\newpage

\section{Introduction}
Since the seminal work of \cite{bloom_ECTA}, a large body of research shows that uncertainty has powerful recessionary effects on a broad spectrum of activity indicators.\footnote{See \cite{bloom_jep14} for a survey.}  However, although recessions do coincide with heightened uncertainty, protracted and elevated uncertainty is not always associated with recessions. Examples include the stock market crash of October 1987, which resulted in enormous losses in stock returns, and the 2011 debt-ceiling crisis, which resulted in increase in U.S. government credit default swaps of 46 basis points without generating a contraction in real activity.\footnote{In November 1998, the Russian crisis and near collapse of the hedge fund LTCM lead various uncertainty proxies to spike above their levels in the 2001 recession without a concomitant slowdown in economic activity.} This paper argues that the dispersion in consumer views about the state of the economy (thereafter consumer disagreement) conveys important information about the systematic effect of uncertainty on economic activity. By developing a new index of consumer disagreement about current and future economic conditions from survey data, we show that spikes in uncertainty during periods of high consumer agreement (``agreed uncertainty'') have the standard depressing effects on activity indicators found in numerous studies (e.g., \citealp{bloom_ECTA}; \citealp{Jurado_etAl_AER2015}; \citealp{GilchristpaperEER}); however, equivalent spikes in uncertainty in periods of high consumer disagreement (``disagreed uncertainty''), have no discernible effects on economic activity.

The starting point of our analysis is a dispersed (and noisy) information framework. \cite{MR2002}, \cite{W2003}, \cite{S2003}, \cite{MW2009}, \cite{MRW2004}, and \cite{Okuda_2021} argue in favour of information frictions manifested in models of sticky and noisy information.  \cite{CG2012} and \cite{CGb2015} (see \citealp{CGK2018a} for a comprehensive survey) establish robust evidence in favor of information rigidities in agents' expectation formation, with the bulk of evidence supporting noisy information models.  This framework allows us to formalize the distinction between agreed and disagreed uncertainty. In this framework, the observed uncertainty -- measured by the conditional volatility of the forecast error -- is a function of both innovations in the \textit{volatility of fundamental disturbances} and innovations in the \textit{volatility of idiosyncratic noise}, inherent in the noisy signals that agents process. This finding opens up the possibility that innovations in both types of volatility may drive changes in the observed uncertainty.  More precisely, the premise of our study is that innovations in the volatility of idiosyncratic noise may increase measured uncertainty, without a change in the volatility of exogenous fundamental disturbances, and the spike in uncertainty may not necessarily exert depressing effects on economic activity.

We provide new evidence from the Michigan Survey of Consumers on the prevalence of dispersion of information manifested in our new index that captures the disparity of consumers' opinions about current and future economic conditions.\footnote{The empirical literature cited above on information frictions focuses primarily on consumer inflation expectations.} We document several new facts. First, consumer disagreement is pervasive and applies to both \emph{current} and \emph{future} economic conditions. Second, it is pro-cyclical and negatively correlated with widely used measures of economic uncertainty. Third, the procyclicality of disagreement is time-varying: it increases in recessions and weakens in periods of robust economic activity. This result evinces the widening in the dispersion of consumer views during economic expansions that diminishes during recessions, leading to more homogeneous views concomitant to the decline in economic activity. In other words, consumers disagree less strongly about current and future conditions (low disagreement) when then economy is in a recession.

The core analysis proceeds in two steps. First, we develop a simple model with noisy and dispersed information that sheds light on the interplay between disagreement and uncertainty. We then use the predictions of the model to formulate simple sign restrictions in a Bayesian VAR model to identify shocks to agreed and disagreed uncertainty in the data. The empirical analysis based on our novel index of consumer disagreement sheds lights on important differences in the effects of agreed and disagreed uncertainty shocks in the data.

In our simple model, agents receive idiosyncratic signals about a fundamental shock and form forecasts about the path of the latter by solving a signal extraction problem.  The resulting dispersion of forecasts is proportional to the variance of the noise in the signal. We derive measures of uncertainty and disagreement in the model that are consistent with their empirical counterparts.  Specifically, uncertainty in the model is the variance of the forecast errors made by the agents (the standard measure of uncertainty in \citealp{Jurado_etAl_AER2015} and several other studies). We show that the measure of uncertainty in the model is an increasing function of both the variance of the fundamental shock and the variance of the idiosyncratic noise.\footnote{The variance of the idiosyncratic shock affects uncertainty in the model because the forecast error itself is a function of a key parameter -- the signal-to-noise ratio -- that controls the updating of forecasts agents make. The latter is inversely related to the variance of idiosyncratic noise.}  At the same time, the model disagreement index is an increasing function of the variance of noise, but a decreasing function of the variance of the fundamental shock. Therefore, an increase in either the variance of fundamental shock or the variance of the noise can increase uncertainty, but with \emph{opposite} shifts in the index of disagreement. More concretely, a rise in the variance of the fundamental shock increases uncertainty and \emph{decreases} the index of disagreement, but a rise in the variance of the noise increases both uncertainty and the index of disagreement. Thus, although uncertainty always raises when the variances of the fundamental shock or the noise rise, the opposite response of the index of disagreement allows us to identify shocks to agreed and disagreed uncertainty. These distinct predictions provide a set of minimal sign restrictions we use in a medium-scale VAR model, with U.S. monthly data from 1977 to 2020, to estimate the dynamic effects of agreed and disagreed uncertainty shocks.

The baseline empirical results can be summarized as follows. \emph{Agreed uncertainty} shocks, identified by a concomitant fall in disagreement and rise in uncertainty indicators, generates large, protracted, contractionary economic effects consistent with the standard negative impact of economic uncertainty on real activity, as reported in seminal studies by \cite{bloom_ECTA}, \cite{Jurado_etAl_AER2015}, and \cite{Ludvigson_etAl_AEJ}. Specifically, a positive innovation in agreed uncertainty is associated with large and  persistent declines in industrial production, and employment. By contrast, \emph{disagreed uncertainty} shocks identified by the joint increase in disagreement and uncertainty indicators exhibit qualitatively different dynamic effects.  Although the rise in uncertainty is strong, significant, and persistent, as in the case of agreed uncertainty shocks, economic activity indicators do not exhibit any depressing effects.  A positive innovation in disagreed uncertainty generates a short-lived positive response of industrial production and employment, after which both activity indicators return to the pre-shock level.  Finally, the contrasting dynamic effects of agreed and disagreed uncertainty shocks are robust to using various measures of consumer disagreement, uncertainty, and they are robust to VAR models that encompass a broader spectrum of macroeconomic activity indicators, as well as to VAR models that distinguish consumer disagreement by education and age cohorts.

These empirical findings contribute to the growing literature on the macroeconomic effects of economic uncertainty, and we are the first study to link uncertainty and consumer disagreement. Our evidence sheds light on a new channel in the propagation of uncertainty to economic activity, showing that high consumer disagreement is a relevant indicator for the dampened effect of uncertainty in the economy. \cite{bloom_ECTA}, \cite{Jurado_etAl_AER2015}, and \cite{Ludvigson_etAl_AEJ} show that uncertainty shocks are strongly contractionary on economic activity. \cite{BakerBloomDavis16} develop an index of economic policy uncertainty and show that innovations in policy uncertainty exert a negative effect on employment, industrial production, and investment. \cite{Bachmann_al_AEJM13} use uncertainty measures from U.S. and German business survey data and find a significant negative effect of uncertainty in production and employment. \cite{GilchristpaperEER} and \cite{Haroon_JME2019}, stress the interaction between financial conditions and uncertainty, providing evidence that the negative impact of uncertainty shocks is amplified when financial conditions worsen. \cite{FV_GQ_RR_U_AER2011}, \cite{FV_GQ_K_RR_AER2015}, \cite{JFV-fiscal,JFV-Matt}, \cite{JFV-technology}, \cite{mumtaz2013impact}, \cite{BORN201468}, \cite{Basu_Bundick_ECTA2017}, \cite{Cascaldi-Garcia_JEL2021}, \cite{melosi-et-al-2022}, and several others show that uncertainty from different sources, such as fiscal and monetary policy, costs of borrowing, and future perceived uncertainty, results in reduced economic activity. We also relate to \cite{caggiano2014uncertainty}, \cite{leduc2016uncertainty}, \cite{theodoridis2016news}, \cite{schaal2017uncertainty}, and \cite{cascaldi2021news} which show a tight link between uncertainty, labor, and production markets.  Earlier literature studies the cyclical effects of \textit{first moment} noise shocks (e.g., \citealp{Lorenzoni_AER09}; \citealp{Blanchard_et_al_AER13}; \citealp{forni_al_aejm17}). Our work contributes to this literature by identifying potential cyclical effects of second-moment noise shocks.
Recent work also shows that episodes of high uncertainty may not have adverse economic effects. For example, \cite{Segal15} distinguish between bad and good uncertainty, and their good uncertainty measure is benign for production and consumption.\footnote{Using measures of low and high uncertainty from quantile factor models, \cite{KorobilisSchroeder2022} show that only high-uncertainty shocks cause a significant fall in industrial production. \cite{Aastveit2017} show that in periods of high uncertainty, monetary policy effects to output are dampened.} \cite{Bergeretal2020} separate contemporaneous shocks in realized stock market volatility from news shocks that they interpret as forward-looking uncertainty, which are benign for economic activity. 

The remainder of the paper is organized as follows. Section \ref{Sec1_Measure} derives our measures of consumer disagreement and studies the time-series properties of our index. Section \ref{sec_model_theory} develops a stylized model to study the links between uncertainty and consumer disagreement. Section \ref{sec_model_empirical} uses predictions  from the model that disentangle the dynamic effects of agreed and disagreed uncertainty.  Section \ref{sec_robustness} explores robustness of the empirical results to alternative modeling assumptions. Section \ref{Sec_Conclusion} concludes the paper.

\section{Measuring consumer disagreement \label{Sec1_Measure}}
In this section we construct a new index of consumer disagreement using the University of Michigan Survey of Consumers. It is a parsimonious index that encapsulates the cross-sectional dispersion of consumer views from different survey questions, and it reveals consumers' information and beliefs on current and future economic conditions.  We then study the cyclical properties of our disagreement index and focus on the link with economic activity and alternative measures of uncertainty.

\subsection{Consumer survey data}
The Michigan Survey of Consumers (hereafter MSC), is produced by the Survey Research Center at the University of Michigan. Each month, it conducts a minimum of 500 interviews, and consumers answer a questionnaire that contains 28 core questions and several subquestions. Survey questions are aggregated over respondents (consumers) to produce approximately 45 monthly and quarterly categorical time series.\footnote{The only exception is the question that asks consumers to forecast a value for inflation one year and five years ahead, which results in a continuous variable. The samples for the Surveys of Consumers are statistically designed to be representative of all American households. For a detailed description of the survey, including questionnaires, see: \texttt{https://data.sca.isr.umich.edu/survey-info.php}.}  To formulate our index, we select questions that capture the views of consumers about current and future economic conditions, summarized in \autoref{Table:MSC data}.

\begin{table}[H]
\centering
\caption{Questions from the Michigan Survey of Consumers} \label{Table:MSC data}
\begin{tabular}{lll} \hline
\textsc{Question} & \textsc{Mnemonic} & \textsc{Topic} \\
Q23 & NEWS & News Heard of Recent Changes in Business Conditions    \\
Q25 & BAGO & Current Business Conditions Compared with a Year Ago    \\
Q26 & BEXP & Expected Change in Business Conditions in a Year    \\
Q28 & BUS12 & Business Conditions Expected During the Next Year    \\
Q29 & BUS5 & Business Conditions Expected During the Next 5 Years    \\ \hline
\end{tabular}
\end{table}

Consumer responses to the survey questions consist of three qualitative categories (``better/about the same/worse,''); the associated time-series measures the proportion of respondents in each category.\footnote{Depending on the question, these answers can also take the form ``favorable/no mention/unfavorable,'' ``good time/uncertain/bad
time,'' or ``more/about the same/less.''} Our benchmark measure is an index of \emph{tail disagreement}, which reflects disagreement between the two polar categories in the distribution of responses. That is, the \emph{tail disagreement} index extracts disagreement from the ``better/worse'' (or ``good time/bad time,'' or ``favorable/unfavorable'') responses. Formally, the definition of the disagreement index is:
\begin{eqnarray}
T_{t}^{(j)} = 1 - \frac{\vert b_{t}^{(j)}-w_{t}^{(j)} \vert }{100}, \label{index_tail}
\end{eqnarray}
where $j=\text{NEWS, BAGO, BEXP, BUS12, BUS5}$ indexes each of the five survey questions, $b_t^j$ is the percentage of respondents in question
$j$ with a positive/optimistic answer, and $w_t^j$ is the percentage of respondents with a negative/pessimistic answer. The disagreement index $T_{t}^{(j)}$ takes values of 0 and 1 by construction. A value equal to zero, which occurs if either $b_{t}^{(j)}$ or $w_{t}^{(j)}$ is equal to 100, indicates all respondents have the same opinion or view about the current and future economic outlook and therefore no disagreement. On the other hand, a value equal to 1 indicates that consumers are evenly split between the two polar responses, reflecting sharp differences in opinions or views and consequently maximal disagreement. This indicator is intuitive but ignores information from the middle category of responses (e.g., ``no mention,'' ``same''). In section \ref{sec_robustness} we compute the Shannons' entropy (\citealp{Shannon48}) measure of disagreement, which considers both the polar and ``middle'' category responses. The entropy can be a measure of uncertainty: consumers are more uncertain about economic conditions when the ``middle'' category has a non zero chance of occurring.  We show that results are robust to this consideration. It is important to stress that the qualitative approach in the reporting of views suggests our measure of consumer disagreement refers to what we can loosely call ``directional'' disagreement. Thus, our concept of disagreement is different from disagreement among professional forecasters. In other words, our index does not convey information about the intensity of the responses (e.g., how much better relative to how much worse). The index also cannot capture disagreement within the proportion of consumers that report better (or worse) economic prospects.

\subsection{Time-series properties of consumer disagreement}
We use monthly data spanning the period 1978M1 to 2020M12, and we derive distinct measures of disagreement by applying the formula in equation (\ref{index_tail}) to each of the five survey questions. We denote the singular disagreement measures related to each survey question in Table \autoref{Table:MSC data} by $T^{NEWS}$, $T^{BAGO}$, $T^{BEXP}$, $T^{BUS12}$, and $T^{BUS5}$. The measures of disagreement based on the mnemonics ``NEWS'' and ``BAGO'' in \autoref{Table:MSC data} (i.e., $T^{NEWS}$ and $T^{BAGO}$, respectively) refer to \emph{current} business conditions and thus directly relate to the information that consumers receive and process about the past and present economic conditions. If all agents could perfectly access all information relevant for assessing current conditions, the degree of disagreement on \emph{past and present} conditions would be absent. The degree of disagreement on \emph{future} economic conditions will still be present, as agents need to make forecasts conditional on potentially different models of the economy. Thus, a good check to ascertain the degree of information dispersion is to focus on disagreement about \emph{current} economic conditions that would be absent if agents have full information on the state of the economy. Our indices $T^{NEWS}$ and $T^{BAGO}$ record substantial disagreement on present and past economic conditions, suggesting substantial disparity of views across consumers and evincing imperfect information about the state of the economy.

To develop a parsimonious indicator of disagreement, we summarize the information in the five different measures by formulating a single, latent, consumer disagreement index using principal component analysis. In line with the literature on macroeconomic diffusion indexes \citep[see for example,][]{StockWatson2002}, our latent index is the first principal component of the five individual disagreement series. The first principal component is a weighted average of all five series, where the weights (loadings) are such that the latent index maximizes the variance explained for each series.\footnote{In order to ensure that the first principal component describes the direction of maximum variance, we standardize the individual disagreement measures (and the index) to have a mean equal to zero and a variance equal to 1. This transformation does not affect the informational content of each series; rather, it affects the scale. However, disagreement as a concept is not an ordinal measure in the sense that an index value of, say, 0.5 implies that consumers disagree ``twice as much'' compared to a value of 0.25. For that reason, we prefer to work with an index that is standardized.} We refer to our latent index as DISAG, and we use it as the benchmark measure of consumer disagreement for the rest of the analysis.

\begin{figure}[h!]
\caption{Measures of disagreement and loading factors}
\label{fig:factor_vs_individual}
\includegraphics[width=\textwidth, trim={4cm 1cm 3cm 0.7cm}]{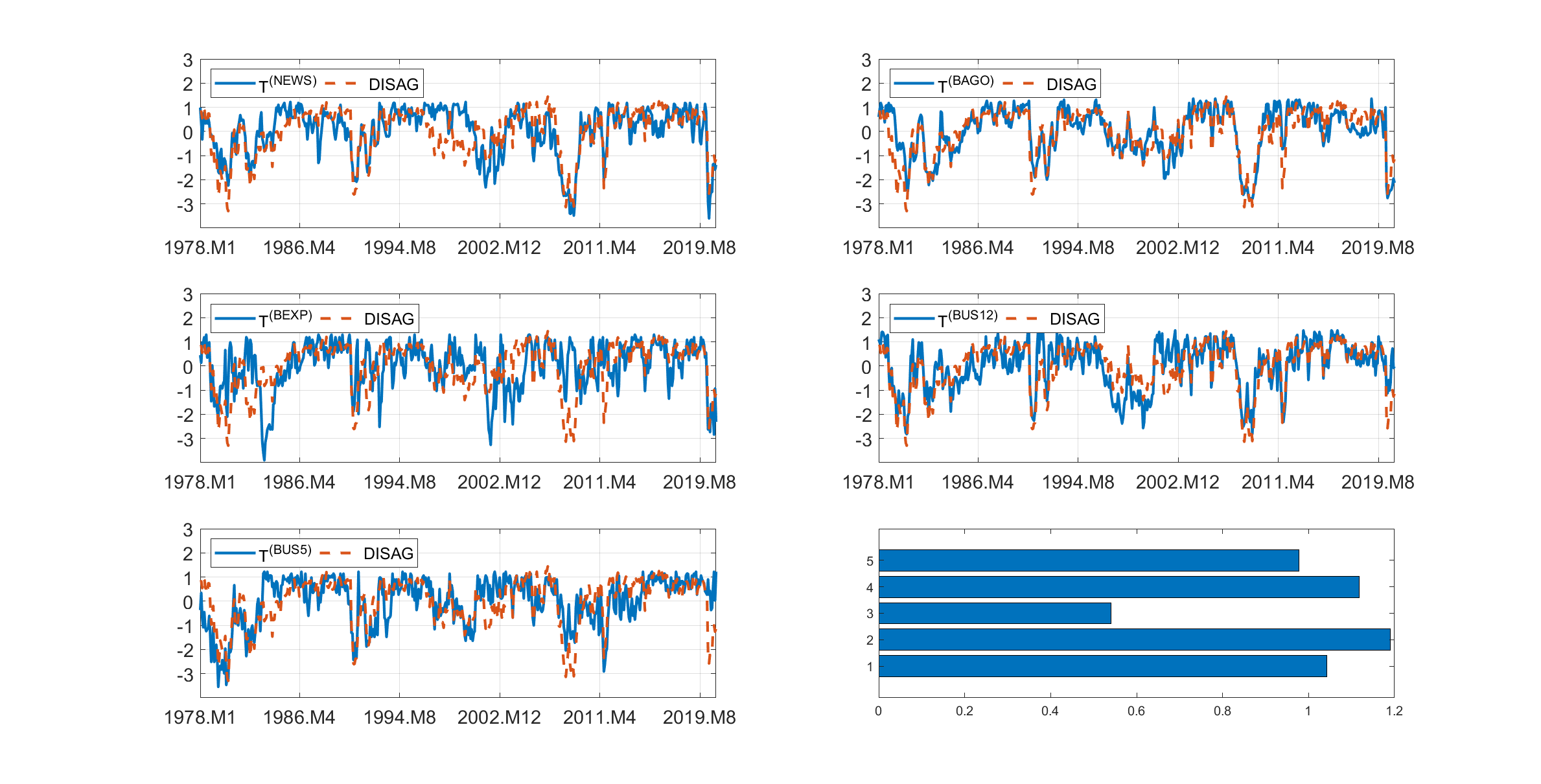}
{\small \emph{Notes: The top four panels and
bottom left panel show individual tails disagreement measures using the Michigan survey questions (solid line) against the aggregate index of consumer disagreement (variable DISAG, dashed line). The bottom right panel shows the estimated weights of each individual series on the principal component. All series are standardized to have a mean of zero and variance of 1.}}
\end{figure}

The top four panels and the bottom left panel of \autoref{fig:factor_vs_individual} show the estimate of the disagreement index (DISAG) against the individual measures of disagreement. A first finding is the large and significant time variation in the disagreement index that also characterizes the individual disagreement series.\footnote{The variability in consumer disagreement remains broadly unchanged across the
full sample period, without displaying a reduction in volatility during the Great Moderation period of 1984-2007 that
characterizes several macroeconomic activity indicators. See \cite{Liu_JBES18} for a discussion of the changes in time-series properties of macroeconomic variables since the 1960s.} The figure shows that the comovement of the disagreement index with each individual series is high. The bottom right panel of \autoref{fig:factor_vs_individual} shows the loadings of each series on the principal component. The values in the figure are the weights with which each individual series contributes to the estimate of our latent disagreement index. The values show that the DISAG index is evenly and strongly correlated with the individual disagreement indexes NEWS, BAGO, BUS12, BUS5, and it is less strongly correlated with the BEXP measure of disagreement.

\begin{figure}[h!]
\caption{Time-varying correlations of DISAG with industrial production and real personal consumption growth}\label{fig:factor_tv_corr}
\includegraphics[width=\textwidth, trim={4cm .5cm 4cm .4cm}]{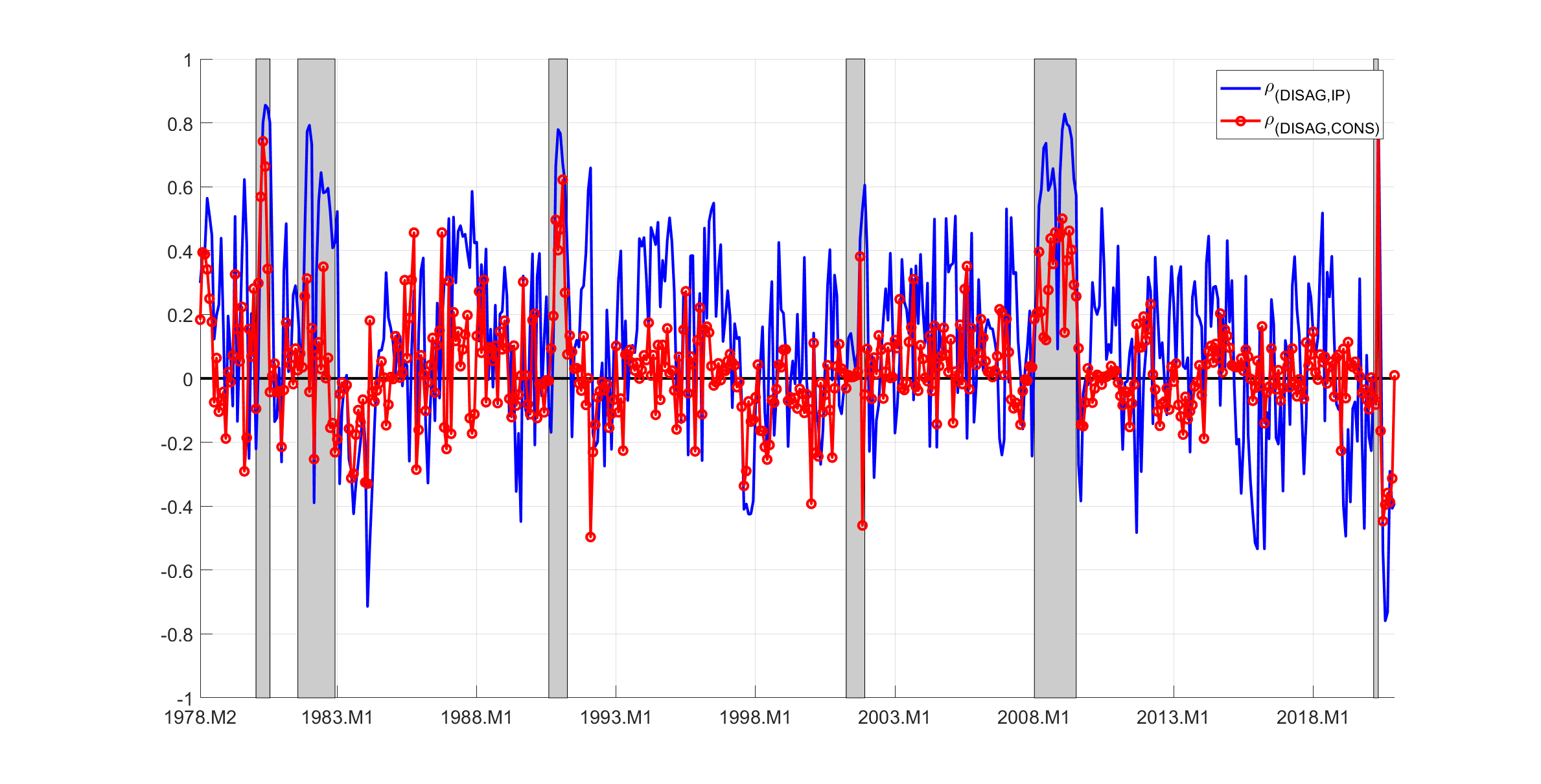}
\small \emph{Notes: IP-Industrial production growth. CONS-real personal consumption growth. The correlation estimates are from a multivariate GARCH model (see \autoref{foot:multi_GARCH} for details).}
\end{figure}

We proceed to study the cyclical properties of DISAG by focusing on the comovement of the disagreement index with two representative measures of economic activity: the monthly Industrial Production Index and Real Personal Consumption Expenditure. Over the entire sample period, DISAG is very weakly correlated with either industrial production (0.19) and real personal consumption growth (0.04); however, the correlation displays significant time variation. \autoref{fig:factor_tv_corr} shows a rolling correlation between the disagreement index, industrial production, and real personal consumption growth: It demonstrates that the correlation coefficient is time varying, covering a wide range of values between -0.65 to 0.85 over the sample period.\footnote{The conditional correlation is obtained from a trivariate BEKK-GARCH(1,1) specification on disagreement and the growth rates of industrial production and consumption. We use Kevin Sheppard's MFE Toolbox for MATLAB (\href{https://www.kevinsheppard.com/code/matlab/mfe-toolbox/}{https://www.kevinsheppard.com/code/matlab/mfe-toolbox/} for the estimation of the BEKK-GARCH(1,1)). The results are qualitatively similar when estimating other multivariate GARCH models, such as the so-called CCC and DCC models. The multivariate GARCH approach is superior to estimating sample correlations in rolling windows of the data sample, because the latter approach discards valuable information in the data and the former uses information in all of the sample when estimating time-varying correlations.\label{foot:multi_GARCH}} Starting from the 1981 recession, the correlation between disagreement and either measure of real activity becomes predominantly positive and peaks during the five subsequent recessions (shown in shaded areas). In other words, the positive correlation between disagreement with economic activity indicators increases significantly during recessions, implying that disagreement falls sharply with declines in real activity.

\begin{figure}[h!]
\begin{center}
\caption{Index of consumer disagreement and uncertainty indicators}\label{fig_uncert_disag}
\includegraphics[width=\textwidth, trim={4cm .5cm 4cm .4cm}]{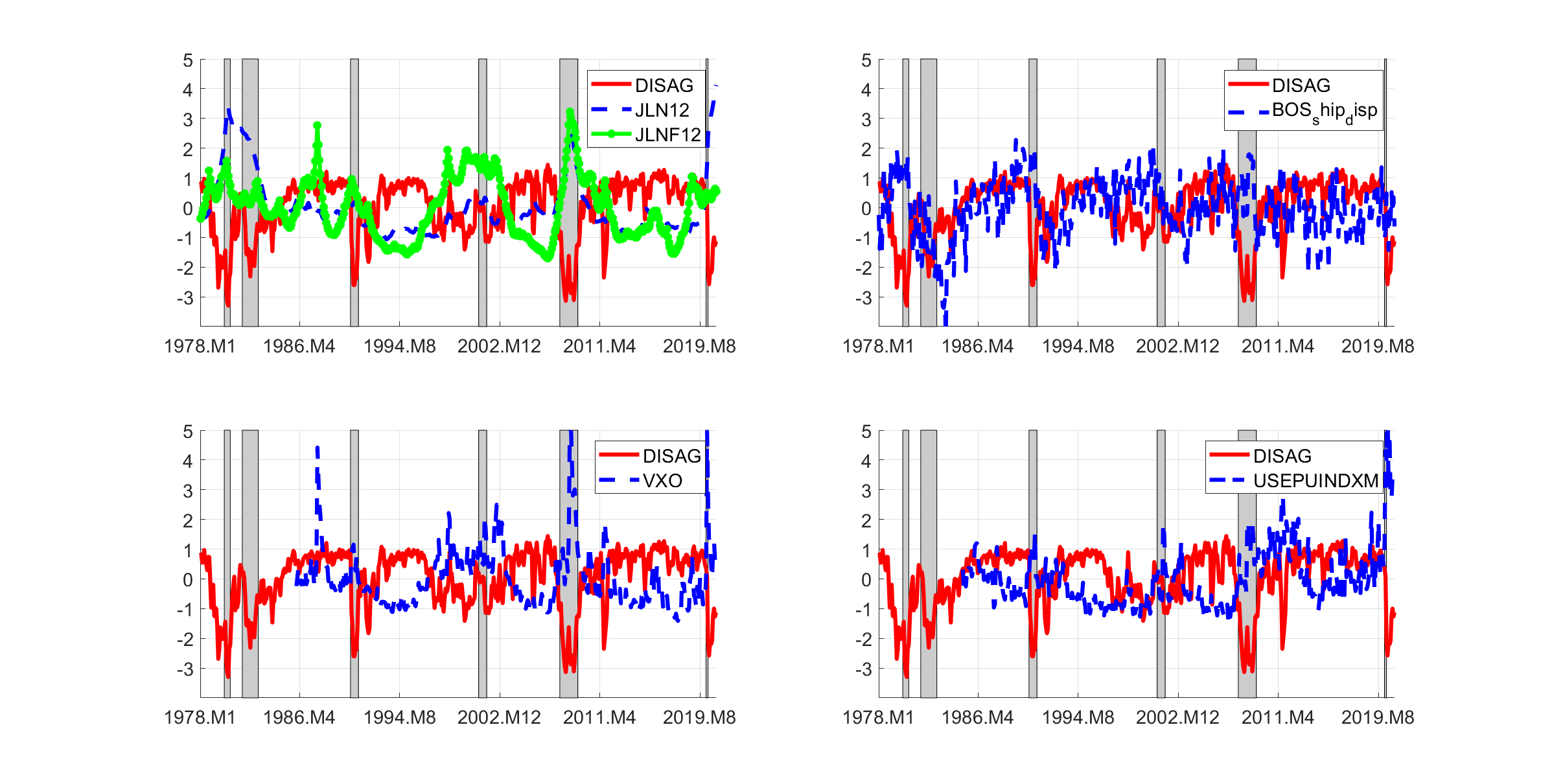}
\end{center}
\small \emph{Notes: The figure plots in a clockwise manner a) consumer disagreement (DISAG) (solid red line) against the 12-month macroeconomic (JLN12) and financial (JLNF12) uncertainty indicators from \cite{Jurado_etAl_AER2015}, b) DISAG (solid red line) against the Business Forecast Dispersion Index (BOS) from \cite{Bachmann_al_AEJM13} (updated to 2019 by the authors), c) DISAG (solid red line) against stock market volatility (CBOE S\&P 100 volatility index (VXO)), and d) DISAG (solid red line) against Economic Policy Uncertainty (EPU) developed by \cite{BakerBloomDavis16}. The series are standardized to have mean zero and unitary variance.}
\end{figure}



We next compare the disagreement index with empirical measures of uncertainty and measures of disagreement derived from business surveys.  \cite{Jurado_etAl_AER2015} develop uncertainty indicators from a large set of macroeconomic and financial time-series data using factor-augmented VAR methods.  The top left panel of \autoref{fig_uncert_disag} displays our disagreement index (solid line) together with the \cite{Jurado_etAl_AER2015} measures of macroeconomic and financial uncertainty (JLN12 and JLNF12, respectively) obtained from a 12-month forecast horizon (dotted and dashed line). The uncertainty indicators are highly countercyclical and exhibit a strong negative comovement with our index of disagreement; the correlations of JLN12 and JLNF12 with the index of consumer disagreement are -0.62 and -0.57, respectively.


The top right panel of \autoref{fig_uncert_disag} compares our disagreement index with the business-level uncertainty index from the Philadelphia Fed Business Outlook Survey (BOS) that encapsulates the cross-sectional-forecast dispersion about six-month-ahead business activity in the manufacturing sector. \cite{Bachmann_al_AEJM13} shows that this index is a good proxy for uncertainty. The correlation of our index with the business dispersion index exhibits a negative yet weak correlation equal to -0.1.\footnote{The survey question in the FED BOS is: \textit{General Business Conditions: What is your evaluation of the level of general business activity six months from now vs. [CURRENT MONTH]: decrease, no change, increase?} To preserve comparability, we compute the BOS forecast dispersion index identically to \cite{Bachmann_al_AEJM13}.}  The bottom left panel of \autoref{fig_uncert_disag} compares our index with the CBOE S\&P 100 volatility index (VXO) measure, the latter being a measure of uncertainty in many previous studies.  VXO exhibits strong negative comovement with our disagreement index, with a correlation coefficient  equal to -0.55. Last, the bottom right panel of \autoref{fig_uncert_disag} compares our index with the measure of Economic Policy Uncertainty developed by \cite{BakerBloomDavis16}.  As in the case of business dispersion, this indicator, capturing a different dimension of uncertainty, is not strongly negatively correlated with our disagreement indicator with a  correlation coefficient equal to -0.36. The key finding from these comparisons is the \emph{negative} comovement of the different uncertainty indicators with consumer disagreement.  In sum, consumer disagreement has fundamentally different cyclical properties compared to indicators of business-level uncertainty, stock market volatility, or uncertainty indicators from forecasts of financial and macroeconomic indicators and economic policy uncertainty.

%


\section{A simple model of information dispersion\label{sec_model_theory}}


We develop a simple model with disagreement arising from imperfect and dispersed information, and uncertainty stemming from changes in the variance of a fundamental shock. We study the effect of information dispersion and the volatility of the fundamental shock on i) the variance of the forecast errors, the empirical proxy for uncertainty, and ii) the model index of disagreement congruous with our empirical measure of disagreement. The model allows us to separately identify shocks to information dispersion and shocks to uncertainty, and it provides simple sign restrictions that enable us to illustrate how disagreement is associated with the different concepts of uncertainty,  which we use to identify the effect of agreed and disagreed uncertainty in the data (see Section \ref{sec_model_empirical}).

The economy is populated by a continuum, large number of $N$ agents defined over the unit interval, indexed by $i$. In each period $t=1,2,...$, the economy experiences the realization of an exogenous process $a_t$ (expressed in logs) whose growth rate ($\Delta a_{t}= a_{t}-a_{t-1}$) follows the invertible moving average (MA) process:
\begin{eqnarray}
a_t-a_{t-1}=\psi_{0}\varepsilon_{t} + \psi_{1}\varepsilon_{t-1} + \psi_{2}\varepsilon_{t-2} +...+ \psi_{n}\varepsilon_{t-n}, \label{a}
\end{eqnarray}
where, $\psi_{0}$, $\psi_{1}$, ..., $\psi_{n}$ are the MA coefficients and $\varepsilon_t\sim N(0,\sigma_{\varepsilon}^2)$ is an i.i.d. fundamental shock with known variance $\sigma_{\varepsilon}^2$.\footnote{The exogenous fundamental process can adopt a variety of interpretations (e.g., productivity or demand shocks that are relevant sources of macroeconomic fluctuations).}

Information is imperfect and dispersed. It is imperfect because agents cannot observe the current fundamental shock $\varepsilon_{t}$ and the current exogenous process $a_{t}$ during each period $t$, while they observe the history $\varepsilon_{t-1}$, ..., $\varepsilon_{t-n}$, and the past exogenous process $a_{t-1}$.

Information is dispersed because each agent $i$ receives a different idiosyncratic signal about the fundamental shock:
\begin{eqnarray}
s_{it}= \varepsilon_{t}+v_{it},\label{eq_signal}
\end{eqnarray}
where $v_{it}\sim N(0,\sigma_{v_{i}}^2)$ is an idiosyncratic, i.i.d. shock with known variance $\sigma_{v_{i}}^2$. The idiosyncratic shock $v_{it}$ blurs the realization of the fundamental shock and generates cross-sectional dispersion in the signals across agents. This formulation implies an innovation to the volatility of the idiosyncratic shock, $\sigma_{v_{i}}^2$, which leads to a greater dispersion of information across agents.  Agents care about the path of the fundamental shock $\varepsilon_{t}$, and they solve a signal extraction problem to infer the fundamental shock from the signal $s_{i}$. Each agent $i$ solves this problem by conditioning on the history and volatilities as follows: $\mathcal{I}_{it}\equiv \{a_{t-1-j},  \varepsilon_{t-1-j},
s_{it-j}, \sigma_{\varepsilon}^2, \sigma_{v}^2 \}_{j=0}^\infty$, where $\mathcal{I}_{it}$ is the agent specific information set.

Formally, each agent $i$ uses equation (\ref{a}) to form expectations about the growth rate of the exogenous process $a$ in future periods $t+1,...,t+n$, which yields:
\begin{eqnarray}
 E\left(\Delta a_{t+k}|\mathcal{I}_{it}\right) &=& \psi_{k} E\left(\varepsilon_{t}|\mathcal{I}_{it}\right) + \psi_{k+1} \varepsilon_{t-1}+...+\psi_{n} \varepsilon_{t+k-n}, \quad \text{for} \quad k = 1,2,... , \label{eq_delta_a_tn}
\end{eqnarray}
as well as in the current period,
\begin{eqnarray}
 E\left(\Delta a_{t}|\mathcal{I}_{it}\right) &=& \psi_{0} E\left(\varepsilon_{t}|\mathcal{I}_{it}\right) + \psi_{1} \varepsilon_{t-1}+...+\psi_{n} \varepsilon_{t-n}, \quad \label{eq_nowcast1}
\end{eqnarray}
where $E$ is the rational expectations operator, and $E\left(\varepsilon_{t}|\mathcal{I}_{it}\right)$ is the expectation on the current fundamental shock $\varepsilon_{t}$ conditional on the information set $\mathcal{I}_{it}$, which can be represented as the linear projection of $\varepsilon_{t}$ on $s_t$ by solving the signal extraction problem. Equations (\ref{eq_delta_a_tn}) and (\ref{eq_nowcast1}) show that the presence of the idiosyncratic signal generates cross-sectional dispersion on current and future growth expectations of the exogenous process $a$, reflected by the dependency of the conditional expectations $E\left(\varepsilon_{t}|\mathcal{I}_{it}\right)$ on the agent-specific information set. By solving the signal extraction problem for agent $i$, we rewrite equation (\ref{eq_delta_a_tn}) as:
\begin{eqnarray}
  E\left(\Delta a_{t+k}|\mathcal{I}_{it}\right)  &=& \psi_{k} \gamma_{i} s_{it} + \psi_{k+1} \varepsilon_{t-1}+...+\psi_{n} \varepsilon_{t+k-n} \label{eq_delta_a_itn}\\
  &=& \psi_{k} \gamma_{i} (\varepsilon_{t} + v_{it}) + \psi_{k+1} \varepsilon_{t-1}+...+\psi_{n} \varepsilon_{t+k-n} \quad k=1,2,... \nonumber
\end{eqnarray}
and equation (\ref{eq_nowcast1}) as:
\begin{eqnarray}
  E\left(\Delta a_{t}|\mathcal{I}_{it}\right)  &=& \psi_{0} \gamma_{i} s_{it} + \psi_{1} \varepsilon_{t-1}+...+\psi_{n} \varepsilon_{t-n} \label{eq_nowcast2}\\
  &=& \psi_{0} \gamma_{i} (\varepsilon_{t} + v_{it}) + \psi_{1} \varepsilon_{t-1}+...+\psi_{n} \varepsilon_{t-n} \quad k=1,2,... \nonumber
\end{eqnarray}
where
\begin{equation}\label{eq_gamma_it}
  \gamma_{i} = \frac{\sigma_{\varepsilon }^2}{\sigma_{\varepsilon }^2+\sigma_{v_i }^2}
\end{equation}
is the agent-specific linear projection coefficient. Equations (\ref{eq_delta_a_itn}) and (\ref{eq_nowcast2}) show that the future and current expected growth rate of the exogenous process $a$ depends on the agent-specific reaction to the signal, controlled by the coefficient $\gamma_{i}$. The response of these expectations to the signal falls with the dispersion of information encapsulated by the variance of the idiosyncratic shock, $\sigma_{v_i}^2$, and it increases with the variance of the fundamental shock, $\sigma_{\varepsilon }^2$, as implied by equation (\ref{eq_gamma_it}). The dispersion of information decreases the content of information contained in the signal received by each agent $i$, and it makes the conditional expectations in equations (\ref{eq_delta_a_itn}) and (\ref{eq_nowcast2}) less responsive to the signal. In the rest of the analysis, we simplify the analytical derivation of the system without loss of generality by assuming an identical variance of the idiosyncratic shock across agents (i.e., $\sigma_{v_i }^2=\sigma_{v }^2$).

\subsection{Interplay between uncertainty and information dispersion}

In this section, we study the interplay between uncertainty and information dispersion. We proxy uncertainty with the variance of the $k$-periods-ahead forecast errors for $\Delta a_{t+k}$, and disagreement with an index derived from simulations of the model that is consistent with our measure of disagreement in Section \ref{Sec1_Measure}. Our aim is to map the effect of dispersed information and the spread of the fundamental shock into the empirical proxies for disagreement and uncertainty.

As a preliminary step, we derive the $k$-periods-ahead \emph{aggregate expectations} by averaging the different expectations of the single agents in equation (\ref{eq_delta_a_itn}) across the $N$ agents in the economy:
\begin{eqnarray}
E\left(\Delta a_{t+k}|\mathcal{I}_{t}\right)&=&\frac{1}{N}\sum_{i=1}^{N} E\left(\Delta a_{t+k}|\mathcal{I}_{it}\right)\nonumber\\
&=&\psi_{k} \gamma\frac{1}{N}\sum_{i=1}^{N}(\varepsilon_{t}+v_{it})  + \frac{1}{n}\sum_{i=1}^{n} \left( \psi_{k+1} \varepsilon_{t-1}+...+\psi_{n} \varepsilon_{t+k-n} \right )\nonumber\\
&=& \psi_{k} \gamma \varepsilon_{t} + \frac{1}{N}\sum_{i=1}^{N} \left ( \psi_{k+1} \varepsilon_{t-1}+...+\psi_{n} \varepsilon_{t+k-n} \right ), \label{eq_agg_delta_a_tn}
\end{eqnarray}
where $\frac{1}{N}\sum_{i=1}^{N} v_{it}$ converges to zero by the law of large numbers, and the average projection coefficient is equal to:
\begin{eqnarray}
\gamma =\frac{\sigma_{\varepsilon }^2}{\sigma_{\varepsilon }^2+\sigma_{v}^2}.\label{projection_aggr}
\end{eqnarray}

\subsubsection{Variance of forecast error}

Our proxy for uncertainty is the variance of the forecast error $k$-periods ahead, which is equal to:
\begin{eqnarray}
  var[\Delta a_{t+k} - E(\Delta a_{t+k}|\mathcal{I}_{t})]  &=& var[\psi_{0} \varepsilon_{t+k} + \psi_{1} \varepsilon_{t+k-1} + \psi_{2} \varepsilon_{t+k-2}+...+\psi_{k} (1-\gamma_t) \varepsilon_t] \nonumber \\
  &=&  \left[ \psi_{0}^{2} + \psi_{1}^{2} +  \psi_{2}^{2} + ... + \psi_{k}^{2}\left(\frac{\sigma_{v}^2}{\sigma_{\varepsilon }^2+ \sigma_{t}^2}\right)^2 \right] \sigma_{\varepsilon }^2   \nonumber \\
  & &  + 2 \psi_{0} \sum_{j=1}^{k-1} \psi_{j} cov(\varepsilon_{t+k}, \varepsilon_{t+k-j})  + 2 \psi_{1} \sum_{j=2}^{k-1} \psi_{j} cov(\varepsilon_{t+k-1}, \varepsilon_{t+k-j})+... \nonumber \\
  & & + 2 \psi_{k-2} \sum_{j=1}^{k-1} \psi_{j} cov(\varepsilon_{t+2}, \varepsilon_{t+k-j}) + 2 \psi_{k}(1-\gamma) \sum_{j=0}^{k-1} \psi_{j} cov(\varepsilon_{t}, \varepsilon_{t+j+1})\nonumber \\
   &=&
  \left[ \psi_{0}^{2} + \psi_{1}^{2} +  \psi_{2}^{2} + ... + \psi_{k}^{2}\left(\frac{\sigma_{v}^2}{\sigma_{\varepsilon }^2+ \sigma_{v}^2}\right)^2 \right] \sigma_{\varepsilon }^2,\label{eq_var_forecast}
\end{eqnarray}
where the covariance terms are equal to zero because the shock $\varepsilon_{t}$ is  i.i.d.  Equation (\ref{eq_var_forecast}) shows two important properties of the effect of uncertainty and information dispersion on the variance of the forecast error. First, the effect of a unitary change in uncertainty on the variance of the forecast error at time $t+k$ is equal to:\footnote{Appendix \ref{proof_inequality} shows that, for an invertible MA process, the sign of equation (\ref{eq_uncert_on_FE}) is always positive.}
\begin{equation}\label{eq_uncert_on_FE}
  \frac{\partial var[\Delta a_{t+k} - E(\Delta a_{t+k}|\mathcal{I}_{t})] }{\partial \sigma_{\varepsilon }^2 } = \psi_{0}^{2} + \psi_{1}^{2} +  \psi_{2}^{2} + ... + \psi_{k}^{2}\left(\frac{\sigma_{v}^2}{\sigma_{v }^2+ \sigma_{\varepsilon }^2}\right)\left(\frac{\sigma^2_v - \sigma_{\varepsilon }^2 }{\sigma_{v }^2+ \sigma_{\varepsilon }^2} \right) >0.
\end{equation}


This positive derivative underpins and supports the prevalent adoption of the variance of the forecast error as a proxy for uncertainty.
Second, the variance of the forecast error increases with a unitary change in information dispersion:\footnote{This finding is consistent with the positive relationship between dispersion in beliefs and aggregate uncertainty,
as outlined in \cite{melosi_bianchi_IER2016}.}
\begin{equation}\label{eq_dispers_on_FE}
\frac{\partial var[\Delta a_{t+k} - E(\Delta a_{t+k}|\mathcal{I}_{t})] }{\partial \sigma_{v }^2 } = 2 \psi_0^2 \sigma_{\varepsilon } \left( \frac{\sigma_v^2}{\sigma_{\varepsilon }^2+\sigma_v^2} \right)
\left( \frac{\sigma_{\varepsilon }^2}{(\sigma_{\varepsilon }^2+\sigma_v^2)^2} \right) >0.
\end{equation}

Our model shows that the empirical proxy for uncertainty, measured by the variance of the forecast error, increases in \emph{both} innovations in the variance of the fundamental uncertainty shock, and in the variance of the idiosyncratic noise. These comovements, complemented by two further restrictions we derive in the next section, allow us to disentangle innovations to the variance of fundamental shocks (agreed uncertainty) and innovations to information dispersion (disagreed uncertainty).

\subsubsection{Mapping information dispersion and volatility of fundamental shocks on disagreement}
We use the model to study the mapping from information dispersion to disagreement, and investigate the relation between disagreement and uncertainty. Because we cannot derive an analytical solution that shows the effect of information dispersion and the variance of fundamental shocks on disagreement, we compute the disagreement index from numerical simulations of the model consistent with our empirical index.

In the model, the cross-sectional expectations of consumers about economic conditions are represented by:
\begin{eqnarray}
E(\Delta a_{t+k}|\mathcal{I}_{it})&=&\psi_{k} \gamma_{it} (\varepsilon_{t}+ v_{it}) + \psi_{k+1} \varepsilon_{t-1}+...+\psi_{n} \varepsilon_{t+k-n}. \label{forecastsim}
\end{eqnarray}

In the MSC, the consumer responses to several questions about \emph{future} business conditions is the natural empirical concept corresponding to the forecast captured by equation (\ref{forecastsim}).  We therefore use the equation to generate artificial survey data consistent with the qualitative responses in MSC by defining the following indexes for individual answers:

\begin{equation}
\text{Expected Conditions}: \left\{
\begin{array}{ll}
b_{it}^{\text{Expected}}=1 \;\;\;\mbox{if}\;\;\; E(\Delta a_{t+k}|\mathcal{I}_{it})>0,\\
w_{it}^{\text{Expected}}=-1  \;\;\;\mbox{if}\;\;\;  E(\Delta a_{t+k}|\mathcal{I}_{it})<0.
\end{array}
\right. \label{index_bus12}
\end{equation}


We apply the standard quantification method for qualitative survey data and code with $b_{it}=1$ a positive forecast, and $w_{it}=-1$ a negative forecast. This coding is equivalent to the responses (``better or worse,'' ``good times or bad times'') reported for the survey questions in the MSC summarized in Table \ref{Table:MSC data}. Consistent with the empirical disagreement index described in Section \ref{Sec1_Measure}, we compute the index of tail disagreement as:
\begin{equation}\label{eq_tail_synt}
\tilde T_{t}=1-\frac{1}{N}\left|\sum_{i=1}^n b_{it}-w_{it}\right|
\end{equation}

We simulate the model as follows. We assume a monthly time period consistent with the frequency in the MSC.  We set the order of the MA process that governs $\Delta a_t$ equal to $n=12$, and we set $k=12$, which corresponds to a forecast one year ahead.\footnote{For robustness we repeat this simulation exercise assuming an annual frequency and changing the order of the MA process to equal 5.}  It is well known that the condition for invertibility of an MA process is the counterpart to the stationarity condition for an AR process. Thus, taking a stationary AR(1) process with AR coefficient equal to $\beta$, we write the MA coefficients as $\psi_{k} = \beta^{k}$.  The variances of the idiosyncratic component, $\sigma^{2}_{v}$, and of the fundamental shock, $\sigma^{2}_{\varepsilon}$, are both allowed to vary in a discrete manner in the set $[1,5]$.\footnote{For each value in the set $[1,5]$ we generate random draws from a normal distribution to compute the forecast in (14) for each economic agent.}  We set the AR coefficient $\beta = 0.5$, although the results are quantitatively very similar to alternative values to this parameter.  We set the number of agents to $N=10000$.  Using this calibration of the model, we compute the tail disagreement index $\tilde T_{t}$ in equation (\ref{eq_tail_synt}).

\begin{figure}
\caption{Disagreement index in the model}
\begin{center}
\includegraphics[width=0.8\textwidth]{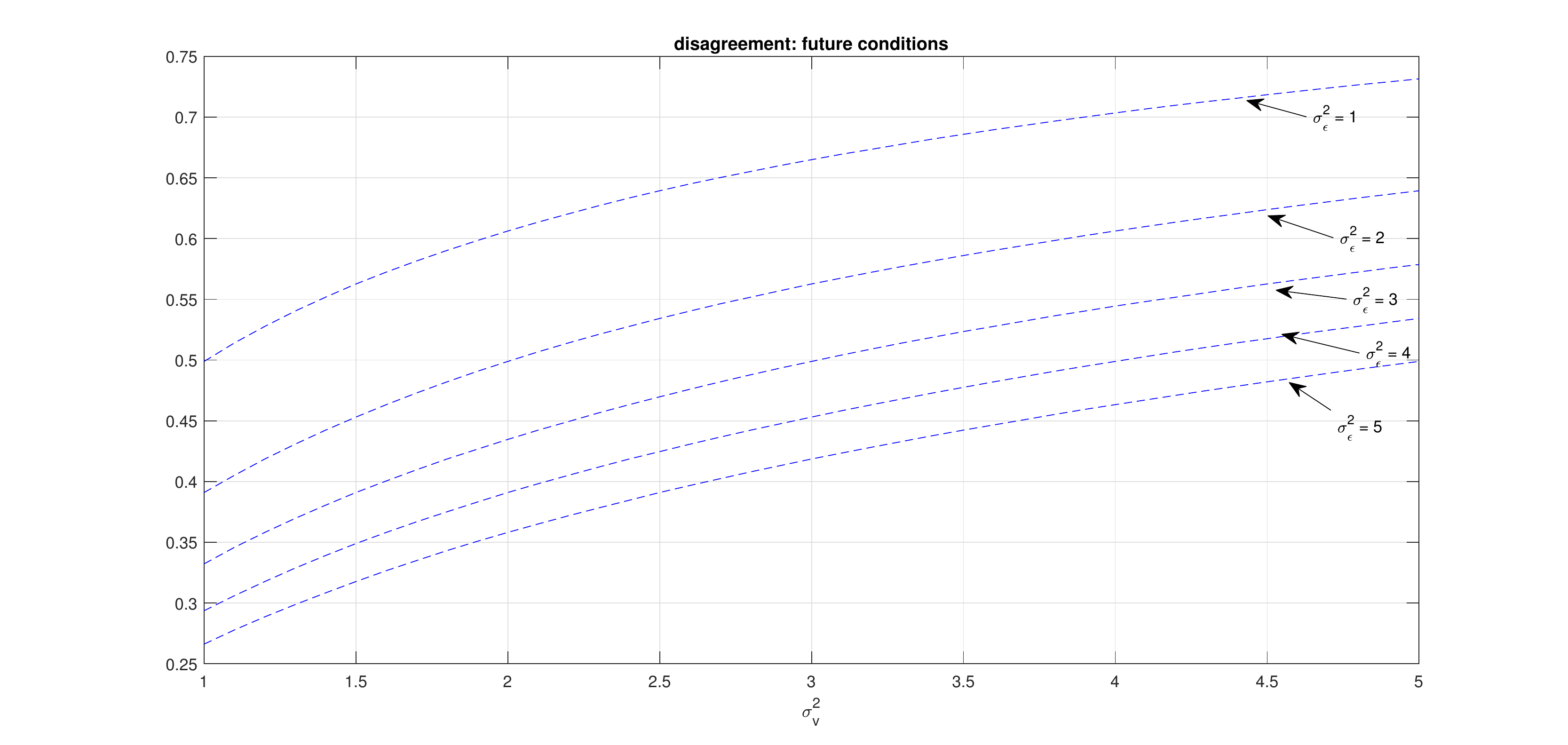}
\label{fig_idio_var_and_disag}
\end{center}
 \small \emph{Notes}:   The disagreement index is obtained from $\tilde T_{t}=1-\frac{1}{N}\left|\sum_{i=1}^n b_{it}-w_{it}\right|$ and plotted as a function of $\sigma^{2}_v$. The Figure shows the disagreement index corresponding 12-months-ahead economic conditions. To ease exposition, the figure plots the index for five alternative values of $\sigma^2_{\varepsilon}$ equal to $1,2,3,4$, and $5$. The disagreement index increases with information dispersion ($\sigma^{2}_v$), but it falls with the increase in the variance of the fundamental shock ($\sigma^2_{\varepsilon}$).
\end{figure}

Each dashed line in Figure \ref{fig_idio_var_and_disag} shows the tail disagreement index from simulating the model (y-axes) as a function of the variance of the idiosyncratic shock $\sigma^2_v$ (x-axes), and we compute it for different values for the variance of the fundamental shock $\sigma^{2}_{\varepsilon}$ equal to $1,2,3,4$, and $5$ (blue-dashed line). The figure illustrates that the disagreement index is an increasing function of information dispersion but decreases with the variance of the fundamental shock $\sigma^{2}_{\varepsilon}$ for any given level of $\sigma^{2}_v$. The intuition from our model is that the signal becomes more precise and agents downplay the idiosyncratic information content of the signal when $\sigma^{2}_{\varepsilon}$ increases, thus agents update expectations more strongly in the direction of the signal and agree more (i.e., agents disagree less). Important to our analysis, the model establishes an inverse comovement between disagreement and uncertainty, consistent with the strong negative correlation between those indicators in the data, as documented in Section \ref{Sec1_Measure}.

%
%


\section{Empirical model}\label{sec_model_empirical}


\paragraph{VAR inference and shock identification.} Our starting point is a Bayesian vector autoregressive (VAR) in the tradition of several recent studies on uncertainty (\citealp{Jurado_etAl_AER2015}, \citealp{Gilchrist2014}). We use the identifying sign restrictions extracted from the simple model of information dispersion in order to disentangle the dynamic effects of agreed uncertainty and disagreed uncertainty shocks in the data using the VARs. The restrictions are (i) those in equations (12) and (13), and (ii) those implied by Figure \ref{fig_idio_var_and_disag}. Table \ref{Table:sign_restrictions} summarizes our identifying restrictions, showing the response of the observed variables (i.e., uncertainty and disagreement) to agreed and disagreed uncertainty, in columns (1) and (2), respectively.

\begin{table}[H]
\centering
\caption{Identifying restrictions} \label{Table:sign_restrictions}
\begin{tabular}{llcc} \hline
 & &\multicolumn{2}{c}{\textsc{Shock} }   \\
 \cline{3-4}
 & & (1) & (2) \\
\textsc{Observed variable} & & $\sigma_\varepsilon$ & $\sigma_v$ \\
&& Agreed Uncertainty & Disagreed Uncertainty\\\hline
Variance of the forecast error &  & + &  +  \\
Index of disagreement &  & $-$ & $+$  \\
\hline
\end{tabular}
\\
\flushleft
\small \emph{Notes}: The entries show the impact response of the variance of the forecast error and the index of disagreement to the shock to agreed uncertainty (column 1) and disagreed uncertainty (column 2).
\end{table}

Column (1) in the table shows that an innovation to the fundamental shock $\sigma^{2}_{\varepsilon}$  is associated with an increase in observed uncertainty (represented by the variance of the forecast error) and a decrease in the index of disagreement.  This is our concept of \emph{agreed uncertainty}. Instead, column (2) in the table shows that an innovation to information dispersion $\sigma^{2}_{v}$ is associated with a simultaneous increase in observed uncertainty and the index of disagreement. This is our concept of \emph{disagreed uncertainty}.\footnote{To rule out the autonomous effect of uncertainty shocks on economic activity that results in overstating the impact economic effects of agreed or disagreed uncertainty shocks, we impose a zero-impact response of activity indicators to the identified shocks.} Using these distinct comovements in observed uncertainty and the index of disagreement, we identify the two distinct concepts of uncertainty shocks in the data.

We tackle estimation of the VAR under these restrictions using the Bayesian Markov chain Monte Carlo (MCMC) algorithm developed in \cite{Korobilis2022}, which allows us to sample sign and zero restrictions in arbitrarily large VARs with high computational efficiency. For the $n \times 1$ vector of time-series variables $\bm y_{t}$, the VAR takes the multivariate regression form:
\begin{equation}
\mathbf{y}_{t} = \mathbf{\Phi} \mathbf{x}_{t} + \bm{\varepsilon}_{t}, \label{VAR}
\end{equation}
where $\mathbf{y}_{t}$ is a $\left( n \times 1 \right)$ vector of observed variables, $\mathbf{x}_{t} = \left( 1,\mathbf{y}_{t-1}^{\prime},...,\mathbf{y}_{t-p}^{\prime} \right)^{\prime}$ a $\left( k \times 1 \right)$ vector (with $k=np+1$) containing a constant and $p$ lags of $\mathbf{y}$, $\mathbf{\Phi}$ is an $(n \times k)$ matrix of coefficients, and $\bm{\varepsilon}_{t}$ is a $\left( n \times 1 \right)$ vector of disturbances distributed as $N\left( \mathbf{0}_{n \times 1},\mathbf{\Omega} \right)$ with $\mathbf{\Omega}$ an $n \times n$ covariance matrix. We further assume the following factor decomposition of $\bm{\varepsilon}_{t}$:
\begin{equation}
\bm{\varepsilon}_{t} = \mathbf{\Lambda} \mathbf{f}_{t} + \mathbf{v}_{t}, \label{factor_model}
\end{equation}
where $\mathbf{\Lambda}$ is an $n \times r$ matrix of factor loadings, $\mathbf{f}_{t} \sim N(\bm{0},\bm{I}_r)$ is an $r \times 1$ vector of factors, and $\mathbf{v}_{t} \sim N(\bm{0},\bm{\Sigma})$ is an $n \times 1$ vector of idiosyncratic shocks with $\bm{\Sigma}$ an $n \times n$ diagonal matrix.

The rationale behind the VAR in equations \eqref{VAR}-\eqref{factor_model} is that the $n$-dimensional vector of VAR disturbances is decomposed into $r$ common shocks $\mathbf{f}_{t}$ ($r<n$) and $n$ idiosyncratic shocks $\mathbf{v}_{t}$.\footnote{\cite{Gorodnichenko2005} also exploits this formulation of the VAR for the identification of monetary policy shocks.} Because $\bm{\Sigma}$ is diagonal, we consider only the $r$ common shocks as structural and the $n$ idiosyncratic shocks as nuisance shocks (e.g., due to measurement error or asymmetric information). Indeed, by left-multiplying the VAR using the generalized inverse of $\mathbf{\Lambda}$, the implied structural VAR form is:\begin{eqnarray}
\mathbf{y}_{t} & = & \mathbf{\Phi} \mathbf{x}_{t} + \mathbf{\Lambda} \mathbf{f}_{t} + \mathbf{v}_{t} \\
\left( \mathbf{\Lambda}^{\prime} \mathbf{\Lambda} \right)^{-1}\mathbf{\Lambda}^{\prime} \mathbf{y}_{t} & = & \left( \mathbf{\Lambda}^{\prime} \mathbf{\Lambda} \right)^{-1}\mathbf{\Lambda}^{\prime} \mathbf{\Phi} \mathbf{x}_{t} +  \mathbf{f}_{t} + \left( \mathbf{\Lambda}^{\prime} \mathbf{\Lambda} \right)^{-1}\mathbf{\Lambda}^{\prime} \mathbf{v}_{t} \\
\mathbf{A}_{1} \mathbf{y}_{t} & = & \mathbf{B}_{1} \mathbf{x}_{t} + \mathbf{f}_{t} + \left( \mathbf{\Lambda}^{\prime} \mathbf{\Lambda} \right)^{-1}\mathbf{\Lambda}^{\prime}\mathbf{v}_{t}, \label{RR_SVAR}
\end{eqnarray}
where $\mathbf{A}_{1}=\left( \mathbf{\Lambda}^{\prime} \mathbf{\Lambda} \right)^{-1}\mathbf{\Lambda}^{\prime}$ and $\mathbf{B}_{1}=\mathbf{A}_{1}\mathbf{\Phi}$. As long as $\bm{\Sigma}$ is diagonal the term $\left( \mathbf{\Lambda}^{\prime} \mathbf{\Lambda} \right)^{-1}\mathbf{\Lambda}^{\prime}\mathbf{v}_{t}$ vanishes asymptotically, meaning that $\mathbf{f}_{t}$ retains the interpretation of structural shocks. Therefore, the desired sign and zero restrictions required for identifying agreed and disagreed uncertainty can take the form of simple parametric restrictions imposed on the respective elements of $\mathbf{\Lambda}$.

Bayesian inference requires specification and tuning of prior distributions for all parameters, and estimation with iterative MCMC algorithms requires further assumptions and tuning parameters. Without dismissing the significance of such choices, we use default, automatic prior choices justified in detail in \cite{Korobilis2022}.\footnote{Our default choice is to iterate the algorithm 600,000 times, discard the first 100,000 iterations, and save every 100$^{th}$ draw from the parameter posterior. In all VARs of different sizes we estimate, this setting ensures low autocorrelation of posterior samples, as well as their good numerical properties.} We provide further technical details in the Appendix.

\paragraph{Data and specifications.}
Because fluctuations in measures of uncertainty and disagreement are short-lived, following \cite{bloom_ECTA}, \cite{Jurado_etAl_AER2015}, and \cite{Bergeretal2020} our benchmark results rely on monthly U.S. macroeconomic data. The sample runs from 1978M1 to 2020M12, where the earliest date is dictated by the availability of the MSC data. Consumer disagreement is measured by our disagreement index DISAG described in Section \ref{Sec1_Measure}. We adopt the 12-month-ahead macro uncertainty indicator developed by \cite{Jurado_etAl_AER2015} as the benchmark measure of uncertainty.  We compute this indicator from estimates of conditional volatilities of $h$-step ahead forecast errors using a monthly dataset of 134 macroeconomic time series and captures broad-based macroeconomic uncertainty.\footnote{Methodologically, the indicator captures broad-based movements in economic uncertainty while filtering out variations in the conditional volatilities of the forecast errors.  This procedure avoids accounting predictable movements in the economy as uncertainty, as shown in \cite{Ludvigson_etAl_AEJ}.} This indicator, being a conditional variance of a forecast error, is therefore the natural counterpart of the model concept. Moreover, because macro-uncertainty is about broad-based future economic conditions, it links more naturally to the concept of information dispersion we use in this paper, which is about dispersion of consumer views about economy-wide business conditions.  The remaining monthly variables in our benchmark VAR specification are: real industrial production index (IP), real personal consumption expenditure index (CONS), total non-farm employment (EMPL), inflation rate based on the personal consumption expenditure price index (INFL), the S\&P 500 index (SP500), and the federal funds effective rate (FEDFUNDS).  We discuss robustness of our results in Section \ref{sec_robustness}.  The VAR models are estimated with 13 lags, and Appendix \ref{app_ecometrix_method} describes the econometric methodology in detail. To conserve space we report IRFs to four key variables from the VAR specification, and the Appendix reports the remaining IRFs.


\paragraph{Benchmark specification.}

\begin{figure}[h!]
\caption{Benchmark model. Agreed $\sigma_{\varepsilon}$ (left) versus disagreed $\sigma_{v}$ (right) uncertainty.}\label{fig_IRF_benckmark}
\begin{center}
\includegraphics[width=0.8\textwidth,, trim={1cm .5cm 0 .4cm}]{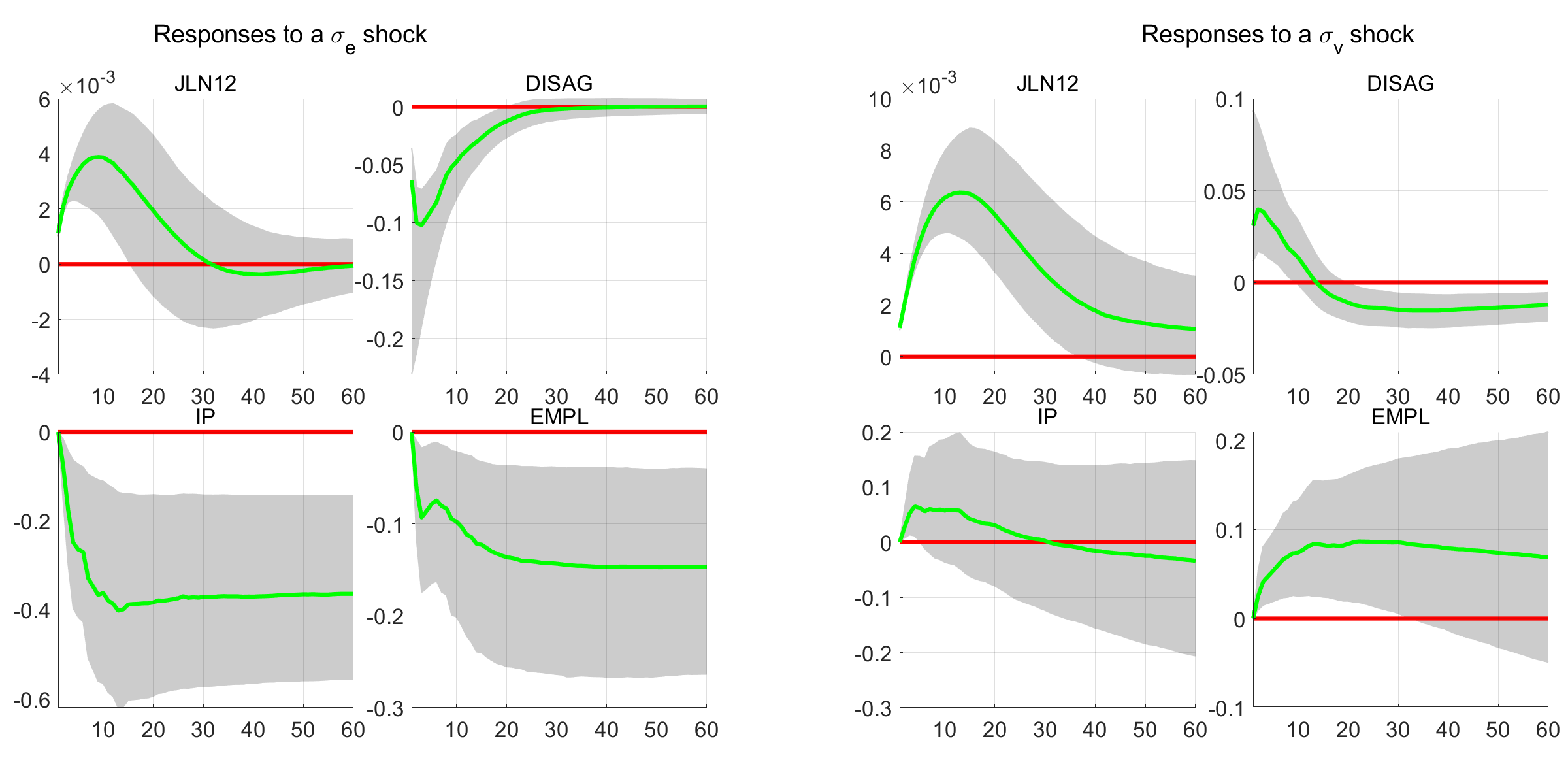}

\end{center}
\small \emph{Notes}: The figure shows impulse responses to the JLN 12-months-ahead uncertainty indicator (JLN12), the disagreement index (DISAG),  industrial production (IP), and employment (EMPL). IRFs from an eight-variable VAR system as described in the text. The shaded gray areas are the 16\% and 84\% posterior bands generated from the posterior distribution of VAR parameters. The units of the vertical axes are percentage deviations, and the horizontal axes report time measured in months.
\end{figure}

The left panel in Figure \ref{fig_IRF_benckmark} shows IRFs to a positive innovation in the variance of the fundamental shock $\sigma_{\epsilon}$ -- \textit{agreed uncertainty} -- identified by imposing the sign restrictions in column (1) of Table \ref{Table:sign_restrictions} on the response of uncertainty and disagreement indicators in the first period after the shock.  The JLN-12 uncertainty indicator rises immediately on impact and remains persistently elevated for approximately 15 months, but the DISAG indicator declines persistently in the short run and remains depressed for approximately 20 months.  Beyond the initial period (where the response of activity indicators are restricted to zero), industrial production, and employment decline sharply and remain depressed even at the sixty month horizon.  Our results echo recent findings in the literature (e.g., \citealp{Jurado_etAl_AER2015}, \citealp{Gilchrist2014}, \citealp{Ludvigson_etAl_AEJ}) using similar empirical methods; they emphasize a significant depressing effect of economic uncertainty on real activity indicators.

The right panel in Figure \ref{fig_IRF_benckmark} shows IRFs to a positive innovation in the variance of idiosyncratic noise, $\sigma_{v}$ -- \textit{disagreed uncertainty}. It is identified by imposing the sign restrictions in column (2) of Table \ref{Table:sign_restrictions} on the response of uncertainty and disagreement indicators in the first period after the shock.  The JLN-12 indicator shows a persistent rise that extends beyond the 30-month horizon, while disagreement displays a short-lived, persistent rise and stays elevated for about 10-months after the shock.  Note that the increase in the JLN-12 indicator is stronger and more persistent than the increase in JLN-12 estimated in the left panel of the figure.  Despite a stronger and more persistent rise in uncertainty, the responses of real activity indicators are qualitatively different than the responses under the agreed uncertainty shock.  Specifically, industrial production exhibits a small positive and statistically significant short-lived response, in contrast to the persistent negative response estimated in the left panel; it then quickly reverts to the pre-shock level. Similarly, employment exhibits a small positive and statistically significant response that persists until month 30, in contrast to the negative persistent response estimated in the left panel.   Thus, disagreed uncertainty shocks are characterized by dynamic effects that are broadly benign for economic activity in the short run, and they differ qualitatively from the strong, adverse, and long-lasting effects on economic activity in the aftermath of shocks to agreed uncertainty. Appendix \ref{app_robustness} reports the complete set of IRFs.

To summarize, identified innovations in \emph{agreed uncertainty} and \emph{disagreed uncertainty} display sharp qualitative differences in the dynamic responses of real activity indicators. \emph{Agreed uncertainty} shocks are robustly contractionary and generate a sustained decline in industrial production and employment. By contrast, \emph{disagreed uncertainty} shocks are broadly benign; they are associated with a small and statistically significant positive response of real activity in the short run. We are the first study to show that disagreement in household views about current and future economic conditions that characterizes disagreed uncertainty is critical for the benign effects of uncertainty on real activity, but agreed uncertainty retains the standard adverse effect on real activity.

\textbf{Forecast error variance decomposition}. We decompose the share of forecast error variance (FEVD) of the benchmark VAR variables into the two identified shocks. This is a useful check to ascertain whether these shocks are important drivers of the empirical indicators of uncertainty and disagreement. 

Figure \ref{fig_var_decomp} below shows the variance decomposition to the identified shocks of agreed uncertainty (green) and disagreed uncertainty (blue); we also report the residual variation, which is not attributed to any identified shock (yellow). Two important findings emerge. First, the share of FEVD in JLN-12 explained by the two shocks is significant, rising to over 70\% after the seven-month horizon. Interestingly, the share of FEVD in JLN-12 accounted for by the innovation in disagreed uncertainty alone exceeds the share of FEVD accounted for by the agreed uncertainty shock at all horizons, suggesting that the former is a significant driver of the variance in JLN-12. The disagreed uncertainty innovation accounts for 50\% of FEVD in JLN-12 after the seven-month horizon and never drops below 20\% of the same FEVD. Second, the two shocks combined account for the majority of the FEVD in disagreement, approximately 70\% at all horizons.  For disagreement, the share in FEVD accounted for by the agreed uncertainty shock exceeds the FEVD share accounted for by the disagreed uncertainty shock by a large margin.  These findings establish that our two different uncertainty shocks jointly explain a significant share of the variation in the indexes of uncertainty and disagreement. Taken together, the results show the key role of innovations to the dispersion of information to explain movements in the different types of uncertainty, supporting the insight from our simple model in Section \ref{sec_model_theory}.  Our exercise also suggests that the combination of two uncertainty shocks accounts for a large share of FEVDs in industrial production, rising from approximately 20\% at the 10-month horizon to 40\% at the 60-month horizon, and innovations to agreed uncertainty account for the majority of this total.  These shocks, however, account for a relatively small share, which is approximately 10\% after 10-months in the FEVD of employment, suggesting that other unidentified shocks are the major drivers of the variation in this variable.\footnote{Even though our sample period and VAR identification is different, the FEVD in IP accounted for by the two uncertainty shocks combined is broadly consistent with estimates of FEVD accounted by macro uncertainty shocks identified via JLN-12 reported in \cite{Jurado_etAl_AER2015}. The FEVD in employment accounted for by the two shocks displayed earlier is somewhat smaller in comparison to estimates of FEVD accounted by the macro uncertainty shocks reported in the same study.}

\begin{figure}[h!]
\caption{Variance Decomposition: Agreed ($\sigma_{\varepsilon}$, green) vs. disagreed ($\sigma_{v}$, blue) uncertainty.}\label{fig_var_decomp}
\begin{center}
\includegraphics[width=0.95\textwidth, trim={5cm .5cm 5cm 1cm}]{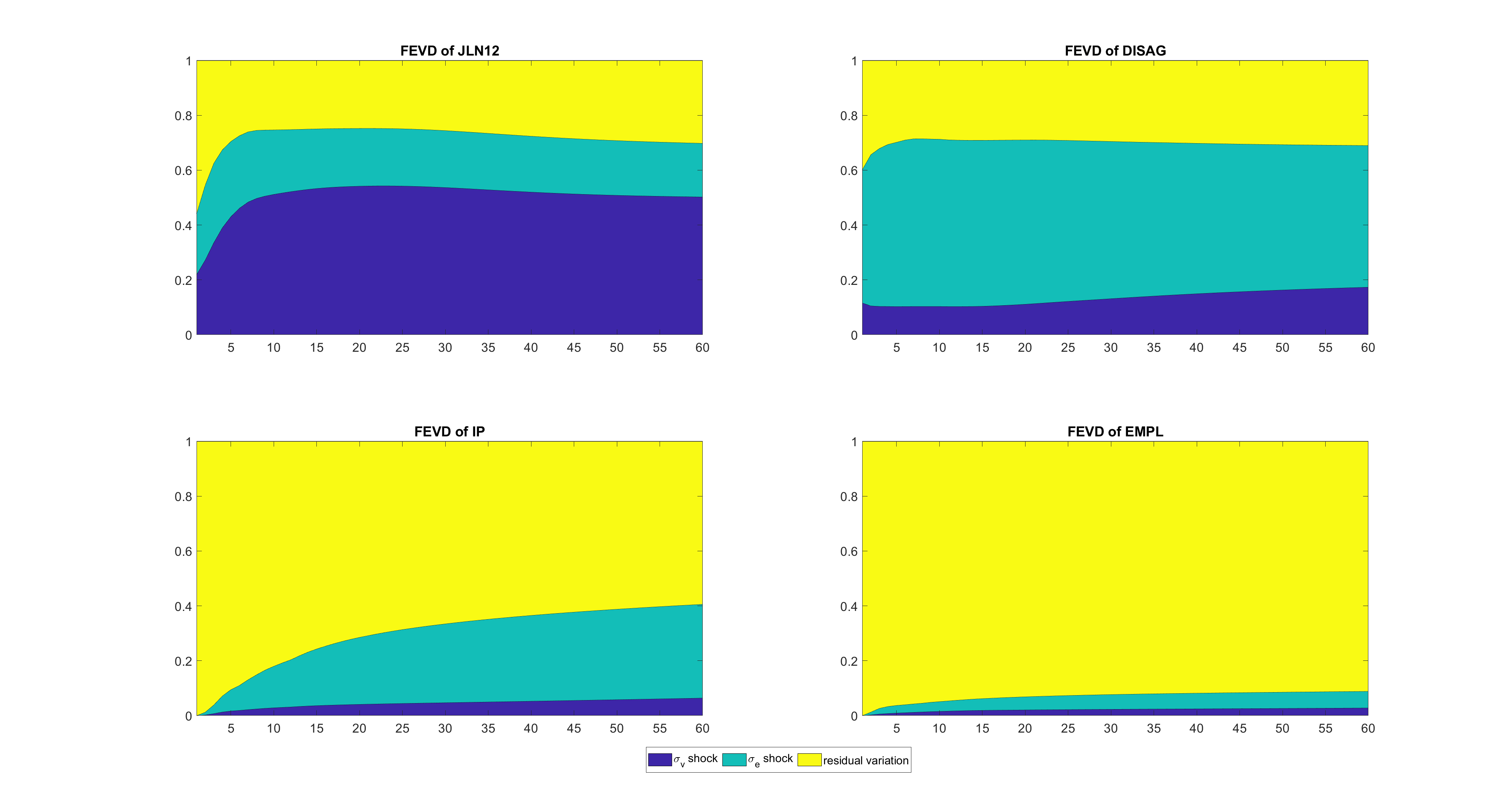}

\end{center}

\small \emph{Notes}: The figure shows the variance decomposition of the variables from the benchmark VAR specification to $\sigma_{\varepsilon}$ (green), $\sigma_{v}$ (blue), and the residual variance not attributed to any other shocks (yellow). The units of the vertical axes are percent, and the horizontal axes reports monthly horizons.

\end{figure}

\section{Robustness analysis\label{sec_robustness}}

In this section we briefly discuss the robustness analysis intended to check the sensitivity of our results to alternative measures of consumer disagreement. We investigate the robustness of our findings when we control for different demographic characteristics of consumers, namely, age and education.  In Appendix \ref{app_robustness} we examine the robustness to alternative proxies to uncertainty used in related studies.\footnote{Specifically, we examine VAR specifications where we switch our benchmark uncertainty indicator to one of the following: the \cite{Jurado_etAl_AER2015} 12-months-ahead financial uncertainty indicator (henceforth JLNF-12), the business dispersion measure (BOS-dispersion) developed in \cite{Bachmann_al_AEJM13}, stock market volatility (CBOE S\&P 100 volatility index VXO), or the Economic Policy Uncertainty index (EPU) developed by \cite{BakerBloomDavis16}.} We also consider VAR specifications that i) estimate the dynamic effects of the two shocks on a broad spectrum of macroeconomic (including labor market) and survey indicators, ii) replace the DISAG index with individual disagreement indices from specific survey questions, iii) estimate a VAR specification using a quarterly macro dataset. These results are also described in Appendix \ref{app_robustness}.

\textbf{Alternative disagreement indicators.} The tails disagreement employed in our benchmark specification is simple and intuitive, but does not fully utilize all the responses from MSC. Specifically, it only considers the two polar categories of responses (better/worse), while ignoring the middle category (depending on the question, this category relates to past/future conditions that are either the ``same'' or ``uncertain''). For that reason we recompute the disagreement index using two alternative measures: ``Entropy disagreement'' using Shannon's (\citealp{Shannon48}) entropy measure, and ``Lacy disagreement'' using the transformation proposed by \cite{Lacy06}. These exploit all possible answers from consumers.

The entropy disagreement is defined as\footnote{This measure we define as disagreement is called the ``Shannon Index'' in ecology and
related sciences, and it is used to measure the diversity and distribution of types of species in a community; see
\cite{Hill1973}.}
$$
H_{t}^j=-\sum_{i=1}^np(x^j_i)\log{p(x^j_i)}
$$
where $x_i^j$ is option $i$ of $n$ possible answers for question $j$, and $p(x^j_i)$ is the proportion of
individuals answering $x_i^j$. This index gives a measure of the cross-sectional uncertainty of consumers about the possible business outcomes that may occur, where $p(x^j_i)$ has an interpretation of probabilities.\footnote{We assume that consumers who have the same view about business conditions do so because they also agree on the probabilities about observing a specific outcome.}  The higher the index the higher the uncertainty and the higher the disagreement. For example, if all consumers shared the same view about the prospects of the economy, the value of the index will be zero, which reflects a situation of zero uncertainty and disagreement. By contrast if consumers are equally divided between the three outcome categories (``better,'' ``worse,'' ``same''), the value of the index attains the maximum value. The second alternative disagreement measure, from \cite{Lacy06}, describes how dispersed or concentrated ordinal data is without requiring further assumptions about
inter-category distances. The Lacy disagreement is defined using,
$$
D_{j}^2 = \sum_{i=1}^{n-1} F_i \left(1 - F_i \right),
$$
where $F_i$ is the cumulative relative frequency for the $i$th category. Note that the sum excludes the last
category, because $F_{n}$ is always 1. This $D_{j}^2$ measure ranges from $0$ to $\left( n - 1 \right) / 4$. When
the value of this measure is zero, all responses fall in the same category. The maximum value of $\left( n - 1
\right) / 4$ denotes completely polarized distribution in which half of the responses are in category $1$ and half
are in category $n$. Values between the minimum and the maximum indicate intermediate levels of dispersion.

We re-estimate the VAR after replacing the disagreement indicator DISAG with the two alternative indicators one at a time, retaining all other variables in the benchmark specification. The results are reported in Figure \ref{fig_disagreement_rob_dis} below.  First, we note that the median IRFs displayed following a shock to agreed uncertainty (left panel) and disagreed uncertainty (right panel) are qualitatively and quantitatively similar when we use either the Lacy (DISAG-L, dashed-green line) or Entropy (DISAG-E, dashed-blue line) concept of disagreement in the VAR, and they are broadly similar to the IRFs we estimate from the benchmark specification (also plotted in the same figure).  This result shows that the different VAR specifications identify the same shocks to agreed and disagreed uncertainty. Moreover, the VAR specifications with the DISAG-L and DISAG-E indicators suggest that the short-run positive response of industrial production and employment following an innovation to disagreed uncertainty are stronger in comparison to the responses in the same variables estimated in the benchmark specification. Overall, this exercise ensures that the DISAG indicator used in the benchmark VAR is robust to including information from those consumers that are more uncertain about the strength or weakness of current and future economic conditions.

\begin{figure}[h!]

\caption{Alternative disagreement indexes. Agreed ($\sigma_{\varepsilon}$, left) vs. disagreed $\sigma_{v}$ (right) uncertainty.} \label{fig_disagreement_rob_dis}
\begin{center}
\includegraphics[width=0.95\textwidth,, trim={1cm .5cm 0 .4cm}]{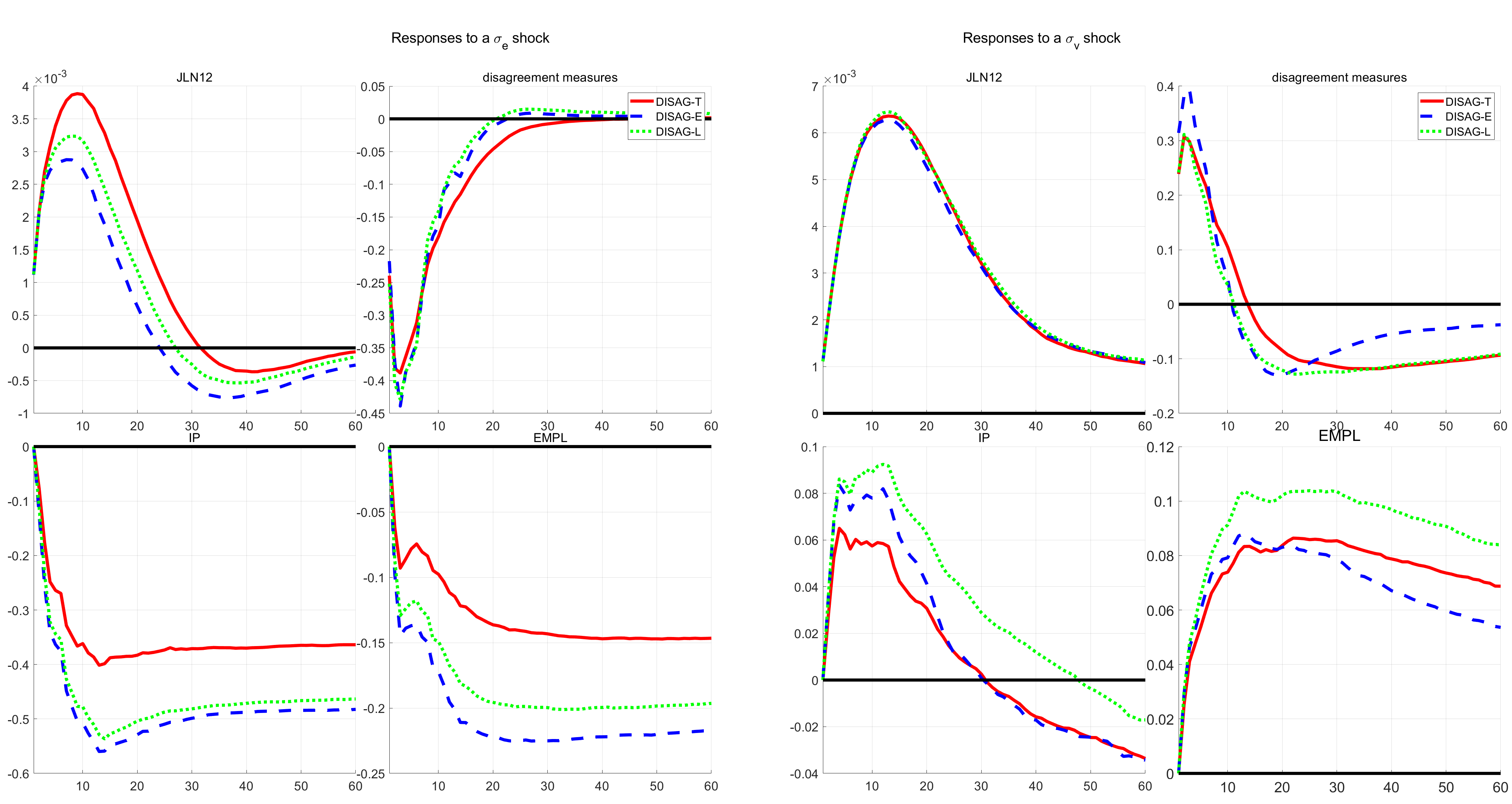}
\end{center}

\small \emph{Notes}: The figure shows (median) impulse responses for alternative disagreement indexes: Tail disagreement (DISAG-T) as used in the benchmark VAR, Entropy disagreement (DISAG-E), and Lacy disagreement (DISAG-L). The units of the vertical axes are percentage deviations, and the horizontal axes reports time measured in months.

\end{figure}

\begin{figure}[h!]
\caption{Benchmark model--education: High school. Agreed $\sigma_{\varepsilon}$ (left) versus disagreed $\sigma_{v}$ (right) uncertainty.}\label{fig_IRF_benckmark_edu1}
\begin{center}
\includegraphics[width=0.8\textwidth, trim={1cm .5cm 0 .4cm}]{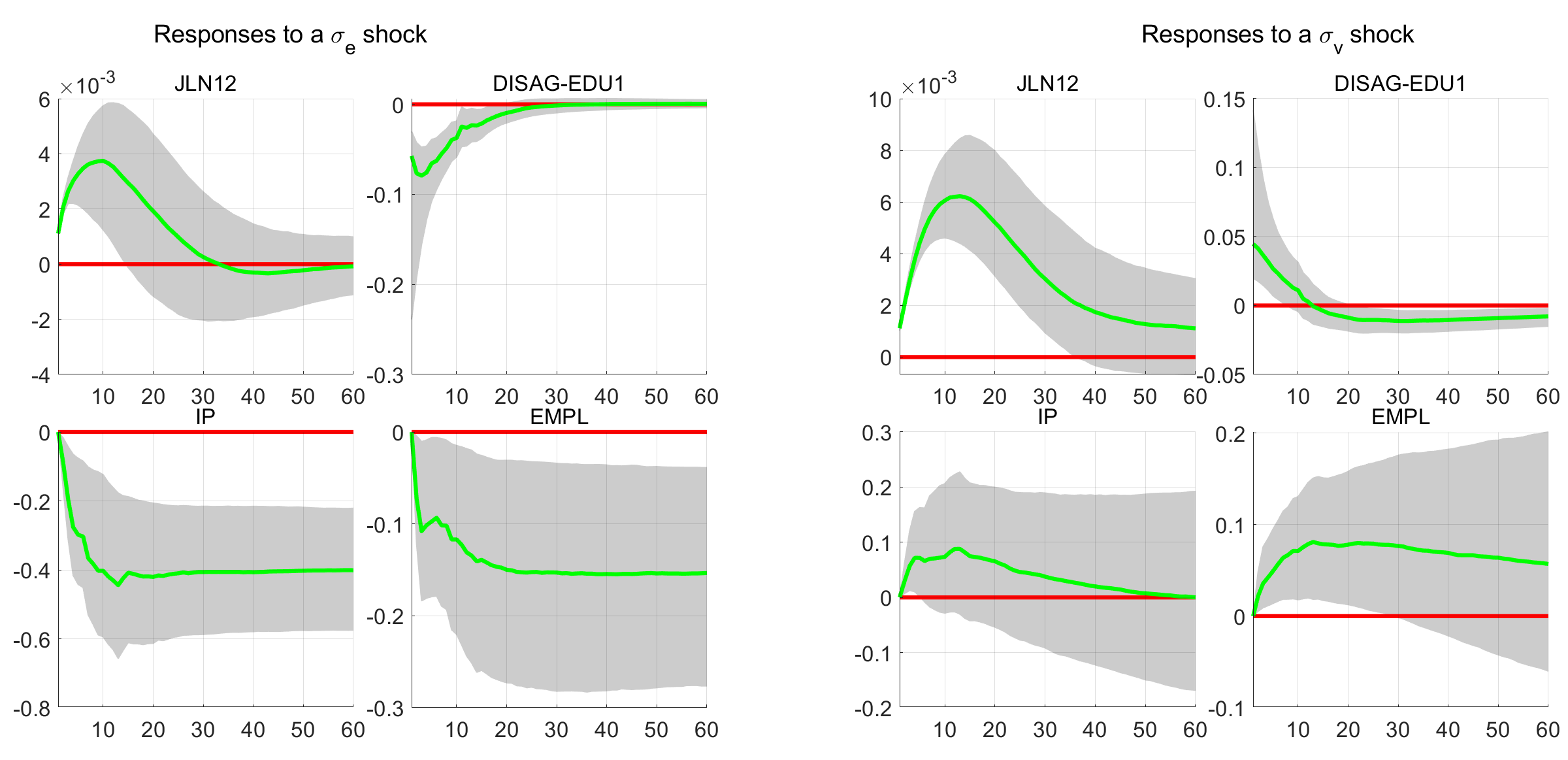}

\end{center}
\small \emph{Notes}: The figure shows impulse responses to the JLN 12-months-ahead uncertainty indicator (JLN12), the disagreement index for high school education level (DISAG-EDU1),  industrial production (IP), and employment (EMPL). We compute IRFs from an eight-variable VAR system as described in the text. The shaded gray areas are the 16\% and 84\% posterior bands generated from the posterior distribution of VAR parameters. The units of the vertical axes are percentage deviations, and the horizontal axes report time measured in months.
\end{figure}

\begin{figure}[h!]
\caption{Benchmark model--education: some college. Agreed $\sigma_{\varepsilon}$ (left) versus disagreed $\sigma_{v}$ (right) uncertainty.}\label{fig_IRF_benckmark_edu2}
\begin{center}
\includegraphics[width=0.8\textwidth, trim={1cm .5cm 0 .4cm}]{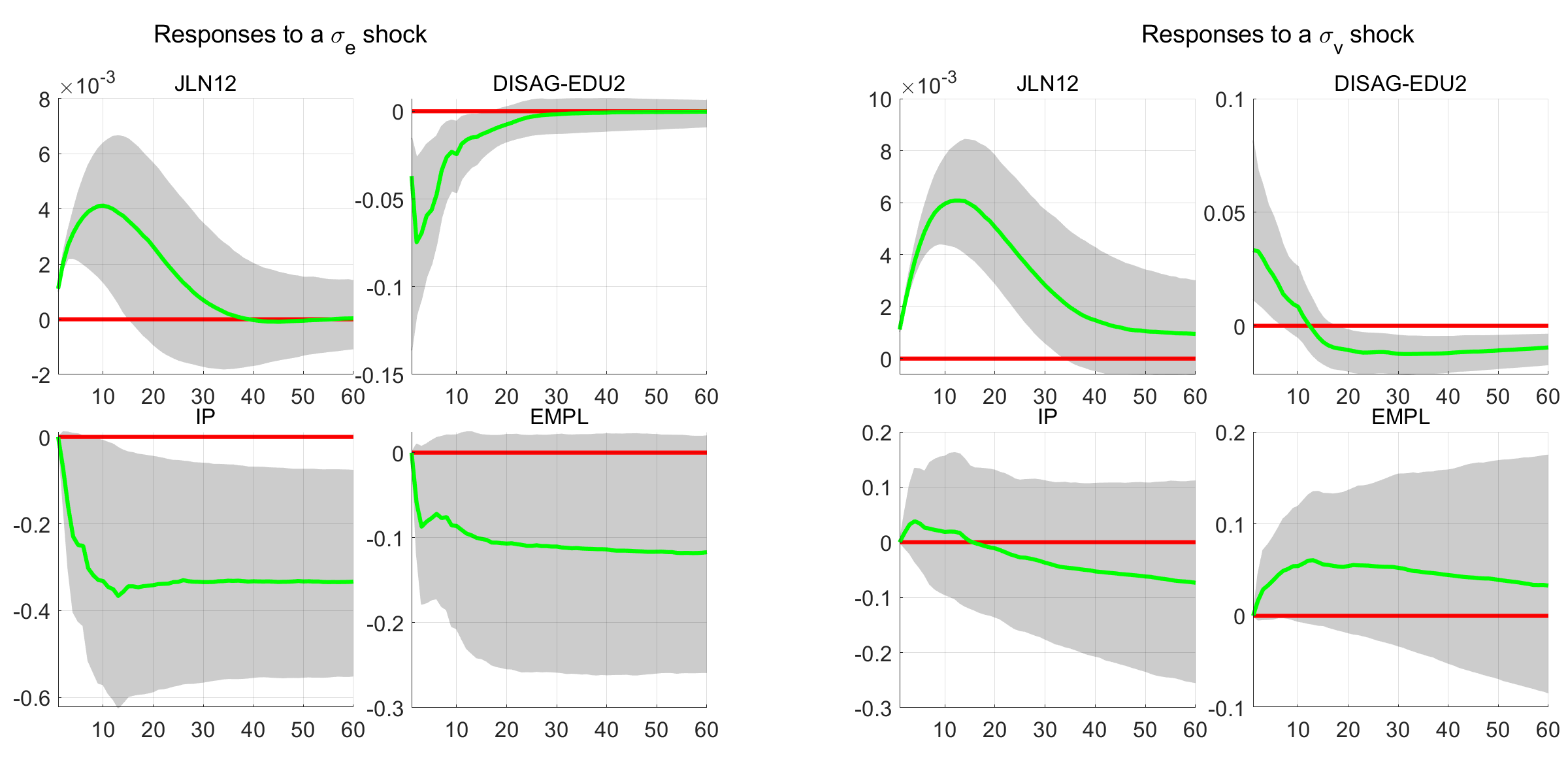}

\end{center}
\small \emph{Notes}: The figure shows impulse responses to the JLN 12-months-ahead uncertainty indicator (JLN12), the disagreement index for college educational level (DISAG-EDU2),  industrial production (IP), and employment (EMPL). We compute the IRFs from an eight-variable VAR system as described in the text. The shaded gray areas are the 16\% and 84\% posterior bands generated from the posterior distribution of VAR parameters. The units of the vertical axes are percentage deviations, and the horizontal axes report time measured in months.
\end{figure}

\begin{figure}[h!]
\caption{Benchmark model--education: College or higher. Agreed $\sigma_{\varepsilon}$ (left) versus disagreed $\sigma_{v}$ (right) uncertainty.}\label{fig_IRF_benckmark_edu3}
\begin{center}
\includegraphics[width=0.8\textwidth, trim={1cm .5cm 0 .4cm}]{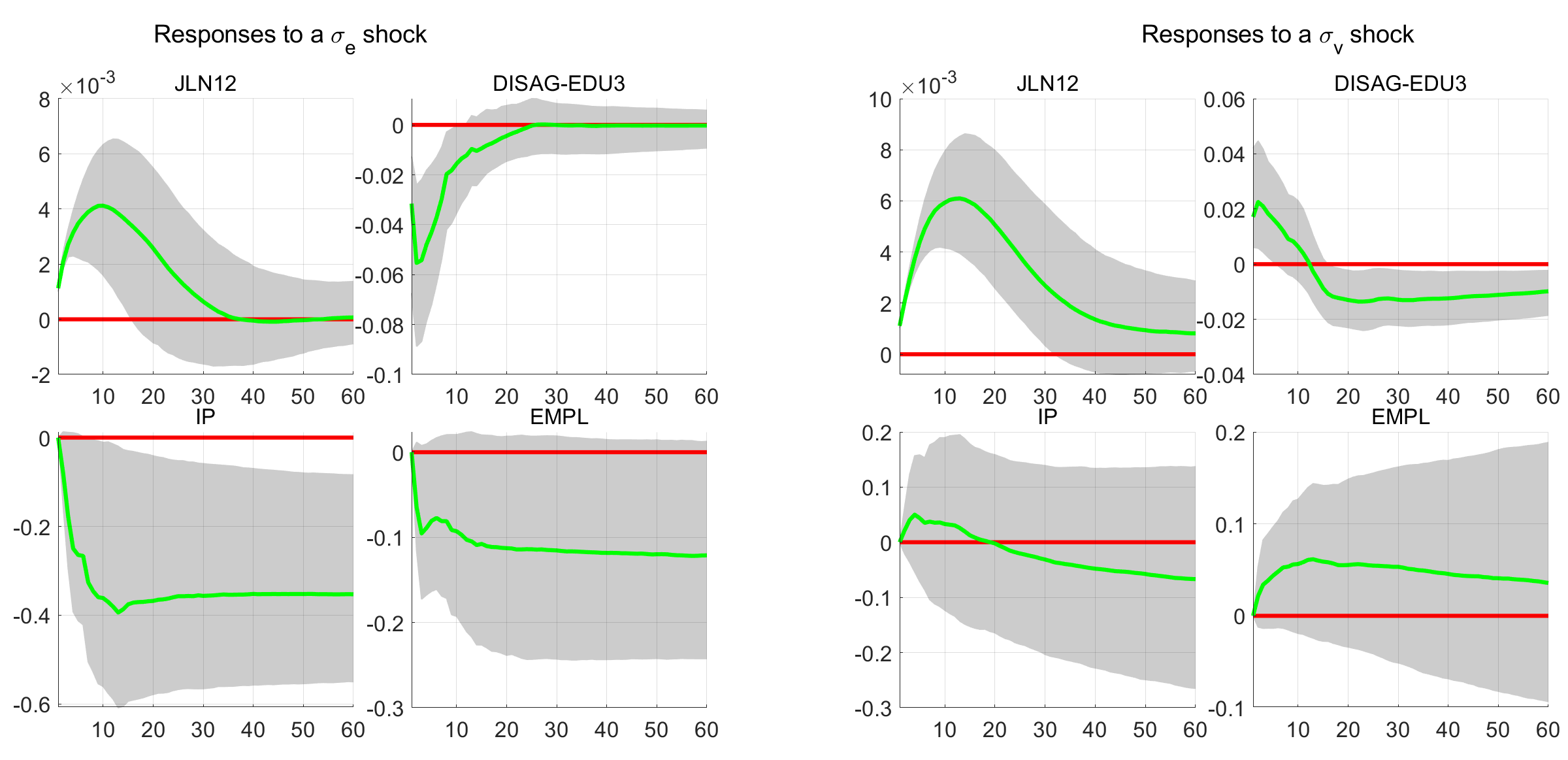}

\end{center}
\small \emph{Notes}: The figure shows impulse responses to the JLN 12-months-ahead uncertainty indicator (JLN12), the disagreement index for college-or-higher education level (DISAG-EDU3),  industrial production (IP), and employment (EMPL). We compute IRFs from an eight-variable VAR system as described in the text. The shaded gray areas are the 16\% and 84\% posterior bands generated from the posterior distribution of VAR parameters. The units of the vertical axes are percentage deviations, and the horizontal axes report time measured in months.
\end{figure}

\begin{figure}[h!]
\caption{Benchmark model--age: 18-34. Agreed $\sigma_{\varepsilon}$ (left) versus disagreed $\sigma_{v}$ (right) uncertainty.}\label{fig_IRF_benckmark_age1}
\begin{center}
\includegraphics[width=0.8\textwidth, trim={1cm .5cm 0 .4cm}]{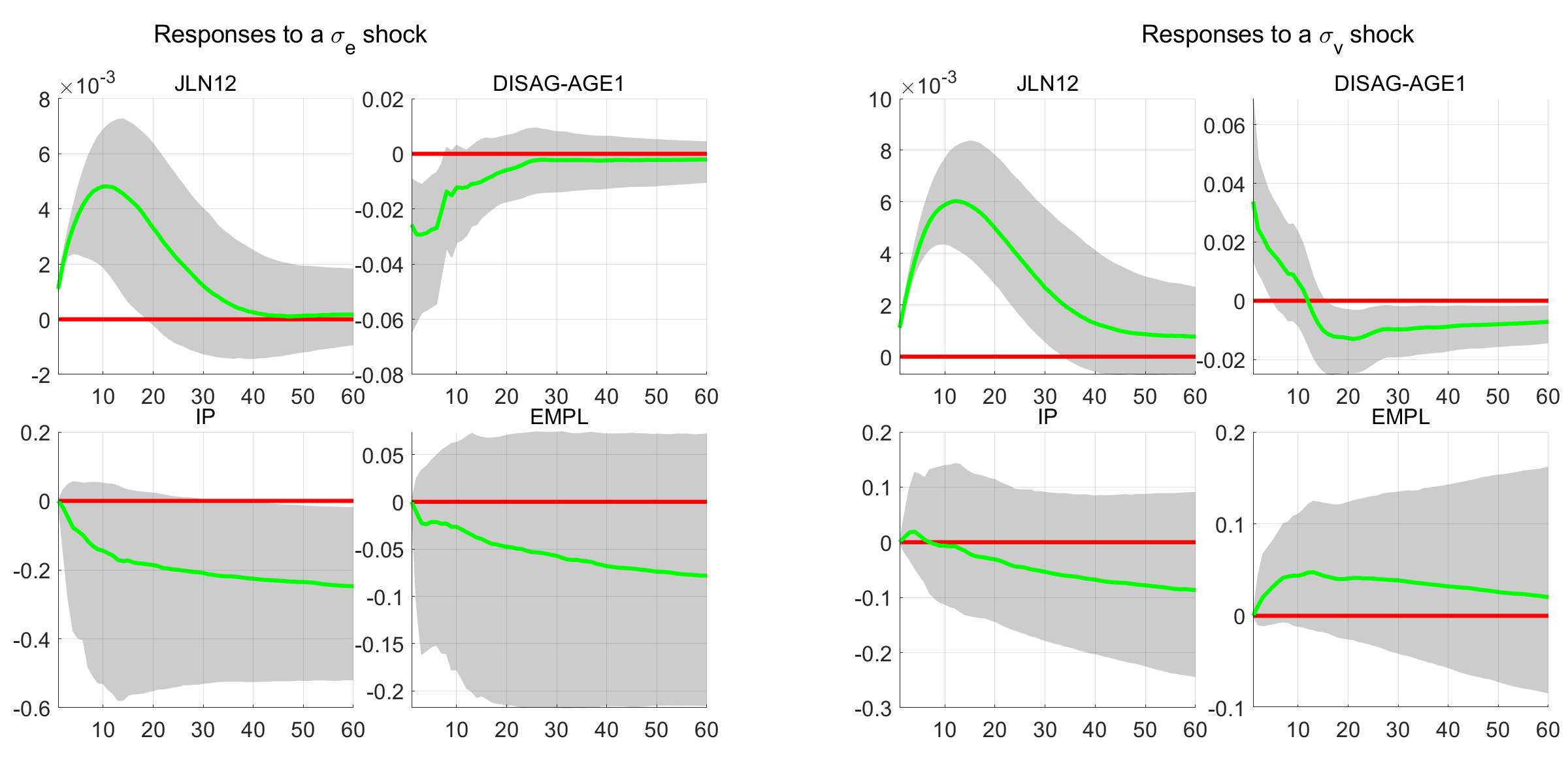}

\end{center}
\small \emph{Notes}: The figure shows impulse responses to the JLN 12-months-ahead uncertainty indicator (JLN12), the disagreement index for 18-34 age group (DISAG-AGE1),  industrial production (IP), and employment (EMPL). We compute IRFs from an eight-variable VAR system as described in the text. The shaded gray areas are the 16\% and 84\% posterior bands generated from the posterior distribution of VAR parameters. The units of the vertical axes are percentage deviations, and the horizontal axes report time measured in months.
\end{figure}

\begin{figure}[h!]
\caption{Benchmark model--age: 35-54. Agreed $\sigma_{\varepsilon}$ (left) versus disagreed $\sigma_{v}$ (right) uncertainty.}\label{fig_IRF_benckmark_age2}
\begin{center}
\includegraphics[width=0.8\textwidth, trim={1cm .5cm 0 .4cm}]{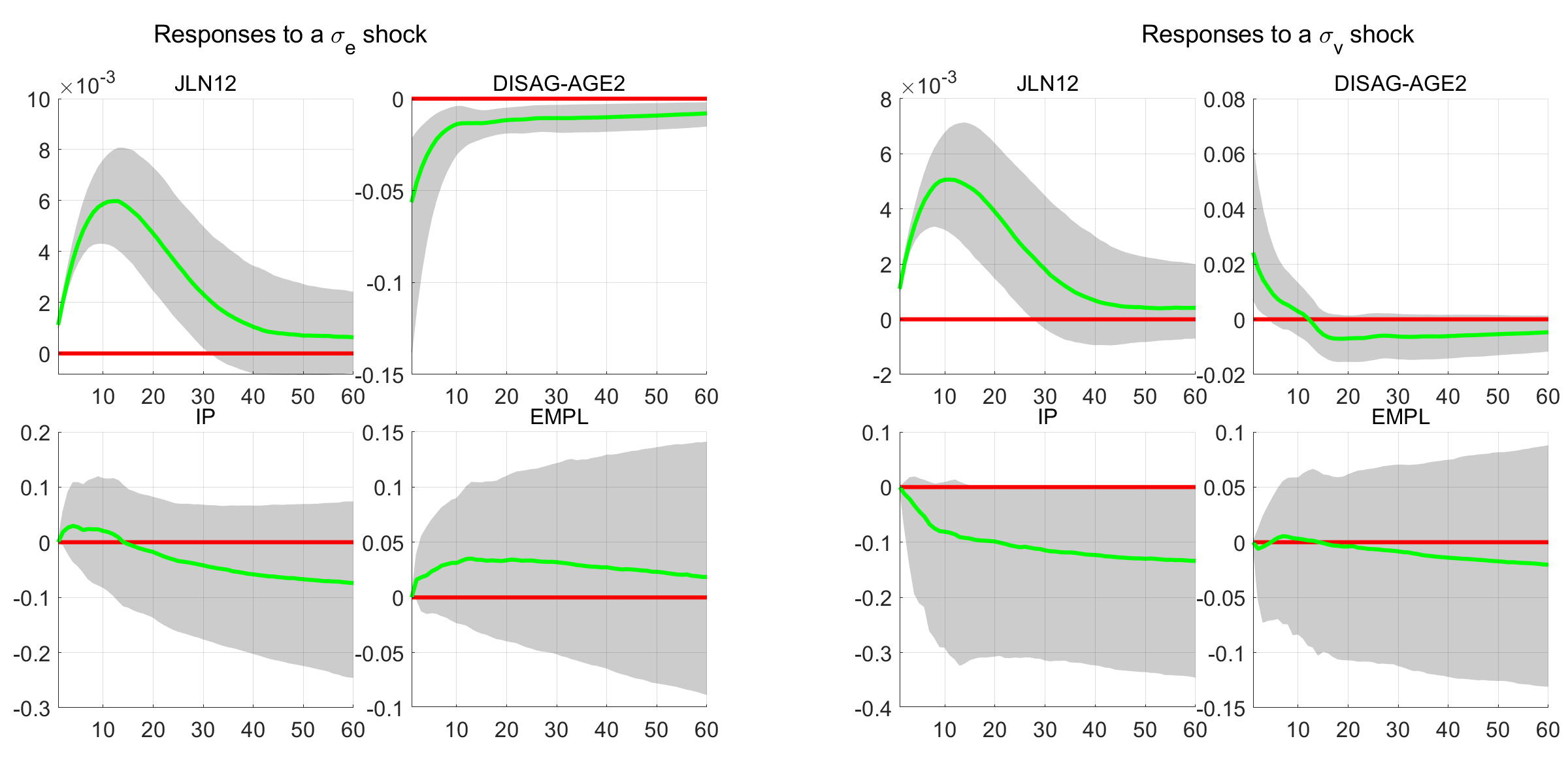}

\end{center}
\small \emph{Notes}: The figure shows impulse responses to the JLN 12-months-ahead uncertainty indicator (JLN12), the disagreement index for age 35-54 (DISAG-AGE2),  industrial production (IP), and employment (EMPL). We compute the IRFs from an eight-variable VAR system as described in the text. The shaded gray areas are the 16\% and 84\% posterior bands generated from the posterior distribution of VAR parameters. The units of the vertical axes are percentage deviations, and the horizontal axes report time measured in months.
\end{figure}

\begin{figure}[h!]
\caption{Benchmark model--age: 55 and above. Agreed $\sigma_{\varepsilon}$ (left) versus disagreed $\sigma_{v}$ (right) uncertainty.}\label{fig_IRF_benckmark_age3}
\begin{center}
\includegraphics[width=0.8\textwidth, trim={1cm .5cm 0 .4cm}]{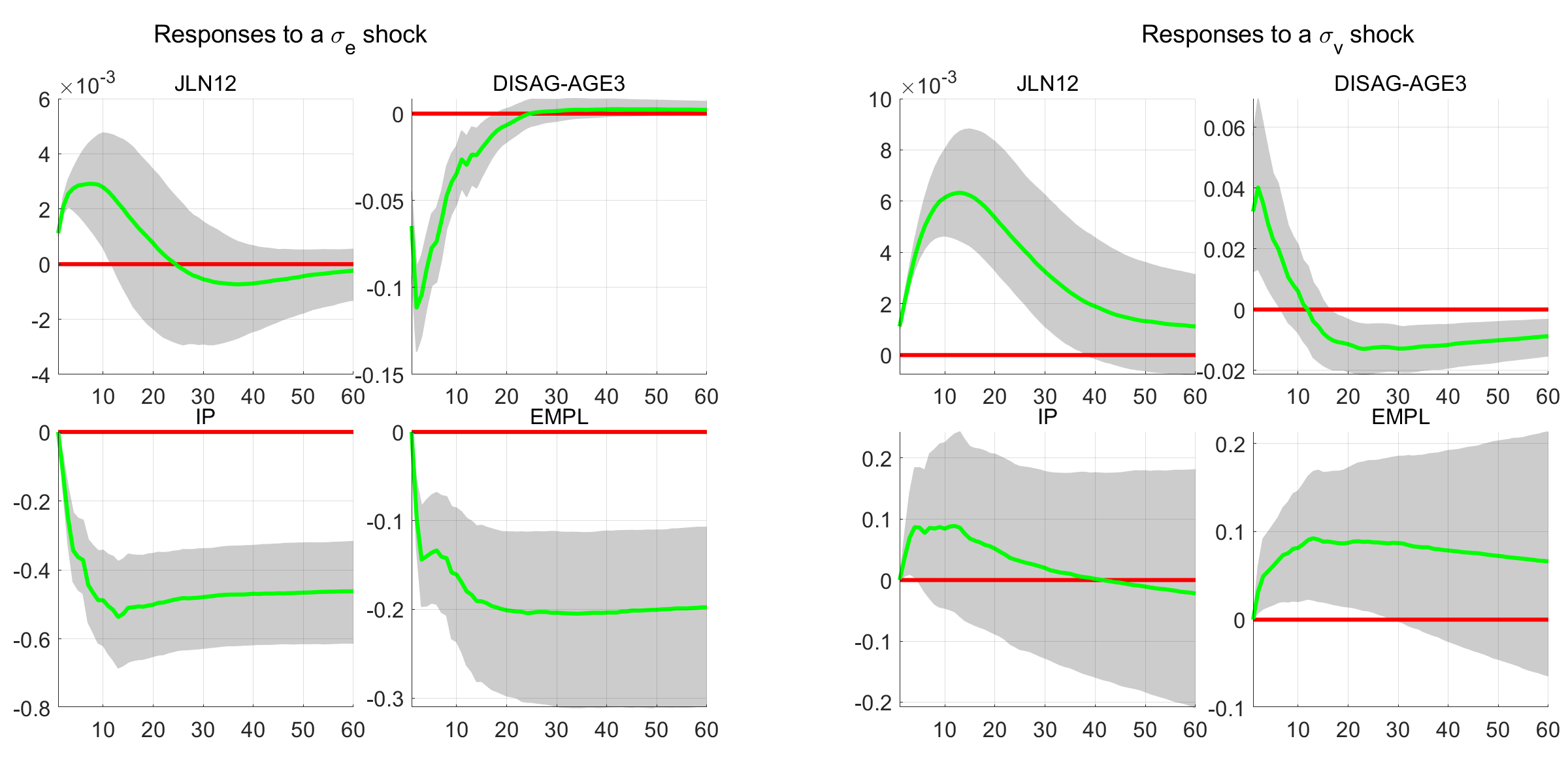}

\end{center}
\small \emph{Notes}: The figure shows impulse responses to the JLN 12-months-ahead uncertainty indicator (JLN12), the disagreement index for age 55 and above (DISAG-AGE3),  industrial production (IP), and employment (EMPL). We compute the IRFs from an eight-variable VAR system as described in the text. The shaded gray areas are the 16\% and 84\% posterior bands generated from the posterior distribution of VAR parameters. The units of the vertical axes are percentage deviations, and the horizontal axes report time measured in months.
\end{figure}

\textbf{Whose disagreement: Education and age.} In addition to the overall aggregate response to the survey questions, the MSC collects  demographic responses from consumers of different education and age status. They collect responses from three education categories, namely: \textit{high-school}, \textit{some college}, and \textit{college degree}. They also collect responses from three age groups: 18-34, 35-54, and 55 and above. In this section we compute disagreement indicators for each of these education and age groups -- six in total -- using the tails concept of disagreement.  We then, re-estimate the benchmark VAR using these indicators one at a time. Figures \ref{fig_IRF_benckmark_edu1}, \ref{fig_IRF_benckmark_edu2}, and \ref{fig_IRF_benckmark_edu3} display dynamic effects from the VARs that condition on the disagreement indicators based on the different education groups.  The dynamic effects of agreed and disagreed uncertainty shocks, when we condition on disagreement of the least educated group (\textit{high school} education, Figure \ref{fig_IRF_benckmark_edu1}), are very similar quantitatively to those dynamic effects reported for the baseline specification.  Interestingly, the IRFs from the specifications that condition on disagreement of the \textit{some college} and \textit{college degree} groups do not exhibit the sharp differences in the estimated real effects following agreed and disagreed uncertainty shocks.  Industrial production is the only variable that appears to respond statistically significantly following an agreed uncertainty shock. Figures \ref{fig_IRF_benckmark_edu2} and \ref{fig_IRF_benckmark_edu3} suggest industrial production and employment do not respond statistically significantly following a disagreed uncertainty shock.  It appears that disagreement from those consumers with \textit{high school} educations matters the most for the real activity effects we estimate in the benchmark specification.  We report results from the VAR specifications conditioned on disagreement indicators based on the three age groups.  When we condition the VAR on disagreement from the age groups, 18-34, and 35-54 age groups (see Figures \ref{fig_IRF_benckmark_age1}, \ref{fig_IRF_benckmark_age2}), the responses to industrial production and employment following agreed and disagreed uncertainty shocks are not statistically different from zero (with the exception of industrial production after the 40-month horizon in Figure \ref{fig_IRF_benckmark_age1}).  By contrast, when we condition on disagreement of the 55-and-over age group, the responses to the real activity indicators following agreed and disagreed uncertainty innovations (Figure \ref{fig_IRF_benckmark_age3}) are strong and statistically significant and very similar to the dynamic effects displayed in \ref{fig_IRF_benckmark}.  Moreover, the dynamic responses of the real activity indicators to agreed and disagreed innovations display the systematic differences we estimate in the benchmark specification. These results suggest that disagreement from the 55-and-over age group appears to be the most relevant driver behind our benchmark results, which are based on the aggregate responses.

\section{Conclusion\label{Sec_Conclusion}}

In this paper we establish two new, distinct concepts of uncertainty shocks, namely, \emph{agreed} and \emph{disagreed} uncertainty shocks. We show that the dispersion of consumer views about current and future economic conditions, measured by consumer disagreement, is an important conditioning factor for the effect of uncertainty on economic activity.  We present a dispersed and noisy information model where agents form expectations by processing idiosyncratic signals about an economic fundamental. We use the model to illustrate the connection between consumer disagreement, which is a manifestation of information dispersion, and uncertainty. The model shows that the change in observed uncertainty, measured by the variance of a forecast error, is a function of both the variance of the fundamental shock and the variance of the idiosyncratic noise.  Thus, a larger dispersion of views on economic conditions (i.e., higher consumer disagreement) may increase the variance of the forecast error without involving any change in the volatility of exogenous
fundamental forces in the economy.

We use the model to formulate simple sign restrictions that disentangle the dynamic effects of innovations to \emph{agreed} and \emph{disagreed} uncertainty on U.S. economic indicators in a medium-scale Bayesian VAR model. In our benchmark specification, innovations in agreed uncertainty foreshadow significant and often long-lasting, depressing effects on economic activity, namely, industrial production and employment, corroborating the evidence from numerous studies. By contrast, innovations in disagreed uncertainty (a rise in uncertainty in periods of high consumer disagreement) are benign for economic activity indicators, and they often lead to a short-run, positive responses in those indicators. Our analysis suggests that shocks to disagreed uncertainty are non-recessionary. Our results imply it is important to distinguish between the two types of uncertainty shocks to study the link between uncertainty and economic activity.

Our study opens up interesting avenues for future research. The analysis implies that the disclosure of information that reduces disagreement may increase the adverse effect of uncertainty. A straightforward extension of our analysis is to study how policy announcements that convey information about the economy may result in lower disagreement and exacerbate the negative effect of uncertainty. It would be interesting to study whether a strategic diffusion of information that maintains a wide range of views could alleviate, or even overturn, the adverse effect of uncertainty. Finally, our results show that the heterogeneity of views is critical for the aggregate effect of uncertainty on output, suggesting that models with heterogenous agents may prove fruitful for the study of expectations and the interplay between uncertainty and economic activity. We plan to pursue some of these ideas in future work.

\newpage

\bibliographystyle{apalike}
\bibliography{biblio_GKTZ}

\newpage

\appendix

\section*{Appendix}

\renewcommand{\theequation}{A.\arabic{equation}} \setcounter{equation}{0}
\renewcommand{\thefigure}{A.\arabic{figure}} \setcounter{figure}{0}
\renewcommand{\thetable}{A.\arabic{table}} \setcounter{table}{0}

\section{Proof of inequality (\ref{eq_uncert_on_FE})}\label{proof_inequality}

We can write equation \eqref{eq_uncert_on_FE} as
\begin{equation}
\psi_{0}^{2} + \psi_{1}^{2} +  \psi_{2}^{2} + ... +  \psi_{k-1}^{2} >  - \psi_{k}^{2}\left(\frac{\sigma_{v}^2}{\sigma_{v }^2+ \sigma_{\varepsilon}^2}\right)\left(\frac{\sigma^2_v - \sigma_{\varepsilon}^2 }{\sigma_{v }^2+ \sigma_{\varepsilon}^2} \right). \label{eq_uncert_on_FE2}
\end{equation}
All variance parameters are positive, hence, the term $A = \left(\frac{\sigma_{v}^2}{\sigma_{v }^2+ \sigma_{\varepsilon}^2}\right)$ can take values in the support $[0,1]$ and the term $B = \left(\frac{\sigma^{2}_{v} - \sigma_{\varepsilon}^2 }{\sigma_{v }^2+ \sigma_{\varepsilon}^2} \right)$ in the support $[-1,1]$. Consequently, the term on the RHS of equation \eqref{eq_uncert_on_FE2} attains a maximum when $A=1$ and $B=-1$, and it is sufficient to prove that
\begin{equation}
\psi_{0}^{2} + \psi_{1}^{2} +  \psi_{2}^{2} + ... +  \psi_{k-1}^{2} >  \psi_{k}^{2}.
\end{equation}
As all $\psi_{i}^{2}$ are non-negative for $i=1,...,k$, it suffices to show that
\begin{equation}
\psi_{0}^{2}  >  \psi_{k}^{2}.
\end{equation}

An MA($k$) process is invertible if the associated polynomial equation
\begin{equation}
\psi(z) = \psi_0 z^k + \psi_1 z^{k-1} + ... + \psi_{k-1} z + \psi_{k},
\end{equation}
has $k$ real or complex-valued roots $z^{\star}_j$, $j=1,...,k$, that are inside the unit circle, that is, $\vert z^{\star}_{j} \vert < 1 $. By Vieta's formulas we have that
\begin{equation}
\prod_{j=1}^{k}z^{\star}_{j} = (-1)^k \frac{\psi_{k}}{\psi_{0}}.
\end{equation}
If we take squares on both sides, and we use the fact that $\prod_{j=1}^{k}\left(z^{\star}_{j}\right)^2 < 1$ because $\vert z^{\star}_{j} \vert < 1$, we have that
\begin{eqnarray}
\left((-1)^k\frac{\psi_{k}}{\psi_{0}}\right)^2 & < & 1 \Rightarrow \\
\frac{\psi_{k}^2}{\psi_{0}^2} & < & 1 \Rightarrow \\
\psi_{0}^{2} & > &  \psi_{k}^{2},
\end{eqnarray}
which means that inequalities \eqref{eq_uncert_on_FE2} and \eqref{eq_uncert_on_FE} hold.

\newpage

\renewcommand{\theequation}{B.\arabic{equation}} \setcounter{equation}{0}
\renewcommand{\thefigure}{B.\arabic{figure}} \setcounter{figure}{0}
\renewcommand{\thetable}{B.\arabic{table}} \setcounter{table}{0}
\section{Data Appendix\label{sec:data_appendix}}
Our benchmark results are based on a VAR with eight monthly time series, but overall we test the robustness of our results using multiple measures of disagreement (based on different statistical measures of qualitative variation, different measures of current and future expectations, and different demographic groupings), multiple measures of uncertainty proposed in the literature, and multiple combinations of macroeconomic variables. All series are presented in \autoref{table:monthly_data} below. There are four main sources indicated in the third column of this table, FRED (Federal Reserve Economic Data, \href{https://fred.stlouisfed.org/}{https://fred.stlouisfed.org/}), UofM (University of Michigan Survey of consumers, \href{https://data.sca.isr.umich.edu/}{https://data.sca.isr.umich.edu/}), JLN2015 (data from \cite{Jurado_etAl_AER2015}, available at \href{https://www.sydneyludvigson.com/macro-and-financial-uncertainty-indexes}{https://www.sydneyludvigson.com/macro-and-financial-uncertainty-indexes}) and Philly Fed (Federal Reserve Bank of Philadelphia, Business Outlook Survey, \href{https://www.philadelphiafed.org/surveys-and-data/regional-economic-analysis/manufacturing-business-outlook-survey}{https://www.philadelphiafed.org/surveys-and-data/regional-economic-analysis/manufacturing-business-outlook-survey}).\footnote{The Business Outlook Survey (BOS) data are used to extract the uncertainty index of \cite{Bachmann_al_AEJM13} based on question 4 of the survey (expectations about shipments from six months from now). As the authors do not provide updates on this index, we use the raw BOS data and apply the transformation $FDISP_{t}$ \citep[see][page 7]{Bachmann_al_AEJM13} to compute its values.} All data were downloaded in different dates throughout July 2021. The fourth column of \autoref{table:monthly_data} shows the stationarity transformations applied to the series, where Tcode = 1 is for levels and Tcode = 5 is for first differences of the natural logarithm. Series that are originally observed at daily or weekly frequencies (e.g. FEDFUNDS) are converted into monthly by taking simple arithmetic averages over the calendar month.

We also estimate a VAR on quarterly data to test the robustness of our findings. This is particularly important as the MSC data are available at quarterly frequency from 1960, instead of 1978 for monthly data. Additionally, at the quarterly frequency we are able to use GDP as the proxy for output, plus we have other important series not available at the monthly frequency (for example, investment). The uncertainty measure is again JLN12 and for disagreement we use our primary tails index (both averaged over the quarter to produce quarterly frequencies). The additional quarterly variables are (FRED mnemonics, followed by Tcode, in parenthesis): GDP (GDPC1, 5), real personal consumption (PCECC96
, 5), GDP deflator (GDPDEF, 5), real gross private domestic investment (GPDIC1
), employment (PAYEMS, 5), Federal funds rate (FEDFUNDS, 1). Data on the S\&P 500 index are quarterly averages of the same monthly series presented in \autoref{table:monthly_data}. We also use a measure of stock market variance, but because the popular VIX index is available only since 1985, we use instead variable SVAR obtained from \href{https://sites.google.com/view/agoyal145}{Amit Goyal}'s webpage.\footnote{See Ivo Welch and Amit Goyal (2008), A Comprehensive Look at The Empirical Performance of Equity Premium Prediction, The Review of Financial Studies, Volume 21, Issue 4, Pages 1455-1508.} All quarterly series are available for the period 1960Q1 - 2020Q4.

\begin{table}
\centering
\caption{Monthly dataset, 1978M1 - 2020M12} \label{table:monthly_data}
\resizebox{.95\textwidth}{!}{
\begin{tabular}{llcc}
Mnemonic	&	Description	&	Source	&	Tcode	\\ \hline
\multicolumn{4}{c}{\underline{\textsc{Variables in benchmark VAR}}}							\\
JLN12	&	JLN Macroeconomic Uncertainty, 12 months	&	JLN2015	&	1	\\
DISAG	&	Tails disagreement total index	&	UofM	&	1	\\
INDPRO	&	Industrial Production: Total Index	&	FRED	&	5	\\
PCEPI	&	Personal Consumption Expenditures (Price Index)	&	FRED	&	5	\\
DPCERA3M086SBEA	&	Real personal consumption expenditures	&	FRED	&	5	\\
PAYEMS	&	All Employees, Total Nonfarm	&	FRED	&	5	\\
SP500	&	S\&P 500	&	Yahoo! Finance	&	5	\\
FEDFUNDS	&	Federal Funds Effective Rate	&	FRED	&	1	\\
\multicolumn{4}{c}{\underline{\textsc{Other macroeconomic variables}}}							\\
DDURRA3M086SBEA	&	Real personal consumption expenditures: Durable goods	&	FRED	&	5	\\
DNDGRA3M086SBEA	&	Real personal consumption expenditures: Nondurable goods	&	FRED	&	5	\\
DSERRA3M086SBEA	&	Real personal consumption expenditures: Services	&	FRED	&	5	\\
AHETPI	&	Total Average Hourly Earnings of Production and Nonsupervisory Employees	&	FRED	&	5	\\
AWHNONAG	&	Total Average Weekly Hours of Production and Nonsupervisory Employees	&	FRED	&	5	\\
FLTOTALSL	&	Total Consumer Credit Owned and Securitized, Flow	&	FRED	&	1	\\
TERMCBPER24NS	&	Finance Rate on Personal Loans at Commercial Banks	&	FRED	&	1	\\
HWI	&	Composite help-wanted index	&	FRED	&	1	\\
PAGO	&	Current Financial Situation Compared with a Year Ago	&	UofM	&	1	\\
PEXP	&	Expected Change in Financial Situation in a Year	&	UofM	&	1	\\
RINC	&	Expected Change in Real Household Income During Next Year	&	UofM	&	1	\\
UMEX	&	Expected Change in Unemployment During the Next Year	&	UofM	&	1	\\
DUR-ALL	&	Buying Conditions for Large Household Durables	&	UofM	&	1	\\
VEH-ALL	&	Buying Conditions for Vehicles	&	UofM	&	1	\\
\multicolumn{4}{c}{\underline{\textsc{Alternative measures of uncertainty}}}							\\
JLNF12	&	JLN Financial Uncertainty, 12 months	&	JLN2015	&	1	\\
BOS DISP	&	Business Outlook Survey uncertainty index (expectations about shipments)	&	Philly Fed	&	1	\\
VXO	&	CBOE S\&P 100 Volatility Index: VXO, Index, Monthly	&	FRED	&	1	\\
USEPUINDXM	&	Economic Policy Uncertainty Index for United States	&	FRED	&	1	\\
\multicolumn{4}{c}{\underline{\textsc{Alternative measures of disagreement}}}							\\
DISAG-E	&	Entropy disagreement total index	&	UofM	&	1	\\
DISAG-L	&	Lacey disagreement index	&	UofM	&	1	\\
DISAG\_HS	&	Tails disagreement total index, High-school	&	UofM	&	1	\\
DISAG\_SC	&	Tails disagreement total index, Some college	&	UofM	&	1	\\
DISAG\_CD	&	Tails disagreement total index, College degree	&	UofM	&	1	\\
DISAG\_A18-34	&	Tails disagreement total index, Age group 18-34	&	UofM	&	1	\\
DISAG\_A35-54	&	Tails disagreement total index, Age group 35-54	&	UofM	&	1	\\
DISAG\_A55+	&	Tails disagreement total index, Age group 55+	&	UofM	&	1	\\
\hline \hline
\end{tabular}
}
\end{table}

\renewcommand{\theequation}{C.\arabic{equation}} \setcounter{equation}{0}
\renewcommand{\thefigure}{C.\arabic{figure}} \setcounter{figure}{0}
\renewcommand{\thetable}{C.\arabic{table}} \setcounter{table}{0}
\section{Econometric Methodology\label{app_ecometrix_method}}

This appendix describes the structural vector autoregression methodology for identifying $\sigma_{e}$ and $\sigma_{v}$ shocks via sign restrictions. The core VAR formulation follows \cite{Korobilis2022}, who develops an efficient algorithm for posterior inference in VARs with sign restrictions. This algorithm allows for estimating VARs of arbitrarily large dimensions, and is particularly suited for the monthly medium-scale VAR models with 13 lags we use in this paper. For the $n \times 1$ vector of time series variables $\bm y_{t}$ the VAR takes the multivariate regression form
\begin{equation}
\mathbf{y}_{t} = \mathbf{\Phi} \mathbf{x}_{t} + \bm{\varepsilon}_{t}, \label{App_VAR}
\end{equation}
where $\mathbf{y}_{t}$ is a $\left( n \times 1 \right)$ vector of observed variables, $\mathbf{x}_{t} = \left( 1,\mathbf{y}_{t-1}^{\prime},...,\mathbf{y}_{t-p}^{\prime} \right)^{\prime}$ a $\left( k \times 1 \right)$ vector (with $k=np+1$) containing a constant and $p$ lags of $\mathbf{y}$, $\mathbf{\Phi}$ is an $(n \times k)$ matrix of coefficients, and $\bm{\varepsilon}_{t}$ a $\left( n \times 1 \right)$ vector of disturbances distributed as $N\left( \mathbf{0}_{n \times 1},\mathbf{\Omega} \right)$ with $\mathbf{\Omega}$ an $n \times n$ covariance matrix. We further assume the following factor decomposition of $\bm{\varepsilon}_{t}$
\begin{equation}
\bm{\varepsilon}_{t} = \mathbf{\Lambda} \mathbf{f}_{t} + \mathbf{v}_{t}, \label{App_factor_model}
\end{equation}
where $\mathbf{\Lambda}$ is an $n \times r$ matrix of factor loadings, $\mathbf{f}_{t} \sim N(\bm{0},\bm{I}_r)$ is an $r \times 1$ vector of factors, and $\mathbf{v}_{t} \sim N(\bm{0},\bm{\Sigma})$ is an $n \times 1$ vector of idiosyncratic shocks with $\bm{\Sigma}$ an $n \times n$ diagonal matrix.

The rationale behind the VAR model in equations \eqref{App_VAR}-\eqref{App_factor_model} is that the $n$-dimensional vector of VAR disturbances is decomposed into $r$ common shocks $\mathbf{f}_{t}$ ($r<n$) and $n$ idiosyncratic shocks $\mathbf{v}_{t}$. Because $\bm{\Sigma}$ is diagonal, we consider only the $r$ common shocks to be structural while the $n$ idiosyncratic shocks can be considered as nuisance shocks e.g. due to measurement error or asymmetric information. Indeed, by left-multiplying the VAR using the generalized inverse of $\mathbf{\Lambda}$, the implied structural VAR form is\begin{eqnarray}
\mathbf{y}_{t} & = & \mathbf{\Phi} \mathbf{x}_{t} + \mathbf{\Lambda} \mathbf{f}_{t} + \mathbf{v}_{t} \\
\left( \mathbf{\Lambda}^{\prime} \mathbf{\Lambda} \right)^{-1}\mathbf{\Lambda}^{\prime} \mathbf{y}_{t} & = & \left( \mathbf{\Lambda}^{\prime} \mathbf{\Lambda} \right)^{-1}\mathbf{\Lambda}^{\prime} \mathbf{\Phi} \mathbf{x}_{t} +  \mathbf{f}_{t} + \left( \mathbf{\Lambda}^{\prime} \mathbf{\Lambda} \right)^{-1}\mathbf{\Lambda}^{\prime} \mathbf{v}_{t} \\
\mathbf{A}_{1} \mathbf{y}_{t} & = & \mathbf{B}_{1} \mathbf{x}_{t} + \mathbf{f}_{t} + \left( \mathbf{\Lambda}^{\prime} \mathbf{\Lambda} \right)^{-1}\mathbf{\Lambda}^{\prime}\mathbf{v}_{t}. \label{RR_SVAR}
\end{eqnarray}
As long as $\bm{\Sigma}$ is diagonal the term $\left( \mathbf{\Lambda}^{\prime} \mathbf{\Lambda} \right)^{-1}\mathbf{\Lambda}^{\prime}\mathbf{v}_{t}$ vanishes asymptotically, meaning that $\mathbf{f}_{t}$ retains the interpretation of structural shocks. \cite{Korobilis2022} shows that structural identifying restrictions are identical to parametric restrictions on $\mathbf{\Lambda}$, and provides an efficient Markov chain Monte Carlo (MCMC) scheme for sampling such restrictions in high-dimensional VARs.\footnote{A VAR can be high-dimensional due to the large number of time series $T$, large number of endogenous variables $n$, large number of identified shocks $r$, large number of lags $p$, or combinations of these.}

Based on the model in equations \eqref{App_VAR}-\eqref{App_factor_model} the joint likelihood function can be written as
\begin{equation}
\left( \bm y \vert \bm x, \bm \Phi, \bm \Lambda, \bm f, \bm \Sigma\right) \sim \prod_{t=1}^{T} N \left(     \mathbf{\Phi} \mathbf{x}_{t}, \mathbf{\Lambda} \mathbf{\Lambda}' + \bm \Sigma \right)
\end{equation}
and we define the following prior distributions
\begin{eqnarray}
\bm{\phi}_{i} \equiv vec\left( \mathbf{\Phi}_{i} \right) & \sim & N_{k} \left( \mathbf{0}, \underline{\mathbf{V}}_{i} \right), \label{phi_prior} \\
\underline{\mathbf{V}}_{i,(jj)} & = & \sigma^{2}_{i} \tau_{i}^{2} \psi_{i,j}^{2}, \\
\psi_{i,j} & \sim & Cauchy^{+} \left(0, 1 \right),  \\
\tau_{i}  & \sim & Cauchy^{+} \left(0, 1 \right), \\
\mathbf{f}_{t} & \sim & N_{r} \left( \mathbf{0}, \mathbf{I} \right), \\
\mathbf{\Lambda}_{ij}  & \sim & \left\lbrace \begin{array}{ll} N\left(0, \underline{h}_{ij} \right) I( \Lambda_{ij} > 0), & \text{if $S_{ij} = 1$}, \\
N\left(0,\underline{h}_{ij} \right) I( \Lambda_{ij} < 0),  & \text{if $S_{ij} = -1$}, \\
\delta_{0} \left( \mathbf{\Lambda}_{ij}\right),  & \text{if $S_{ij} = 0$}, \\
N\left(0,\underline{h}_{ij} \right),  & \text{otherwise},
\end{array} \right.  \label{lam_prior}\\
\sigma^{2}_{i} & \sim & inv-Gamma \left(\underline{\rho}_{i},\underline{\kappa}_{i} \right), \label{sigma_prior}
\end{eqnarray}
for $i=1,...,n$, $j=1,...,r$, where $\mathbf{\Phi}_{i}$ is the $i^{th}$ row of $\mathbf{\Phi}$, $\sigma_{i}^{2}$ is the $i^{th}$ diagonal element of the matrix $\mathbf{\Sigma}$, and $\delta_{0} \left( \mathbf{\Lambda}_{ij}\right)$ is the Dirac delta function for $\mathbf{\Lambda}_{ij}$ at zero (i.e. a point mass function with all mass concentrated at zero). The hyperparameters $\psi_{i,j}$ and $\tau_{i}$ are components of a Horseshoe prior, which is a tuning-free shrinkage priors with excellent statistical properties \citep[see][for explanation and references to the statistics literature justifying the excellent theoretical properties of this prior]{Korobilis2022}. Therefore, we only need to select parameters with an underline, namely $\underline{h}_{ij},\underline{\rho}_{i},\underline{\kappa}_{i}$. As we typically do not have substantial prior information on these hyperparameters, it is fairly trivial to choose noninformative values. Following standard norms in Bayesian inference, we set $\underline{h}_{ij}=10$ and $\underline{\rho}_{i},\underline{\kappa}_{i}=0.01$ such that the priors in equations \eqref{lam_prior} and \eqref{sigma_prior} become locally Uniform. Posterior computation and impulse response inference follows \cite{Korobilis2022} and the reader should refer to this paper for technical details.

\renewcommand{\theequation}{D.\arabic{equation}} \setcounter{equation}{0}
\renewcommand{\thefigure}{D.\arabic{figure}} \setcounter{figure}{0}
\renewcommand{\thetable}{D.\arabic{table}} \setcounter{table}{0}
\section{Additional results} \label{app_robustness}

In this Appendix we report i), the complete set of IRFs estimated from the VAR specifications in the main body of the paper and ii), results from VAR specifications that use various economic, financial, and survey indicators, iii), results from VAR specifications that use alternative proxies for uncertainty, iv) VAR specifications with disagreement indices derived from specific questions.  These VAR specifications serve to examine the robustness of the main finding in the main body of the paper, namely, the different dynamic effects of agreed and disagreed uncertainty shocks.

\paragraph{\textbf{Complete IRFs from benchmark model.}} Figure \ref{fig_IRF_macro1_full} below displays the complete set of IRFs from the benchmark specification.  Private consumption displays a negative effect following an agreed uncertainty innovation consistent with the negative responses estimated for industrial production and employment. Following a disagreed uncertainty innovation private consumption exhibits a small short run increase that is nevertheless not statistically significant. There are also systematic differences in the responses of S\&P 500, Federal Funds rate and consumer price inflation following agreed and disagreed uncertainty innovations, suggesting that the qualitatively different dynamic responses following the two shocks are broad based.

\begin{figure}[h!]
\caption{Benchmark model. Agreed  $\sigma_{\varepsilon}$ (left) versus disagreed $\sigma_{v}$ (right) uncertainty.}\label{fig_IRF_macro1_full}
\begin{center}
\includegraphics[width=0.95\textwidth,, trim={1cm .5cm 0 .4cm}]{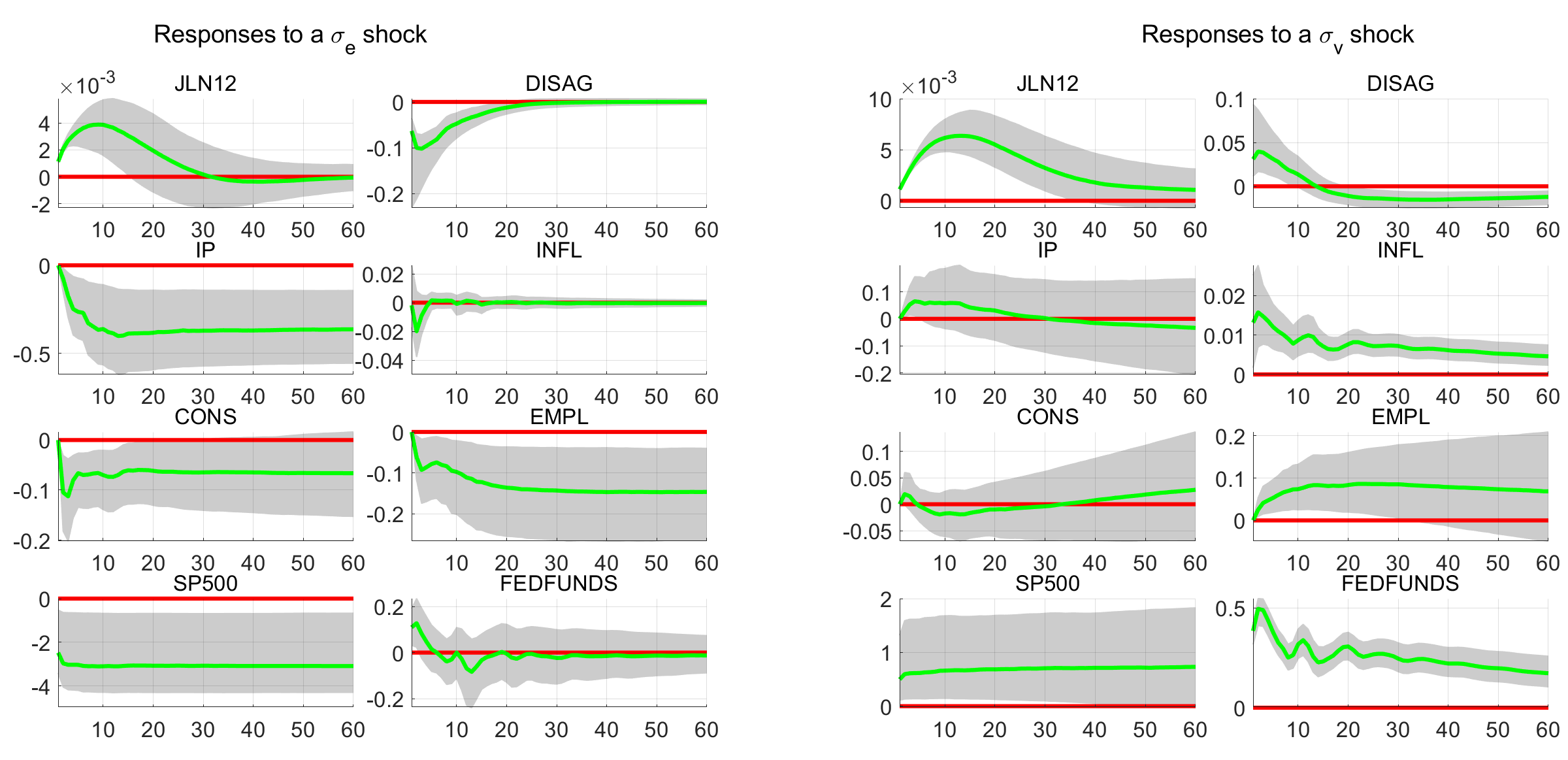}

\end{center}
\small \emph{Notes}: The figure shows impulse responses from a eight-variable VAR system on JLN 12-month ahead uncertainty indicator (JLN12), disagreement index (DISAG),  Industrial production (IP), private consumption (CONS), Consumer price inflation (INFL), employment (EMPL), S\&P 500 index (SP500), Federal funds rate (FEDFUNDS). The shaded gray areas are the 16\% and 84\% posterior bands generated from the posterior distribution of VAR parameters. The units of the vertical axes are percentage deviations, while the horizontal axes reports time measured in months.
\end{figure}

The benchmark specification in section \ref{sec_model_empirical} used the  macro uncertainty measure (JLN-12) as the baseline measure of uncertainty. In this section we replace JLN-12 in the benchmark VAR with four alternative uncertainty measures used in earlier work. \cite{Jurado_etAl_AER2015} developed the 12-month ahead financial uncertainty indicator (henceforth JLNF-12) using estimates of conditional volatilities of $h$-step ahead forecast errors from 147 financial time series.  \cite{Ludvigson_etAl_AEJ} suggest this indicator is a preferable measure of uncertainty as it is less likely to be confounded by exogenous shocks --and hence can be treated as an exogenous source of variation in uncertainty-- in comparison to JLN-12.  Beyond this measure we also use the business dispersion measure (BOS-dispersion), developed in \cite{Bachmann_al_AEJM13}, and a popular measure of stock market volatility (CBOE S\&P 100 volatility index VXO) used in several important studies as a proxy for uncertainty (\citealp{bloom_ECTA}, \citealp{Gilchrist2014}, \citealp{Basu_Bundick_ECTA2017}). Finally, we use the Economic Policy Uncertainty index (EPU) developed by \cite{BakerBloomDavis16}. The latter is developed using text mining methods and captures uncertainty, broadly speaking, about future fiscal, monetary, trade, regulatory policy actions.

\paragraph{\textbf{Complete IRFs from VAR model with JLNF-12.}} Figure \ref{fig_IRF_macro2_full} below displays the complete set of IRFs from the VAR specification with JLNF-12 used as the uncertainty indicator discussed in section \ref{sec_robustness} of the main body.  The IRFs are broadly consistent with the IRFs displayed in the Figure \ref{fig_IRF_macro1_full} above. The left panel which plots the IRFs to the agreed uncertainty shock displays very similar --qualitatively-- depressing effects on industrial production and employment in comparison to the effects estimated when JLN12 is used as the uncertainty proxy.  The negative effects on industrial production and employment are however significantly smaller in magnitude (evaluated at the peak median response) in comparison to the negative effects estimated in the benchmark specification displayed in Figure \ref{fig_IRF_macro1_full}. This appears to be a consequence of the smaller increase of the JLNF-12 uncertainty indicator in comparison to the response of the JLN-12 indicator displayed in Figure \ref{fig_IRF_macro1_full}.  The depressive effects on economic activity are consistent with the evidence in \cite{Ludvigson_etAl_AEJ} who also use JLNF-12 as the uncertainty indicator in their empirical analysis.  The right panel displays the IRFs following an innovation to the disagreed uncertainty shock. Qualitatively the dynamic effects estimated are very much in line with those displayed in \ref{fig_IRF_macro1_full}.  Both industrial production and employment exhibit a short run positive and statistically significant response. Thus, both the benchmark and this alternative specification suggest that innovations to disagreed uncertainty display a benign effect on economic activity, strikingly different from the strong depressing effect on activity estimated under agreed uncertainty.

\begin{figure}[h!]
\caption{JLNF-12 measure. Agreed  $\sigma_{\varepsilon}$ (left) versus disagreed $\sigma_{v}$ (right) uncertainty.}\label{fig_IRF_macro2_full}
\begin{center}
\includegraphics[width=0.95\textwidth,, trim={1cm .5cm 0 .4cm}]{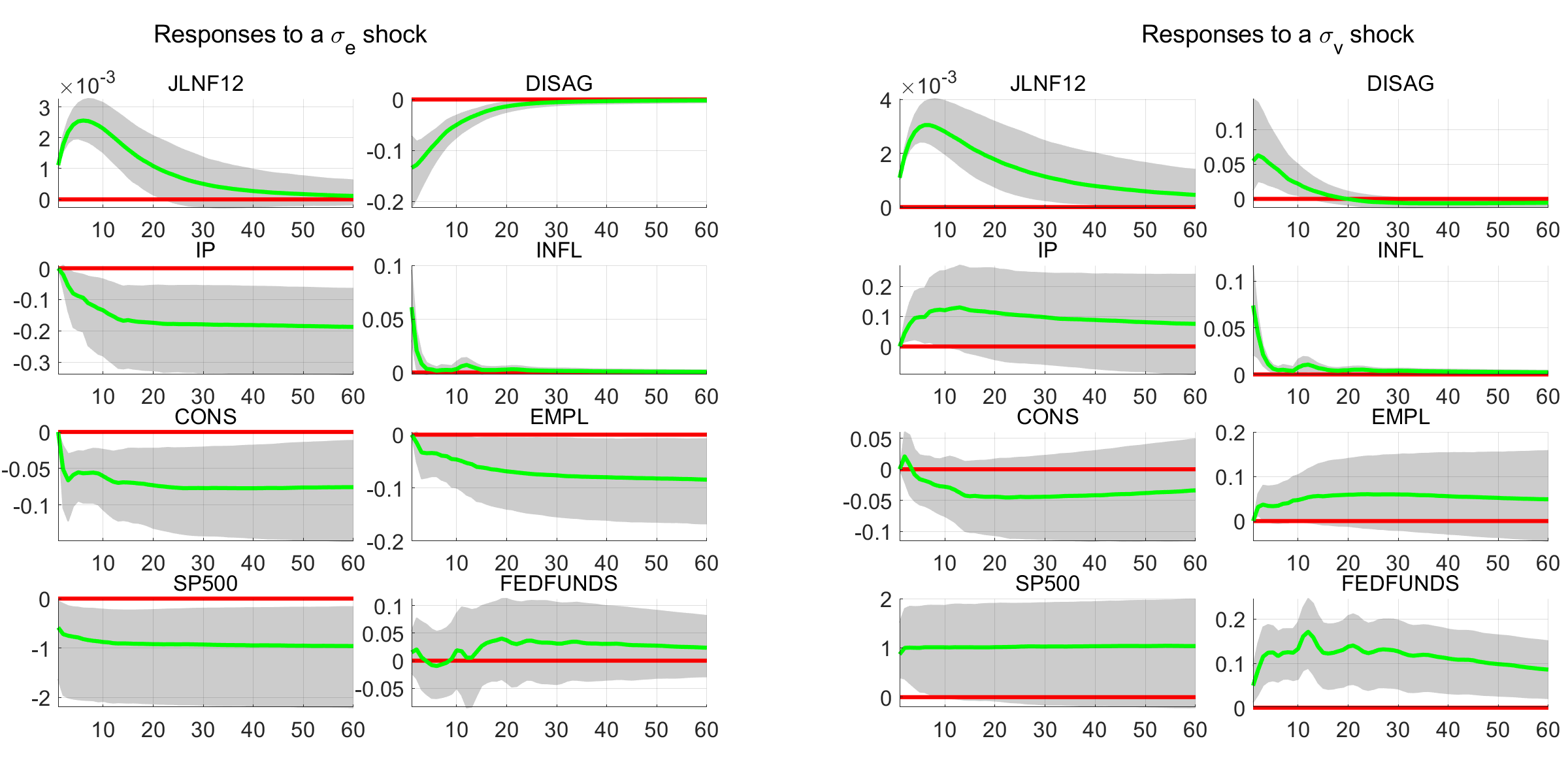}

\end{center}
\small \emph{Notes}: The figure shows impulse responses from a eight-variable VAR system on JLNF 12-month ahead uncertainty indicator (JLNF12), disagreement index (DISAG),  Industrial production (IP), private consumption (CONS), Consumer price inflation (INFL), employment (EMPL), S\&P 500 index (SP500), Federal funds rate (FEDFUNDS). The shaded gray areas are the 16\% and 84\% posterior bands generated from the posterior distribution of VAR parameters. The units of the vertical axes are percentage deviations, while the horizontal axes reports time measured in months.
\end{figure}

\paragraph{\textbf{Complete IRFs from VAR model with BOS-DISPERSION.}} Figure \ref{fig_IRF_macro3_full} below displays the complete set of IRFs from a VAR specification with the BOS-DISPERSION used as the uncertainty indicator.  A key difference in this Figure in comparison to Figure is the different dynamic response of uncertainty: while uncertainty rises under both agreed and disagreed shocks, the business dispersion measure displays more short-lived and non-persistence dynamics in comparison to JLN-12 or JLNF-12. This may not be surprising given this dispersion measure is based on a very different information set --firms in the manufacturing sector, in comparison to the broad spectrum of variables considered in JLN-12 aand JLNF-12. Nevertheless the dynamic effects following an agreed uncertainty shock identified from this alternative indicator suggest a strong and long lasting period of depressed activity, very much in line with conventional wisdom and our findings above.  Focussing on the dynamic effects following a disagreed uncertainty shock, broadly speaking, real activity indicators do not respond in a statistically significant manner.  This suggests the response of real activity indicators is non-contractionary under disagreed uncertainty shocks, identified from this uncertainty proxy, and importantly there is a distinct quantitative difference between the dynamic effects estimated under disagreed and agreed uncertainty shocks.

\begin{figure}[h!]
\caption{Business Dispersion measure. Agreed  $\sigma_{\varepsilon}$ (left) versus disagreed $\sigma_{v}$ (right) uncertainty.}\label{fig_IRF_macro3_full}
\begin{center}
\includegraphics[width=0.95\textwidth,, trim={1cm .5cm 0 .4cm}]{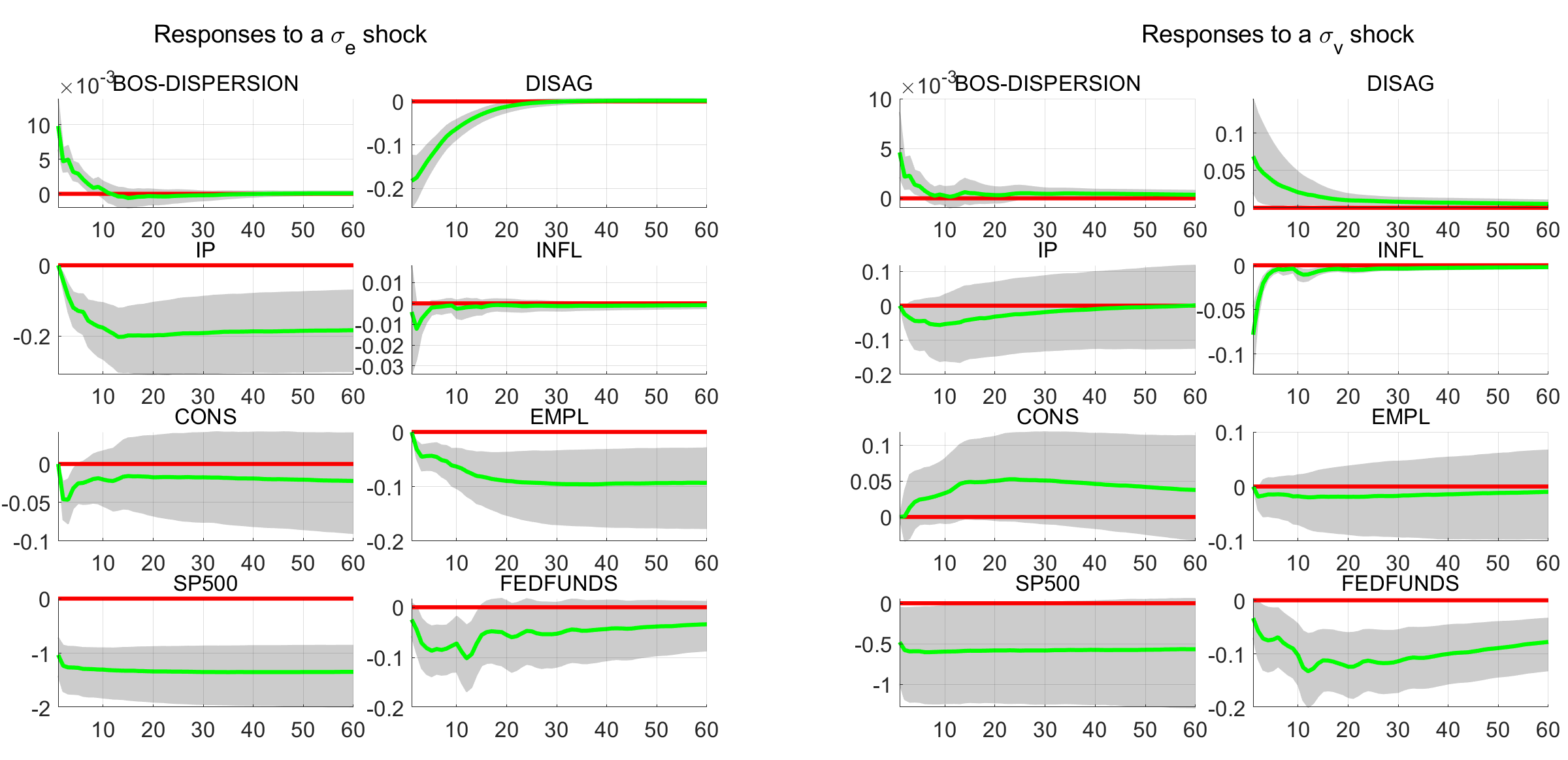}

\end{center}
\small \emph{Notes}: The figure shows impulse responses from a eight-variable VAR system on Business dispersion indicator (BOS-DISPERSION), disagreement index (DISAG),  Industrial production (IP), private consumption (CONS), Consumer price inflation (INFL), employment (EMPL), S\&P 500 index (SP500), Federal funds rate (FEDFUNDS). The shaded gray areas are the 16\% and 84\% posterior bands generated from the posterior distribution of VAR parameters. The units of the vertical axes are percentage deviations, while the horizontal axes reports time measured in months.
\end{figure}

\paragraph{\textbf{Complete IRFs from VAR model with VXO.}} Figure \ref{fig_IRF_macro4_full} below displays the complete set of IRFs from the VAR specification with VXO used as the uncertainty indicator. Uncertainty responds sharply and it is short-lived.  The IRFs to an agreed uncertainty innovation are consistent with a depressing and long lasting effect on real activity --this very similar to what we have estimated in all other specifications.  The dynamic effects estimated following a disagreed uncertainty shock suggest an initial short run but barely statistically significant depressing effect on industrial production which reverts very quickly to the pre-shock level.  The employment response suggests a statistically significant decline. The response of employment is qualitatively different to the findings reported so far.  We do however emphasize two important caveats with this specification. First, the sample period is different, due to the availability of the VXO indicator, beginning in 1986M1.  Second, and more importantly the IRFs suggest that the identification of the disagreed uncertainty shock is potentially problematic.   The positive sign restriction on disagreement that identifies this shock is not satisfied in a statistical sense. Specifically, note that the response of disagreement is not statistically significant different from zero following an innovation to disagreed uncertainty.  We therefore, do not consider this joint response of uncertainty and disagreement, strictly speaking, as genuinely identifying a disagreed uncertainty shock.  Because we want to be as conservative as possible, our sign restrictions put very minimal constraints on the dynamics.  This suggests that more identifying restrictions may be fruitful in order to clearly separate the two types of uncertainty shocks, when using this uncertainty indicator. Nevertheless, there are some quantitative differences in the responses of the real activity indicators; the economic effects following this type of innovation are significantly smaller in magnitude in comparison to the economic effects estimated following an agreed uncertainty innovation.

\begin{figure}[h!]
\caption{Stock market implied volatility (VXO) measure. Agreed  $\sigma_{\varepsilon}$ (left) versus disagreed $\sigma_{v}$ (right) uncertainty.}\label{fig_IRF_macro4_full}
\begin{center}
\includegraphics[width=0.95\textwidth,, trim={1cm .5cm 0 .4cm}]{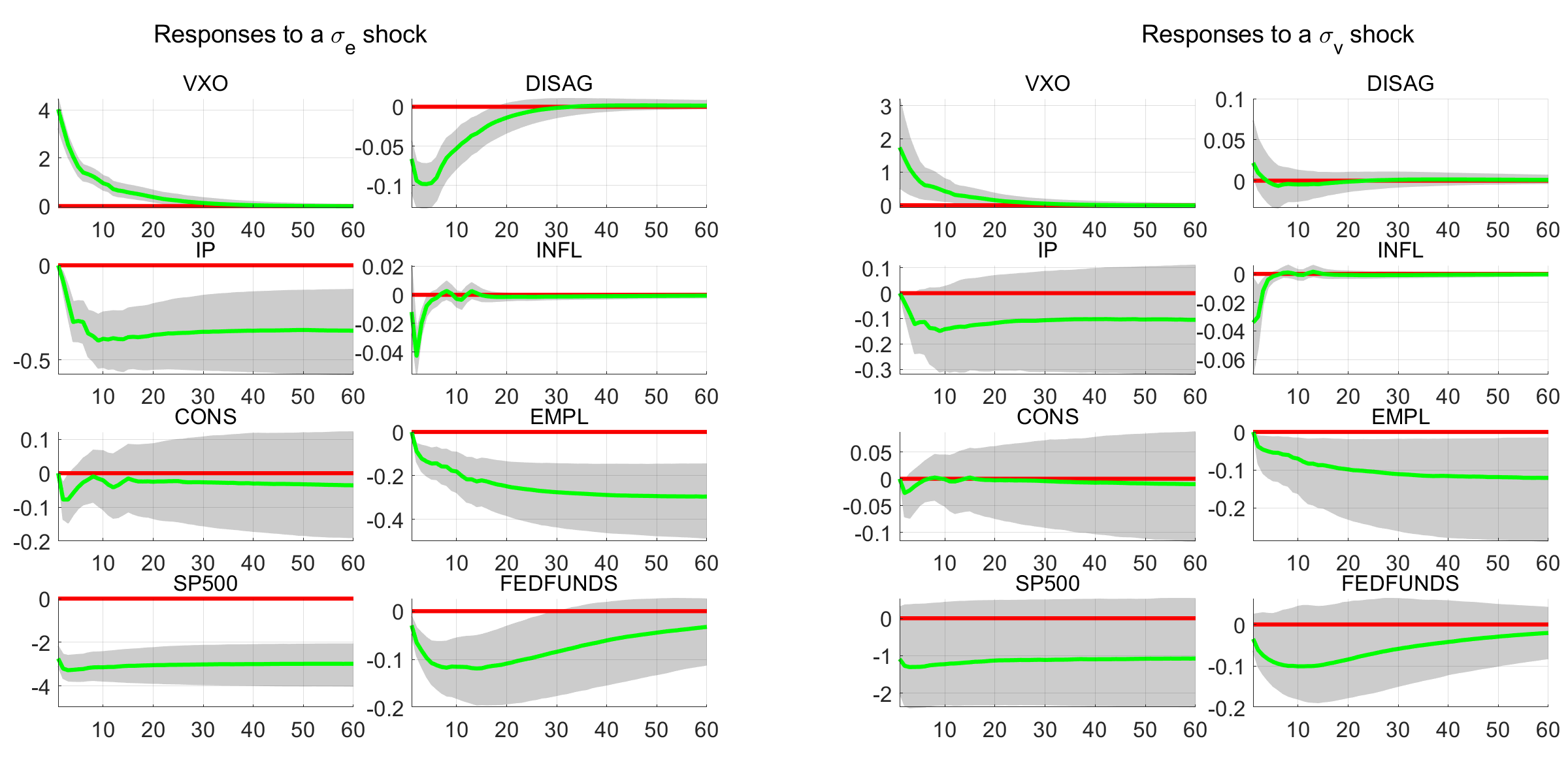}

\end{center}
\small \emph{Notes}: The figure shows impulse responses from a eight-variable VAR system on CBOE S\&P 100 volatility index (VXO), disagreement index (DISAG),  Industrial production (IP), private consumption (CONS), Consumer price inflation (INFL), employment (EMPL), S\&P 500 index (SP500), Federal funds rate (FEDFUNDS). The shaded gray areas are the 16\% and 84\% posterior bands generated from the posterior distribution of VAR parameters. The units of the vertical axes are percentage deviations, while the horizontal axes reports time measured in months.
\end{figure}

\paragraph{\textbf{Complete IRFs from VAR model with EPU.}} Figure \ref{fig_IRF_macro5_full} below displays the complete set of IRFs from the VAR specification with EPU used as the uncertainty indicator.  Its not straightforward to map the connection of this concept of uncertainty to the broad based macro or financial uncertainty indicators examined above. Moreover, the EPU indicator is not clearly related to our central measure of consumer disagreement as the latter refers to business conditions and the former is focussed on economic policy. Therefore its not straightforward to relate a change in information dispersion to the volatility of this indicator which is derived from text mining methods.  The sample period, 1985M1 to 2020M12, for this specification is different to the benchmark due to the availability of the EPU index. We nevertheless wanted to examine the behavior of the real activity indicators using the concepts of agreed and disagreed uncertainty identified via this measure. Figure \ref{fig_IRF_macro5_full} suggests a broad similarity to our findings when considering the dynamic effects following the agreed uncertainty shock --both industrial production and employment exhibit long lasting and depressing effects.  While the response of industrial production is negative following both agreed and disagreed shocks, the response of employment is not statistically significant in the case of disagreed uncertainty shock (except barely so in month two only). We suggest this may be partly due to the fact that, similar to the VXO specification above, the identification of disagreed uncertainty shocks appears to be problematic since the disagreement index barely moves in a statistically significant manner in the case of the disagreed uncertainty shock.

\begin{figure}[h!]
\caption{EPU measure. Agreed  $\sigma_{\varepsilon}$ (left) versus disagreed $\sigma_{v}$ (right) uncertainty.}\label{fig_IRF_macro5_full}
\begin{center}
\includegraphics[width=0.95\textwidth,, trim={1cm .5cm 0 .4cm}]{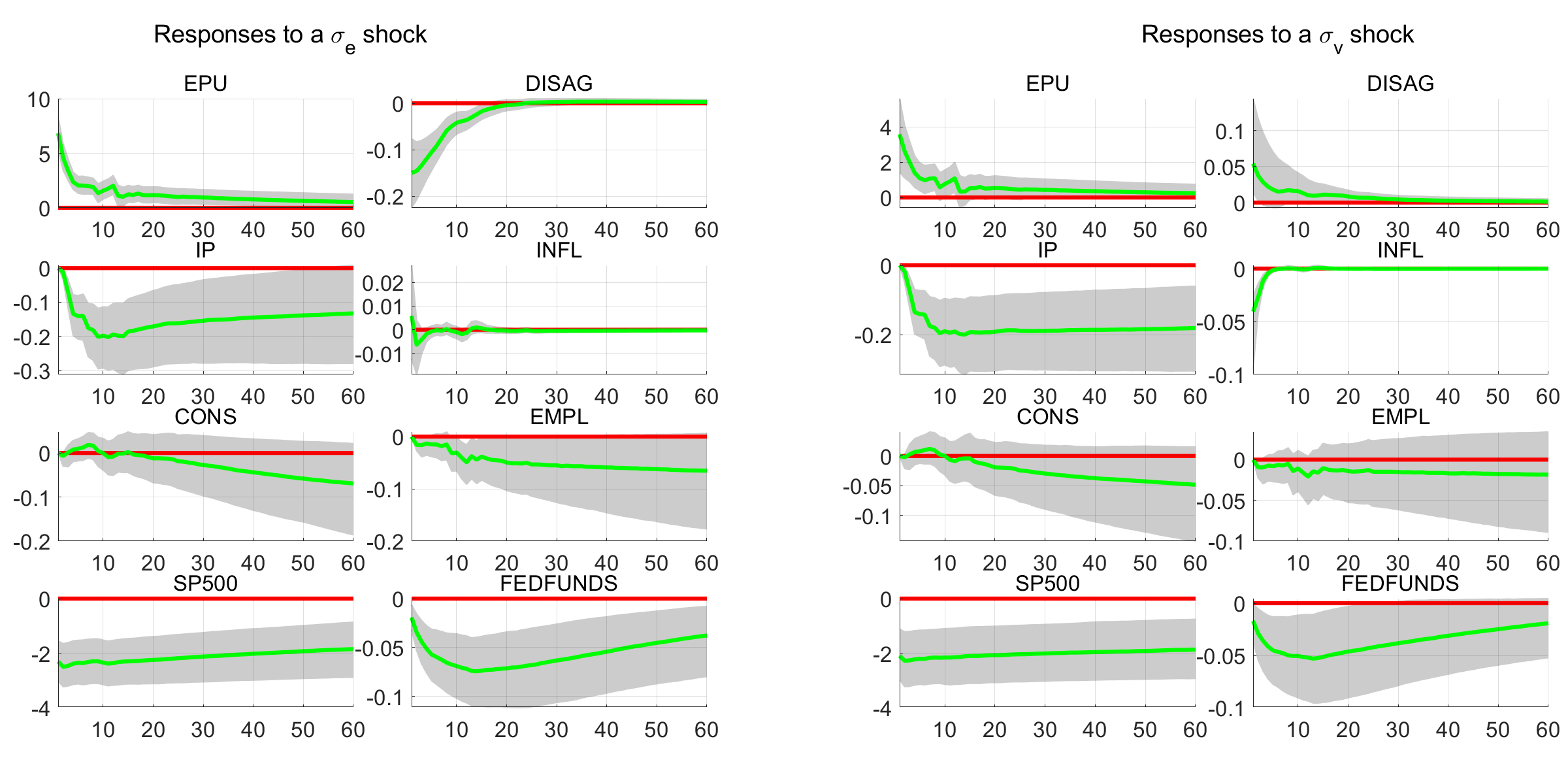}

\end{center}
\small \emph{Notes}: The figure shows impulse responses from a eight-variable VAR system on Economic Policy Uncertainty (EPU), disagreement index (DISAG),  Industrial production (IP), private consumption (CONS), Consumer Inflation (INFL), employment (EMPL), S\&P 500 index (SP500), Federal funds rate (FEDFUNDS). The shaded gray areas are the 16\% and 84\% posterior bands generated from the posterior distribution of VAR parameters. The units of the vertical axes are percentage deviations, while the horizontal axes reports time measured in months.
\end{figure}

\paragraph{\textbf{Complete IRFs from VAR model with DISAG-E.}} Figure \ref{fig_IRF_macro6_full} below displays the complete set of IRFs from the VAR specification with the entropy measure, DISAG-E, used as the disagreement indicator and discussed in section \ref{sec_robustness} of the main body.

\begin{figure}[h!]
\caption{Disagreement entropy measure. Agreed  $\sigma_{\varepsilon}$ (left) versus disagreed $\sigma_{v}$ (right) uncertainty.}\label{fig_IRF_macro6_full}
\begin{center}
\includegraphics[width=0.95\textwidth,, trim={1cm .5cm 0 .4cm}]{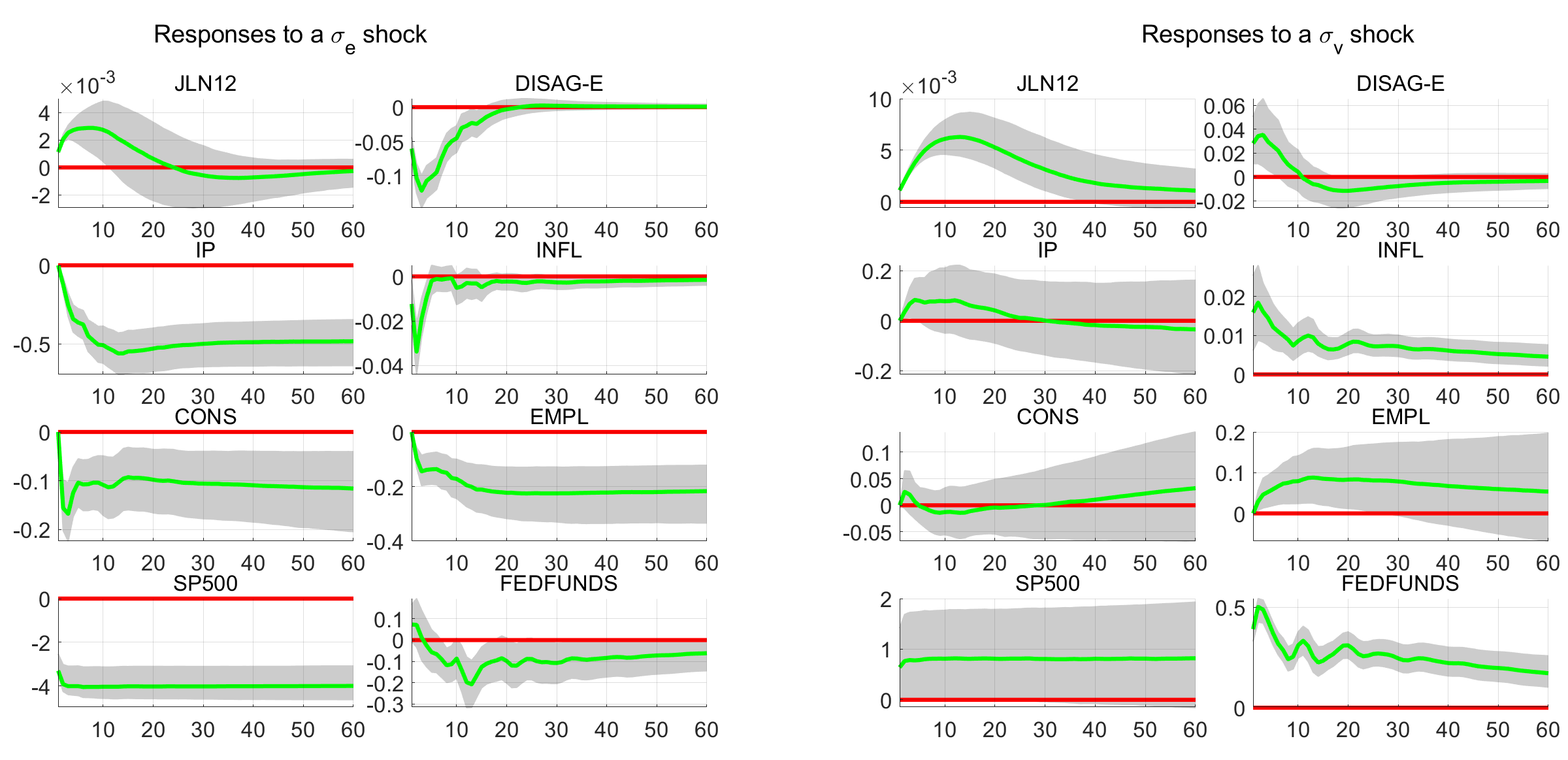}

\end{center}
\small \emph{Notes}: The figure shows impulse responses from a eight-variable VAR system on JLN 12-month ahead uncertainty indicator (JLN12), disagreement index (DISAG-E),  Industrial production (IP), private consumption (CONS), Consumer price inflation (INFL), employment (EMPL), S\&P 500 index (SP500), Federal funds rate (FEDFUNDS). The shaded gray areas are the 16\% and 84\% posterior bands generated from the posterior distribution of VAR parameters. The units of the vertical axes are percentage deviations, while the horizontal axes reports time measured in months.
\end{figure}
\paragraph{\textbf{Complete IRFs from VAR model with DISAG-L.}} Figure \ref{fig_IRF_macro7_full} below displays the complete set of IRFs from the VAR specification with Lacy measure, DISAG-L, used as the disagreement indicator and discussed in section \ref{sec_robustness} of the main body.

\begin{figure}[h!]
\caption{Disagreement Lacy measure. Agreed  $\sigma_{\varepsilon}$ (left) versus disagreed $\sigma_{v}$ (right) uncertainty.}\label{fig_IRF_macro7_full}
\begin{center}
\includegraphics[width=0.95\textwidth,, trim={1cm .5cm 0 .4cm}]{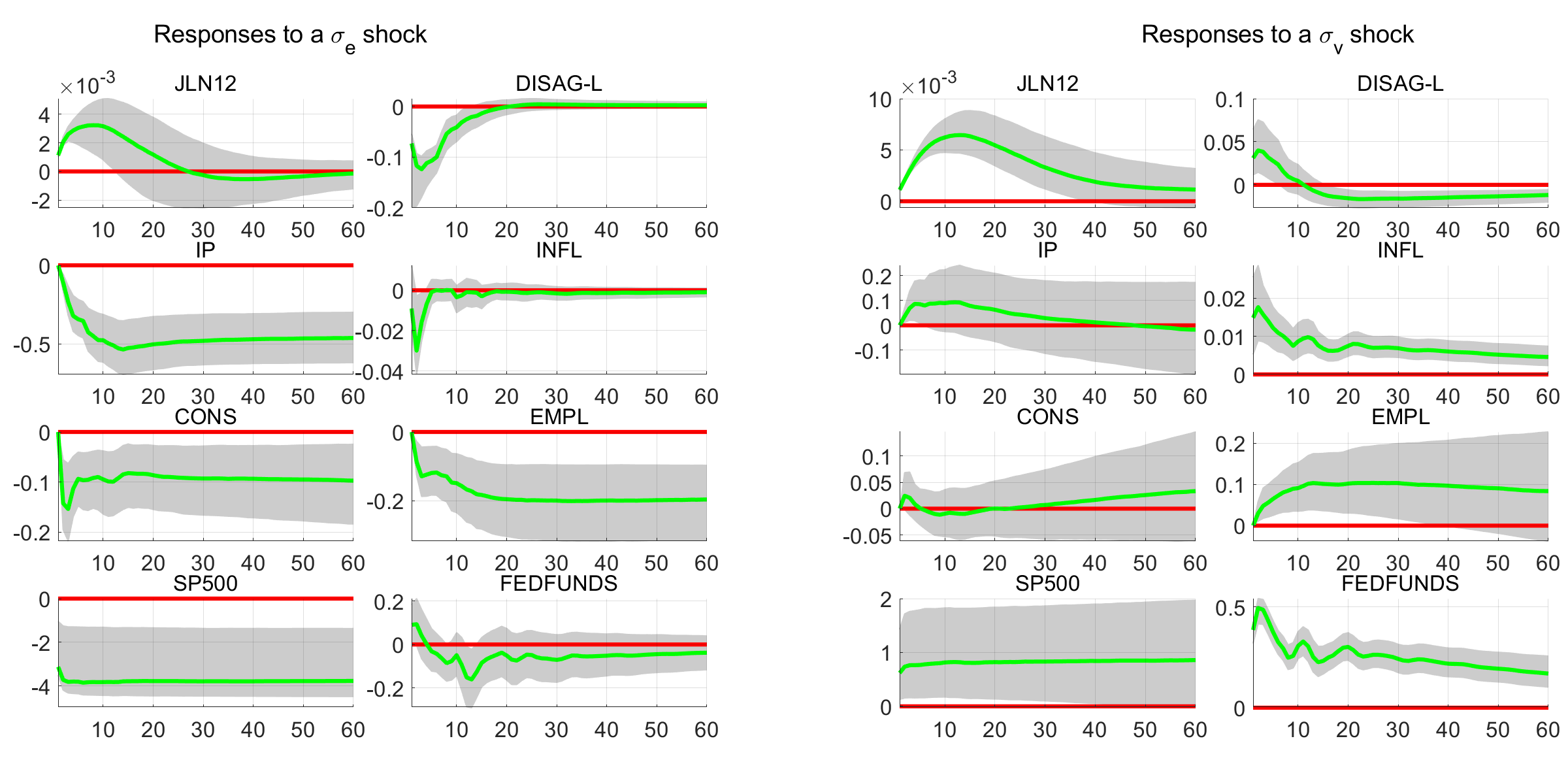}

\end{center}
\small \emph{Notes}: The figure shows impulse responses from a eight-variable VAR system on JLN 12-month ahead uncertainty indicator (JLN12),disagreement index (DISAG-L),  Industrial production (IP), private consumption (CONS), Consumer price inflation (INFL), employment (EMPL), S\&P 500 index (SP500), Federal funds rate (FEDFUNDS). The shaded gray areas are the 16\% and 84\% posterior bands generated from the posterior distribution of VAR parameters. The units of the vertical axes are percentage deviations, while the horizontal axes reports time measured in months.
\end{figure}

\begin{figure}[h!]
\caption{Disagreement NEWS. Agreed  $\sigma_{\varepsilon}$ (left) versus disagreed $\sigma_{v}$ (right) uncertainty.}\label{fig_IRF_macro_news_full}
\begin{center}
\includegraphics[width=0.95\textwidth,, trim={1cm .5cm 0 .4cm}]{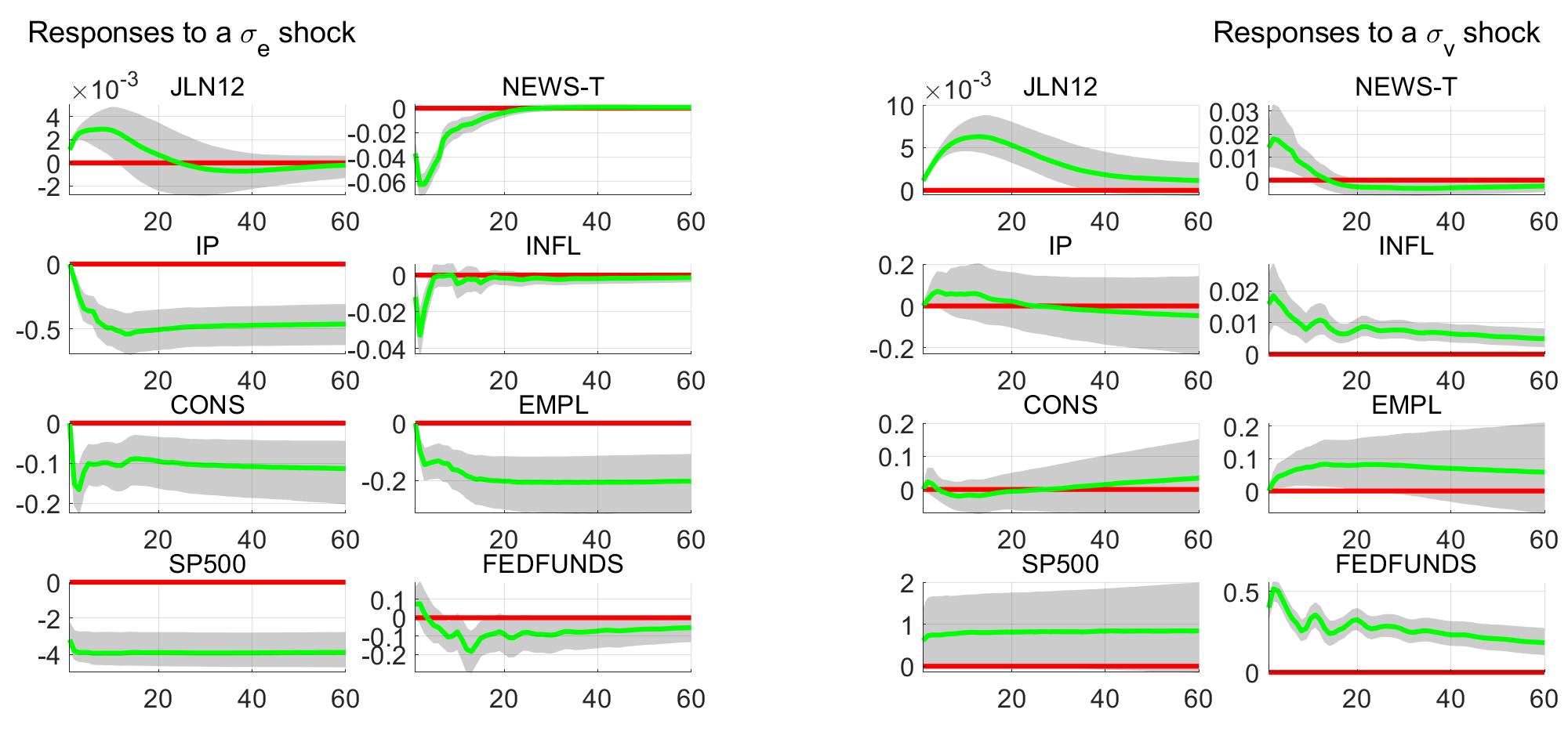}

\end{center}
\small \emph{Notes}: The figure shows impulse responses from a eight-variable VAR system on JLN 12-month ahead uncertainty indicator (JLN12),disagreement index about NEWS (NEWS-T),  Industrial production (IP), private consumption (CONS), Consumer price inflation (INFL), employment (EMPL), S\&P 500 index (SP500), Federal funds rate (FEDFUNDS). The shaded gray areas are the 16\% and 84\% posterior bands generated from the posterior distribution of VAR parameters. The units of the vertical axes are percentage deviations, while the horizontal axes reports time measured in months.
\end{figure}

\begin{figure}[h!]
\caption{Disagreement BUS5. Agreed  $\sigma_{\varepsilon}$ (left) versus disagreed $\sigma_{v}$ (right) uncertainty.}\label{fig_IRF_macro_bus5_full}
\begin{center}
\includegraphics[width=0.95\textwidth,, trim={1cm .5cm 0 .4cm}]{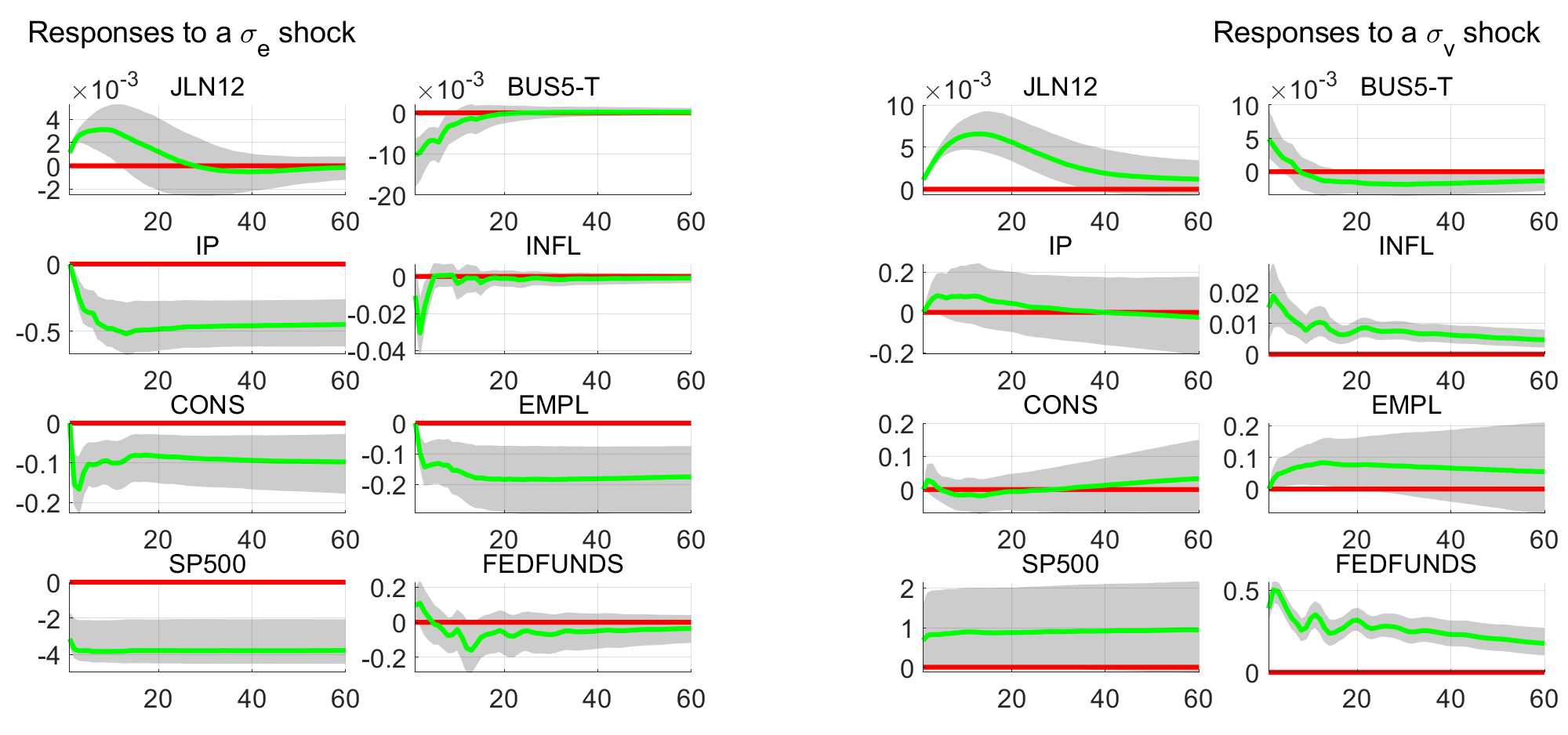}

\end{center}
\small \emph{Notes}: The figure shows impulse responses from a eight-variable VAR system on JLN 12-month ahead uncertainty indicator (JLN12),disagreement index about BUS5 (BUS5-T),  Industrial production (IP), private consumption (CONS), Consumer price inflation (INFL), employment (EMPL), S\&P 500 index (SP500), Federal funds rate (FEDFUNDS). The shaded gray areas are the 16\% and 84\% posterior bands generated from the posterior distribution of VAR parameters. The units of the vertical axes are percentage deviations, while the horizontal axes reports time measured in months.
\end{figure}

\paragraph{\textbf{Individual disagreement (BUS5 and NEWS).}} Our benchmark DISAG index is the first principal component of the five individual disagreement series, described in section 2. This aggregate index captures dispersed consumer views about current and future business conditions. We examine the robustness of our findings when we instead focus on individual disagreement indices. Figures \ref{fig_IRF_macro_news_full} and \ref{fig_IRF_macro_bus5_full} display complete set of IRFs estimated from two specifications where we replace the DISAG indicator in the benchmark VAR with disagreement about NEWS (\textit{News Heard of Recent Changes in Business Conditions}) and BUS5 (\textit{Business Conditions Expected During the Next 5 Years
}), one at a time. The estimated IRFs from those specifications are broadly similar to the those from the benchmark and we do not discuss them further.

\paragraph{\textbf{Labor market conditions and unemployment expectations.}} We enrich our benchmark specification with the following indicators of labor market conditions in addition to employment: the help wanted index --a proxy for labor market tightness, average hourly earnings, and the Michigan consumer expectations about unemployment in the next twelve months.  The left panel in Figure \ref{fig_IRF_labor_markets} shows the IRFs of the enriched VAR model following an innovation to agreed uncertainty. The response of the variables is similar to our benchmark specification, evinced by the significant fall in industrial production and employment; the responses of the additional labor market variables show that following an agreed uncertainty shock depresses the help wanted index for more than sixty months.  Total private consumption displays a depressing effect with an estimated effect which becomes statistically significant in the ten month horizon --this is slightly weaker in comparison to the benchmark specification. The strong positive response of unemployment expectations over the next twelve months signal weak employment prospects expected by households.  Despite the adverse labor market conditions and depressed expectations on labor market prospects, the response of earnings remains statistically insignificant.

The right panel in Figure \ref{fig_IRF_labor_markets} shows that IRFs from an innovation in disagreed uncertainty. The responses of real activity indicators, industrial production, employment and consumption, are statistically insignificant. The dynamic responses of the remaining labor market variables are also statistically insignificant, suggesting this shock has a muted effect on the labor market. Thus, accounting for labor market dynamics in the propagation of uncertainty does not alter our central finding: agreed uncertainty innovations retain the standard adverse effect on economic activity, while disagreed uncertainty innovations remains non-contractionary for economic activity.

\begin{figure}[h!]
\caption{VAR model with additional labor market variables. Agreed $\sigma_{v}$ (left) versus disagreed $\sigma_{\varepsilon}$ (right) uncertainty.}\label{fig_IRF_labor_markets}
\begin{center}
\includegraphics[width=0.95\textwidth,, trim={1cm .5cm 0 .4cm}]{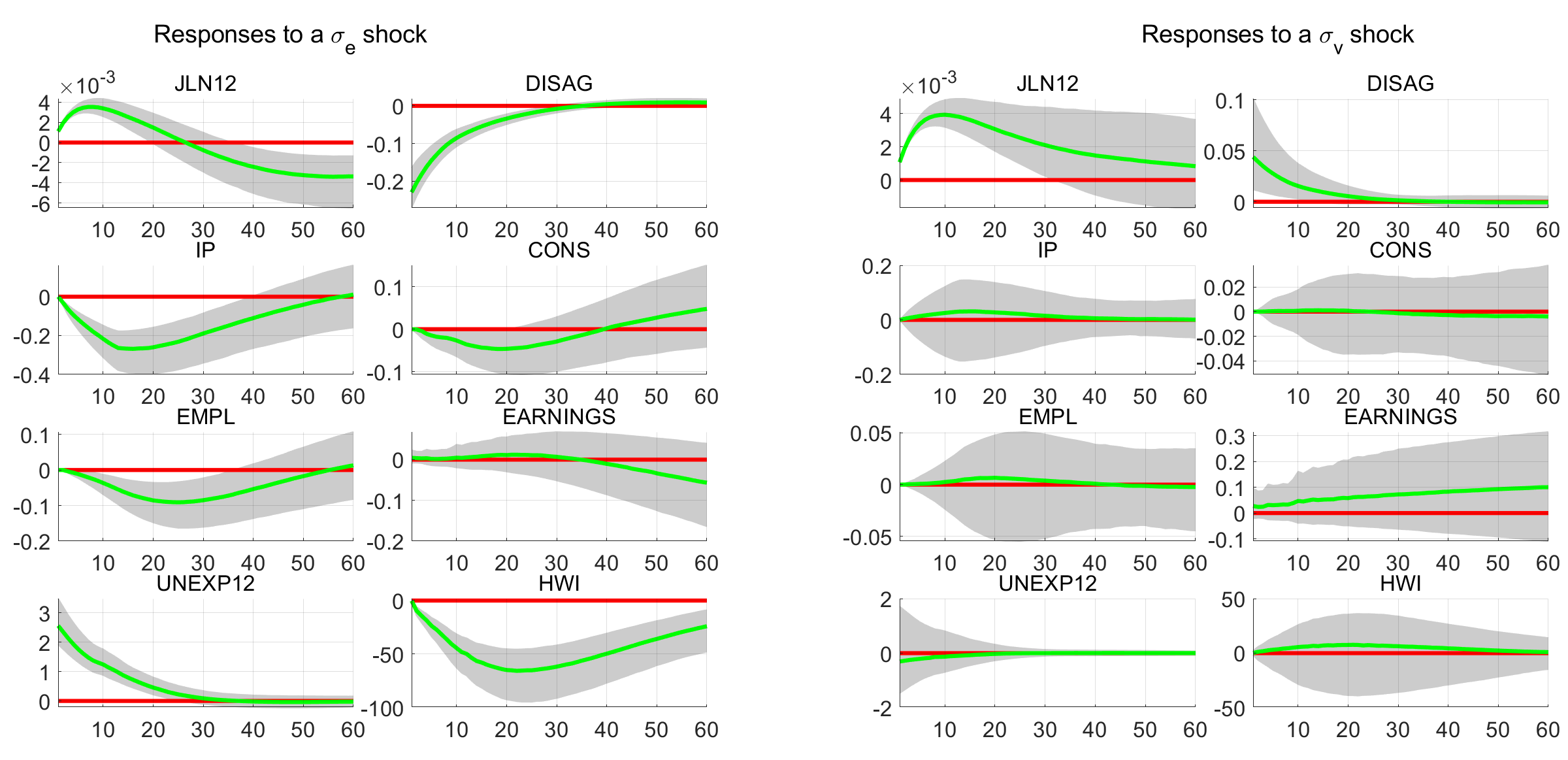}

\end{center}
\small \emph{Notes}: The figure shows impulse responses from a eight-variable VAR system on JLN 12-month ahead uncertainty indicator (JLN12), disagreement index (DISAG),  Industrial production (IP), private consumption (CONS), employment (EMPL), hourly earnings (EARNINGS), 12 month ahead unemployment expectations (UNEXP12), Help wanted index (HWI). The shaded gray areas are the 16\% and 84\% posterior bands generated from the posterior distribution of VAR parameters. The units of the vertical axes are percentage deviations, while the horizontal axes reports time measured in months.
\end{figure}

\begin{figure}[h!]
\caption{Benchmark model. Agreed  $\sigma_{\varepsilon}$ (left) versus disagreed $\sigma_{v}$(right) uncertainty.}\label{fig_IRF_benckmark_discons}
\begin{center}
\includegraphics[width=0.95\textwidth,, trim={1cm .5cm 0 .4cm}]{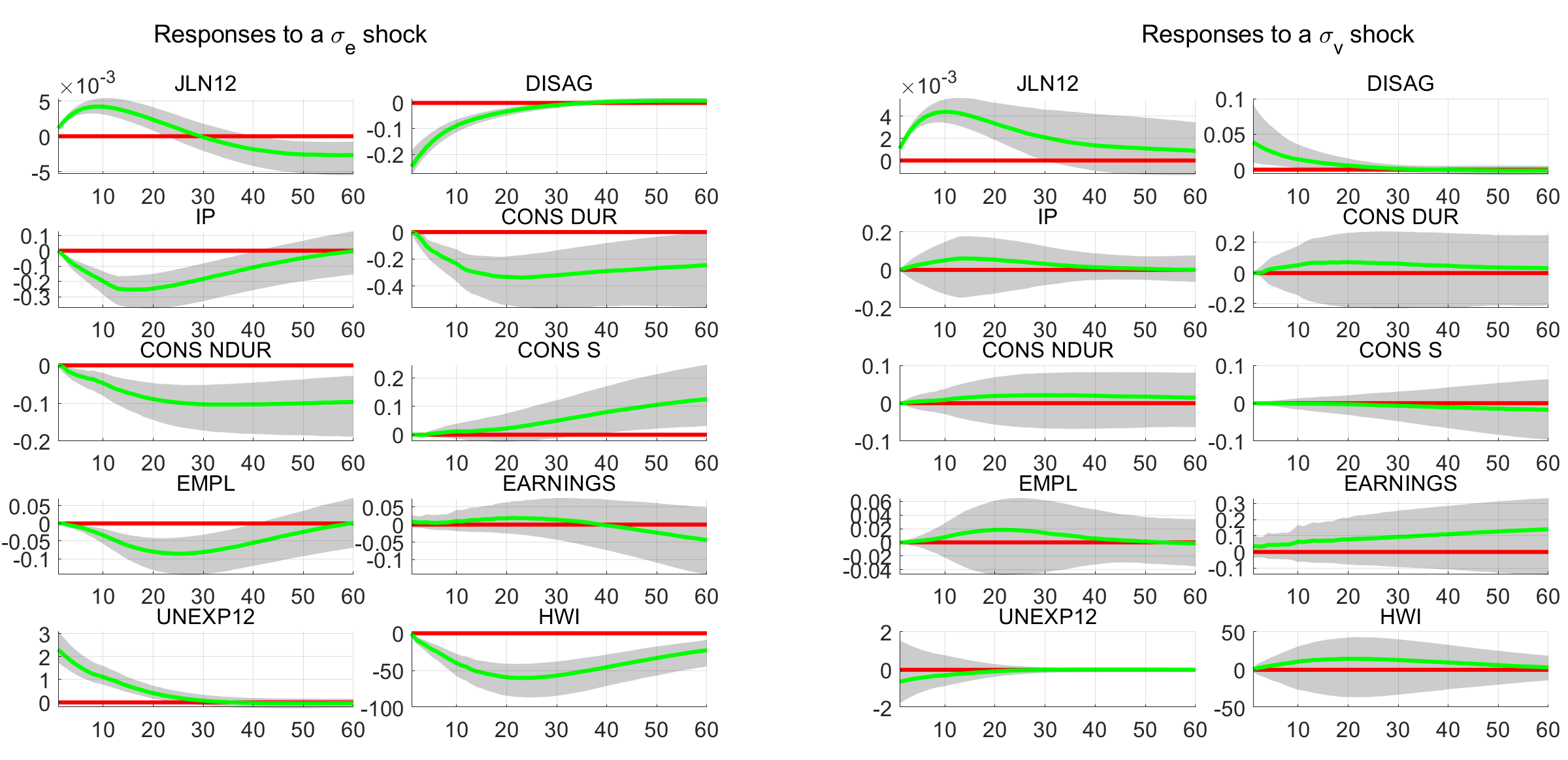}

\end{center}
\small \emph{Notes}: The figure shows impulse responses from a ten-variable VAR system on JLN 12-month ahead uncertainty indicator (JLN12), disagreement index (DISAG),  Industrial production (IP), consumption durables (CONS DUR), consumption non-durables (CONS NDUR), consumption services (CONS S), employment (EMPL), hourly earnings (EARNINGS), 12 month ahead unemployment expectations (UNEXP12), Help wanted index (HWI). The shaded gray areas are the 16\% and 84\% posterior bands generated from the posterior distribution of VAR parameters. The units of the vertical axes are percentage deviations, while the horizontal axes reports time measured in months.
\end{figure}

\paragraph{\textbf{Disaggregated consumption.}} Our benchmark specification includes total private consumption. It is interesting to examine the dynamic responses of different consumption components.  To this end we estimate a VAR specification which builds on the one estimated with the additional labor market variables above, where we introduce real services, non-durables and durables consumption. One would expect that uncertainty would mostly impact large durables purchases.  For example, \cite{Eberly94} emphasizes the option to delay purchases of durable goods in an environment of elevated uncertainty, which in theory would depress durables spending, although the effects on non-durables and services might be weaker.  Similarly, \cite{bernanke1983irreversibility} and \cite{romer1990great} show that uncertainty significantly delays consumer spending on durable purchases by increasing the option value of waiting. \cite{bertola2005uncertainty} provide extensive evidence on the sensitivity of durable goods spending to uncertainty. Figure \ref{fig_IRF_benckmark_discons} displays the IRFs from this specification. Consumption non-durables and durables display a significant depressing effect following an innovation in agreed uncertainty. By contrast, the response of consumer services does not display any negative effects, but interestingly a positive response which becomes statistically significant after the forty month horizon.  Thus, it appears services consumption mitigates the depressing effects on total consumption following an agreed uncertainty shock.  By contrast the responses of the disaggregated consumption components are not statistically significant following a disagreed uncertainty innovation.

\paragraph{\textbf{Results based on a VAR specification using quarterly data.}} We estimate a VAR specification using a sample with a quarterly frequency. This allows to examine the dynamic responses to GDP and investment that are only available in this frequency. The quarterly sample is 1960Q1 to 2020Q4.  The results from this specification are broadly consistent with our benchmark results which are based on a monthly sample. A shock to agreed uncertainty depresses the economic activity indicators, namely real GDP, consumption, employment and private non residential investment. By contrast, real GDP, consumption and employment, do not respond in a statistically significant manner following a disagreed uncertainty shock. Private non residential investment does not respond in the short term but responds negatively after the six month horizon. 

\begin{figure}[h!]
\caption{Benchmark model. Agreed $\sigma_{v}$ (left) versus disagreed $\sigma_{\varepsilon}$ (right) uncertainty.}\label{fig_IRF_labor_markets}
\begin{center}
\includegraphics[width=0.95\textwidth,, trim={1cm .5cm 0 .4cm}]{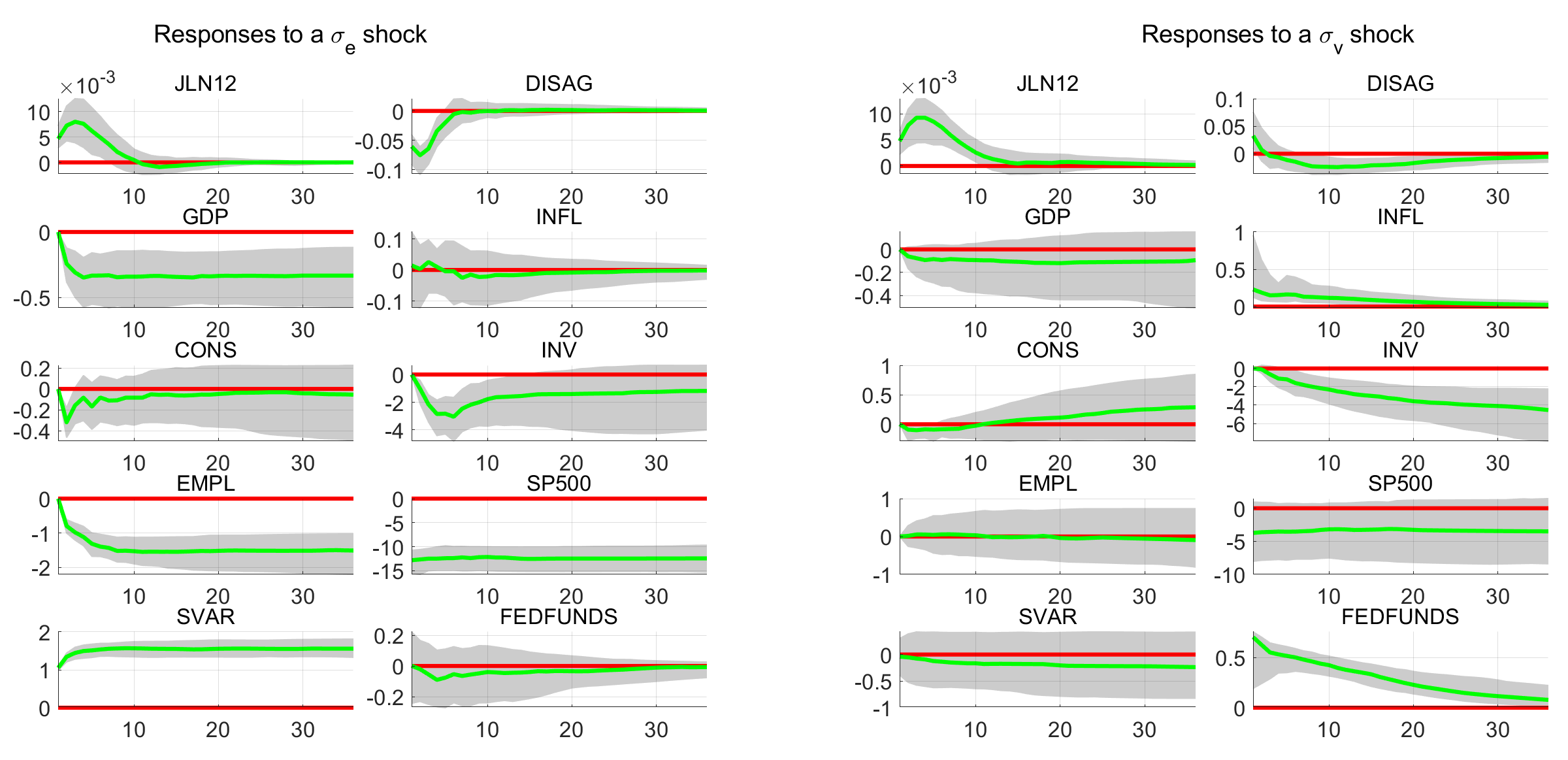}

\end{center}
\small \emph{Notes}: The figure shows impulse responses from a quarterly ten-variable VAR system using JLN 12-month ahead uncertainty indicator (JLN12), disagreement index (DISAG),  GDP, private consumption (CONS), inflation based on the GDP deflator (INFL), private non residential fixed investment (INV), employment (EMPL), S\&P 500 index, Federal funds rate (FEDFUNDS), and the stock market variance from the S\&P Index (SVAR) computed as the sum of squared daily returns on S\&P 500. Sample period 1960Q1 to 2020Q4. The shaded gray areas are the 16\% and 84\% posterior bands generated from the posterior distribution of VAR parameters. The units of the vertical axes are percentage deviations, while the horizontal axes reports time measured in quarters.
\end{figure}

\end{document}